\documentclass{article}
\pdfoutput=1

\usepackage{neurips_data_2023}

\usepackage{amsmath}
\usepackage[utf8]{inputenc} 
\usepackage[T1]{fontenc}    
\usepackage{url}            
\usepackage{booktabs}       
\usepackage{amsfonts}       
\usepackage{nicefrac}       
\usepackage{microtype}      
\usepackage{xcolor}         
\usepackage{multirow}
\usepackage{ulem}
\usepackage{mathabx}
\usepackage[notransparent]{svg}
\usepackage{subcaption}

\usepackage{longtable}
\usepackage{tabu}
\usepackage{tabularx}
\usepackage{setspace}
\usepackage{lipsum}
\usepackage{color}
\usepackage{relsize}%
\usepackage{graphicx}
\usepackage{wrapfig}
\usepackage{enumitem}
\usepackage{hhline}
\usepackage{boldline}
\usepackage{bm}
\usepackage{bbm}
\usepackage{import}

\usepackage[export]{adjustbox}

\usepackage{hyperref}       

\usepackage{algorithm} 
\usepackage{algpseudocode}
\hypersetup{
    colorlinks,
    linkcolor={red!75!black},
    citecolor={blue!75!black},
    urlcolor={blue!80!black}
}

\newcommand{\revision}[1]{{\color{black}#1}}

\title{Boosting Studies of Multi-Agent Reinforcement Learning  on Google Research Football Environment: the Past, Present, and Future}

\author{{\bf Yan Song\thanks{Equal contribution}$^{~~1}$ , He Jiang\footnotemark[1]$^{~~2}$, Haifeng Zhang\thanks{Correspondence to <haifeng.zhang@ia.ac.cn>}$^{~~1}$, Zheng Tian$^{3}$, Weinan Zhang$^{4}$, Jun Wang$^{5}$} \\
\vspace{0.05 cm}\\
{ $^1$Institute of Automation, Chinese Academy of Science, Beijing, China}\\
{ $^2$Digital Brain Lab, Shanghai, China } \\
{ $^3$Shanghai Tech University, Shanghai, China}\\
{ $^4$Shanghai Jiao Tong University, Shanghai, China}\\
{ $^5$University College London, London, UK}}

\begin{document}

\maketitle

\begin{abstract}

Even though Google Research Football (GRF) was initially benchmarked and studied as a single-agent environment in its original paper \cite{kurach2020google}, recent years have witnessed an increasing focus on its multi-agent nature by researchers utilizing it as a testbed for Multi-Agent Reinforcement Learning (MARL). However, the absence of standardized environment settings and unified evaluation metrics for multi-agent scenarios hampers the consistent understanding of various studies. Furthermore, the challenging \textit{5 vs 5} and \textit{11 vs 11} full-game scenarios have received limited thorough examination due to their substantial training complexities. To address these gaps, this paper extends the original environment by not only standardizing the environment settings and benchmarking cooperative learning algorithms across different scenarios, including the most challenging full-game scenarios, but also by discussing approaches to enhance football AI from diverse perspectives and introducing related research tools.  Specifically, we provide a distributed and asynchronous population-based self-play framework with diverse pre-trained policies for faster training, two football-specific analytical tools for deeper investigation, and an online leaderboard for broader evaluation. The overall expectation of this work is to advance the study of Multi-Agent Reinforcement Learning on Google Research Football environment, with the ultimate goal of benefiting real-world sports beyond virtual games.
\end{abstract}

\setcounter{footnote}{0}

\section{Introduction}
\label{sec: intro}

\begin{figure}[htbp]
    \vspace{-3pt}
  \centering
   \begin{subfigure}{0.3\linewidth}
        \includegraphics[width=\linewidth]{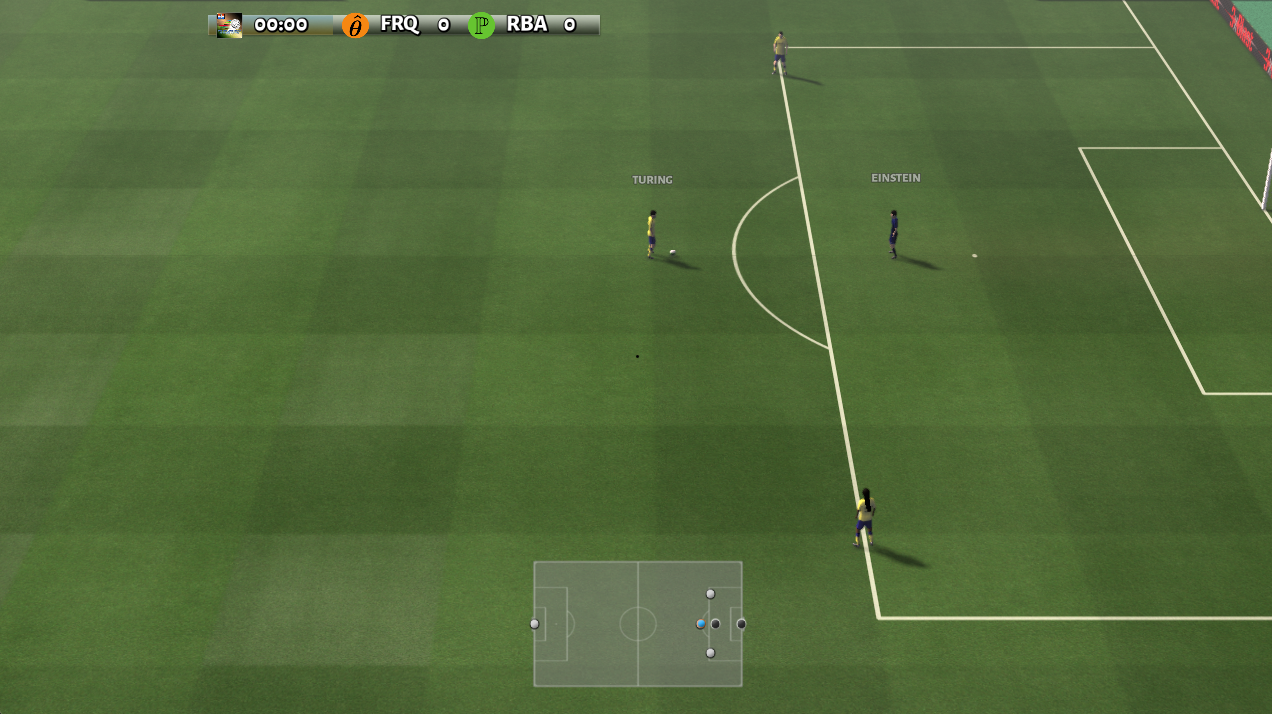}
        \caption{\textit{3 vs 1 with keeper}}
   \end{subfigure}
   \begin{subfigure}{0.3\linewidth}
        \includegraphics[width=\linewidth]{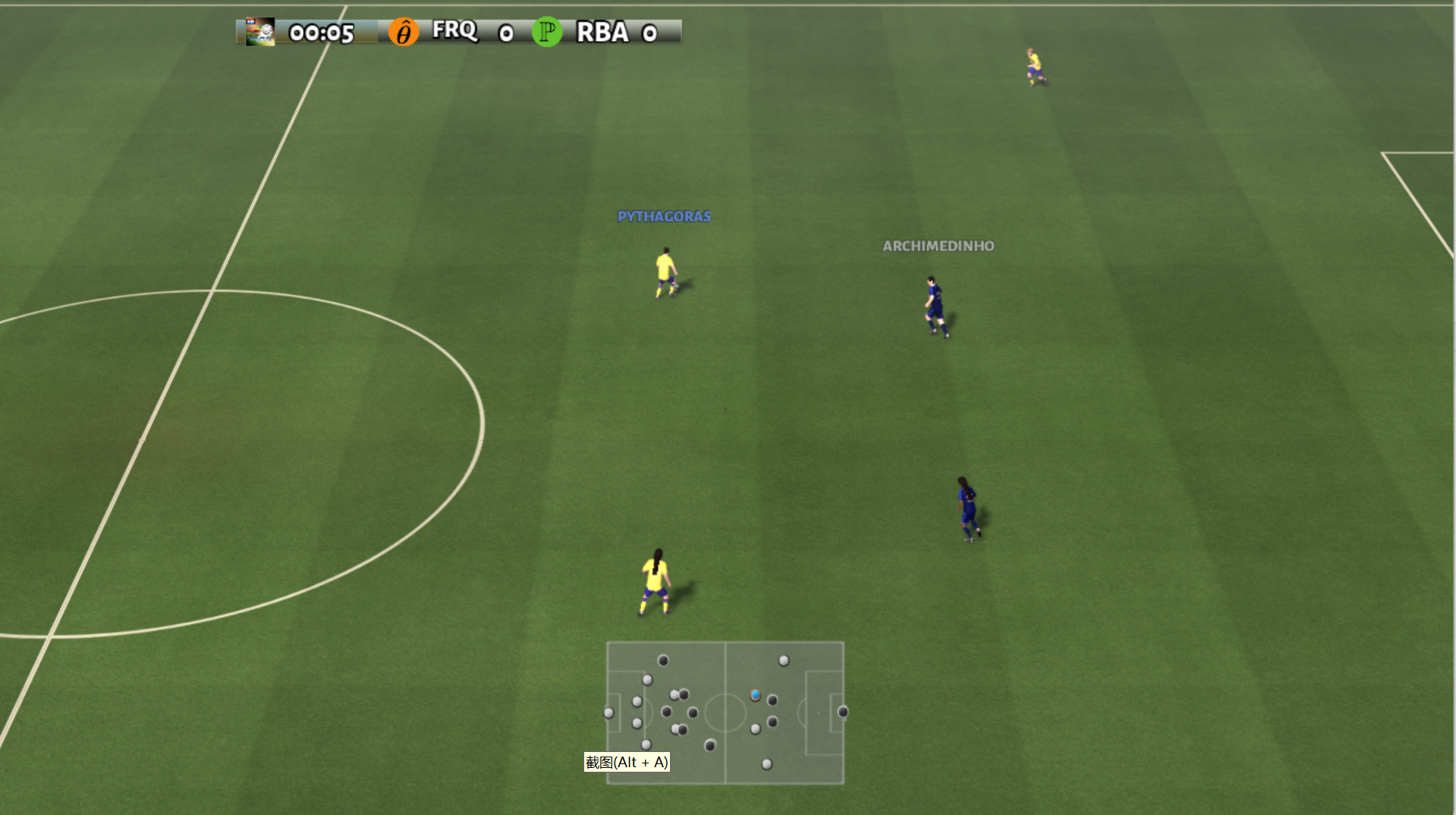}
        \caption{\textit{counterattack hard}}
   \end{subfigure}
    \begin{subfigure}{0.3\linewidth}
      \includegraphics[width=\textwidth]{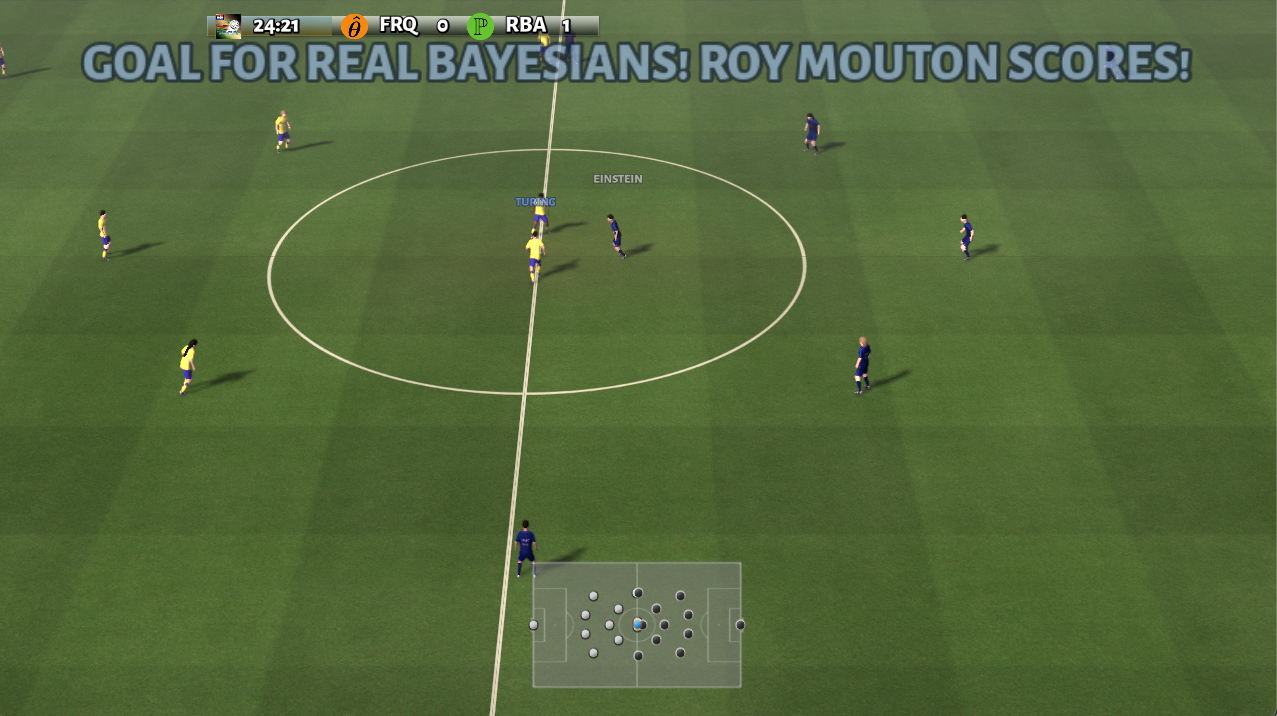}
      \caption{\textit{11 vs 11 full-game}}
      \label{fig:snapshot 11v11}
   \end{subfigure}
  \caption{Snapshots of different Google Research Football scenarios}
  \label{figure:snapshot}
  \vspace{-3pt}
\end{figure}

Soccer is a universally enjoyed sport that is played worldwide. It has proven to be valuable for exploring multi-agent reinforcement learning through research conducted in various environments, including \textit{Markov Soccer Game}\cite{littman1994markov}, the \textit{RoboCup Soccer Simulator} \cite{Kalyanakrishnan2006HalfFO}, \textit{Google Research Football (GRF)} \cite{kurach2020google}, \textit{rSoccer} \cite{Martins2021rSoccerAF}, \textit{DeepMind MuJoCo Multi-Agent Soccer Environment} \cite{Liu2021FromMC}. Among them, GRF stands out due to its ability to emulate realistic soccer games (Figure \ref{figure:snapshot}) like FIFA and Real Football, while allowing algorithms to control all players on the field. It places emphasis on high-level actions rather than low-level controls, enabling the examination of challenging scenarios, such as the real-world \textit{11 vs 11} scenario (Figure \ref{fig:snapshot 11v11}). Consequently, GRF presents an appealing platform for studying both multi-agent cooperation and competition problems.

The GRF environment provides support for both single-agent and multi-agent settings. However, its original paper merely benchmarks single-agent scenarios \cite{kurach2020google}. In this setting, only one player is controlled at a time, and the player to control is determined by an underlying heuristic. The significance of the single-agent setting was highlighted during the Kaggle competition, \textit{Google Research Football with Manchester City F.C.}, in 2020 \cite{kaggle2020}. This competition attracted over a thousand teams to compete online. Top-performing teams such as \textit{Wekick}, \textit{Saltyfish} and \textit{liveinparis} made a substantial impact on subsequent studies of multi-agent scenarios through their ideas, code, and datasets.


In recent years, more researchers have directed their attention towards the multi-agent nature of
GRF, considering it an essential testbed for their algorithms \cite{li2021celebrating,yu2022surprising,wen2022multi,wang2023order}. Inspired by the Kaggle
Competition, \textit{IEEE Conference on Games Football AI Competition} \cite{cog2022} was held in 2022, which focused on multi-agent scenarios. The growing popularity of GRF may also be attributed to the extensive exploration and saturation of many other multi-agent environments, which are no longer suitable for further academic research \cite{ellis2022smacv2}. For instance, the well-known MARL benchmark, \revision{StarCraft Multi-Agent Challenge (SMAC)}, has been extensively investigated and reported to achieve near-optimal performance in numerous studies \cite{Papoudakis2020BenchmarkingMD, wen2022multi,wang2023order}. Some criticism has also been directed towards SMAC due to its lack of stochasticity \cite{ellis2022smacv2}. On the contrary, GRF has strong stochasticity as discussed in its original work \cite{kurach2020google}. Consequently, the MARL community has expressed a need for a new and challenging testbed, and Google Research Football emerges as a desirable choice.


Researchers employ various approaches to evaluate their novel algorithms on GRF. The majority of
studies assess their algorithms on 1 to 5 simplified soccer tasks known as academy scenarios, such as
corner or counterattack \cite{niu2021multi,roy2020promoting,pu2022cognition, li2021celebrating,wen2022multi,ruan2022gcs,fu2022revisiting,hao2023boosting,zhang2023stackelberg}. Some researchers even develop their own customized tasks for evaluation purposes \cite{liu2021semantic,wang2022individual,li2021celebrating}. More recently, a few studies have ventured into
tackling the demanding full-game scenarios involving more players, larger state spaces, and longer
horizons \cite{huang2021tikick, lin2023tizero, wang2023order, song2023empirical}, indicating a growing interest in this set of challenges.


However, due to the considerable variations in scenario settings and evaluation metrics employed by
different studies, it becomes challenging to comprehensively understand and compare performance
across works. Therefore, this paper aims to standardize the scenario settings and provide benchmark
results for frequently utilized multi-agent GRF scenarios. Additionally, as pointed out in \cite{song2023empirical}, overfitting to a fixed opponent does not give a generally strong football AI. In order
to advance future research on cooperative and competitive MARL, we discuss potential ways of improving our football AI and introduce corresponding research tools.

In summary, our contributions can be outlined as follows:

\begin{enumerate}[topsep=0pt,itemsep=0ex,partopsep=0ex,parsep=0ex]
    \item Standard Settings: We establish standardized settings for both academy and full-game scenarios in the multi-agent context of Google Research Football, following some best practices.
    \item Benchmark and Analysis: We conduct an empirical comparison of representative
MARL algorithms on the standardized tasks, accompanied by a comprehensive analysis of
various algorithmic designs. 
    \item Research Tools: To the best of our knowledge, we are the first to release a comprehensive set of research tools for GRF studies\footnote{All our code, including experiment settings and research tools, are available at \url{https://github.com/jidiai/GRF_MARL.git}.}. These tools encompass an efficient distributed population-based self-play training framework, diverse pre-trained models, two football analytical tools, and an online leaderboard.
\end{enumerate}

Overall, we expect to boost studies of MARL on Google Research Football environment. 
Hopefully, one day, related research could go beyond virtual games and brings benefits to real-world sports.

\revision{
The overall structure of our paper is as follows:
\begin{enumerate}[topsep=0pt,itemsep=0ex,partopsep=0ex,parsep=0ex]
    \item The Past: Early works on GRF mainly study the single-agent setting and the later research on the multi-agent scenarios uses inconsistent environmental settings. In Section \ref{sec: intro} and \ref{sec: related work}, we discuss the previous studies on GRF and argue for the need of a unified evaluation.
    \item The Present: More recent studies start to use GRF as a testbed for cooperative MARL but a benchmark is still lacking. In Section \ref{section:Scenarios} and \ref{section:Experiments}, we provide a fully-reproducible MARL benchmark for cooperative tasks on GRF with detailed experiments and analysis.
    \item The Future: We aim to go beyond cooperative tasks to competitive tasks and learn sophisticated human-like strategies from real-world matches. In Section \ref{section:Building Stronger Football AI} and \ref{sec: conclusion}, we provide convenient tools for future studies and discuss potential limitations and research directions.
\end{enumerate}

}

\section{Related Work}
\label{sec: related work}

The primary work related to our research is the original GRF paper \cite{kurach2020google}, which introduces the environment but only studies single-agent scenarios, where only a single active player of a team is controlled. There are some works focusing on solving competitive tasks in this single-agent setting. For example, \revision{Liu et.al \cite{liu2021unifying} beat the strongest built-in AI with a novel PSRO \cite{psro} algorithm that combines both behavior diversity and response diversity. Though such a competitive task in a single-agent scenario is also multi-agent by definition, our work mainly focuses on multi-agent scenarios, where multiple players of a single team are controlled by agents, which renders the cooperation between teammates important.} Our research is motivated by three key factors.

Firstly, we recognize the necessity of benchmarking multi-agent cooperative learning due to its
growing popularity, coupled with the absence of standardized benchmarks. Several recently proposed
multi-agent reinforcement learning algorithms, such as Multi-Agent PPO (MAPPO)\cite{yu2022surprising}, Multi-Agent Transformer (MAT) \cite{wen2022multi} and A2PO \cite{wang2023order}, all choose to validate their superior performance on GRF multi-agent tasks. Works on MARL communications \cite{niu2021multi,ruan2022gcs} and behavioral diversity \cite{li2021celebrating,wang2022individual} also utilize GRF multi-agent tasks as a testbed. However, these studies often adopt different experimental settings, including varying sets of scenarios, environment configurations, and evaluation metrics, making it challenging to compare and interpret results consistently. Our work addresses these challenges by first standardizing the test suite and subsequently benchmarking representative algorithms on it.


Furthermore, we find the need for more attention to the full-game scenarios. Currently, only a few studies have explored methods in these challenging settings. TiKick \cite{huang2021tikick} claims to be the first work tackling the \textit{11 vs 11} scenario, but it actually allows agents to use a default built-in action, which simply delegates the control of players to the underlying heuristic. \revision{Recently, TiZero \cite{lin2023tizero} and Fictitious Cross-Play (FXP) \cite{xu2023fictitious} both beat the strongest built-in AI in the \textit{11 vs 11} setting by respectively learning from a curriculum of self-play and setting a counter-policy population.} Meanwhile, Wang et al. \cite{wang2023order} also outperformed the toughest built-in AI in full games learning training from scratch but with a complex reward function and asynchronous training. Additionally, these papers primarily focus on their new algorithms, providing limited analysis of GRF. In contrast, our work extensively benchmarks and analyzes full-game scenarios using only the official SCORING and CHECKPOINT rewards provided by the game, without employing any additional tricks.


Lastly, we realize the importance of sharing research tools to facilitate further studies in cooperative and competitive learning on GRF. Most related studies, such as \cite{li2021celebrating,niu2021multi,ruan2022gcs,yu2022surprising,fu2022revisiting}, only release
codes limited to simple academy scenarios. Other papers like \cite{huang2021tikick,lin2023tizero} have only released evaluation scripts with trained models so far. In contrast, we provide a comprehensive toolkit covering
training, evaluation, and analysis.


\section{Scenarios}
\label{section:Scenarios}

We carefully keep scenarios that are suitable for multi-agent cooperation from the original set of GRF scenarios to form our benchmark suite (Table \ref{table:scenarios}). We strive to make minimal modifications to these scenarios while adhering to best practices. 
\revision{These selected ones are those commonly used in previous MARL research work to examine the performance of their proposed algorithm. Meanwhile, they also cover different levels of cooperation. For instance, \textit{3 vs 1 with keeper}, \textit{pass and shoot with keeper}, and \textit{run pass and shoot with keeper} are three small-scaled tasks focusing on short-horizon offensive strategy learning and only involve a small group (3-4) of players on the pitch. Whereas scenarios like \textit{corner}, \textit{counterattack easy} and \textit{counterattack hard}, require controlling all players in a team, though only a fraction of them may actually play a critical role. On the other hand, full-game settings such as \textit{5 vs 5} and \textit{11 vs 11} emphasize more on long-horizon planning, and agents are required to consider both offense and defense strategies. Besides, We also exclude academy scenarios involving only one attacking player because there is no cooperation involved. }

There are also some guidelines governing our modification to these scenarios:

\begin{enumerate}[topsep=0pt,itemsep=0ex,partopsep=0ex,parsep=0ex]
    \item To save computational resources, we choose one difficulty level for each scenario, because the difficulty level mainly controls the
reaction time of the built-in AI rather than its strategy. 
    \item Given the distinct action set of the goalkeeper (GK) compared to other non-GK players,
we delegate GK control to the built-in AI in all scenarios. We utilize the default action
set with 19 actions for all other players. We do not recommend using alternative official
action sets, such as action set v2 and the full action set adopted in works \cite{huang2021tikick} and \cite{fu2022revisiting} respectively. These alternative sets include a default built-in action that delegates action selection to the underlying heuristic, which contradicts the purpose of using reinforcement learning to learn
complex controls.
    \item In academy scenarios primarily designed for studying attacks, we end the episode immediately upon possession exchange. This approach alleviates the need for algorithms to consider defensive strategies. For full-game scenarios, we ensure fairness by forcing the two teams to exchange sides at halftime.
\end{enumerate}

Regarding evaluation metrics, we advocate for win rates as they are better normalized and more frequently used in previous studies. Further,
winning is typically the primary objective in a match, albeit with various strategies to achieve it.

\begin{table}[htbp]
  \vspace{-5mm}
  \caption{Scenario name, description and line-up illustration for each benchmark scenario (\textcolor{red}{\huge$\blacktriangleup$}: left-team player; \textcolor{blue}{\huge$\blacktriangleup$}: right-team player; {\huge$\bullet$}: ball).  }\label{table:scenarios}
  \centering
  \begin{spacing}{1.5}
    \begin{tabularx}{\textwidth} {
        | >{\centering\arraybackslash\hsize=.24\hsize}X
        | >{\centering\arraybackslash\hsize=.24\hsize}X
        | >{\centering\arraybackslash\hsize=.24\hsize}X
        V{2} >{\centering\arraybackslash\hsize=.02\hsize}X
        | >{\centering\arraybackslash\hsize=.24\hsize}X
        V{2}}
        \hlineB{2}
        \begin{minipage}[c][0.15\columnwidth][c]{0.2\columnwidth}
		\centering
		\raisebox{-.5\height}{\includegraphics[width=\linewidth]{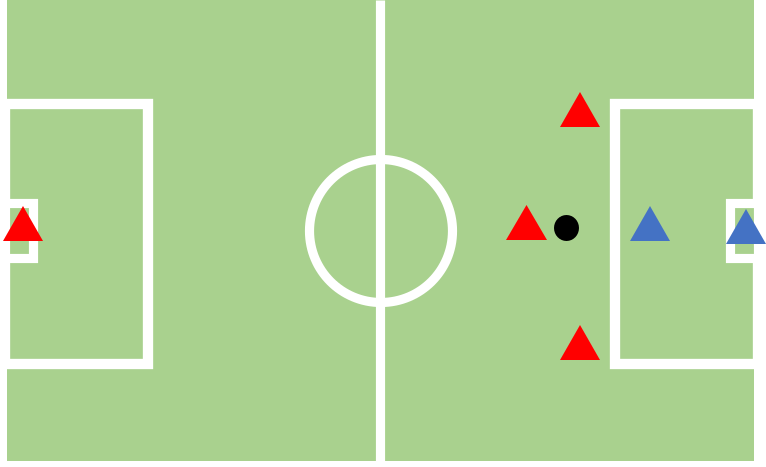}}
	  \end{minipage}
        &
        \begin{minipage}[c][0.15\columnwidth][c]{0.2\columnwidth}
		\centering
		\raisebox{-.5\height}{\includegraphics[width=\linewidth]{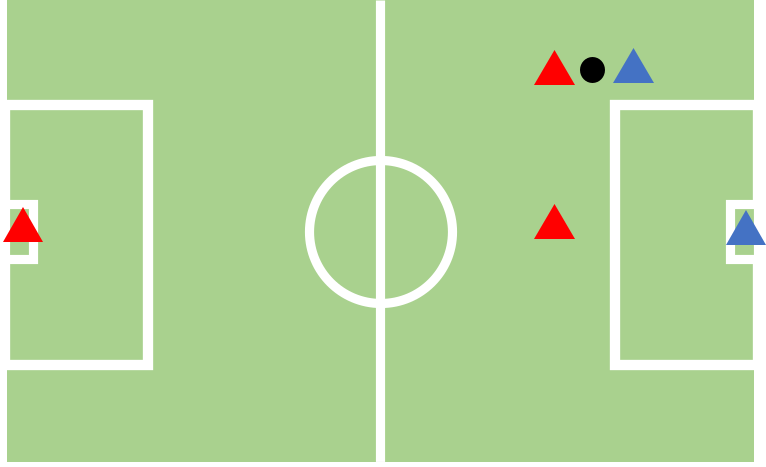}}
        \end{minipage}
        &
        \begin{minipage}[c][0.15\columnwidth][c]{0.2\columnwidth}
		\centering
		\raisebox{-.5\height}{\includegraphics[width=\linewidth]{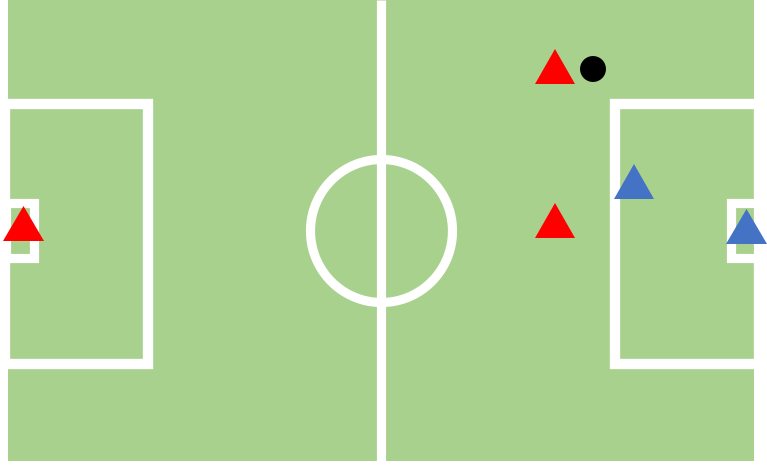}}
        \end{minipage}
        &
         \multirow{7}{*}{\shortstack{}}    
        &
        \begin{minipage}[c][0.15\columnwidth][c]{0.2\columnwidth}
		\centering
		\raisebox{-.5\height}{\includegraphics[width=\linewidth]{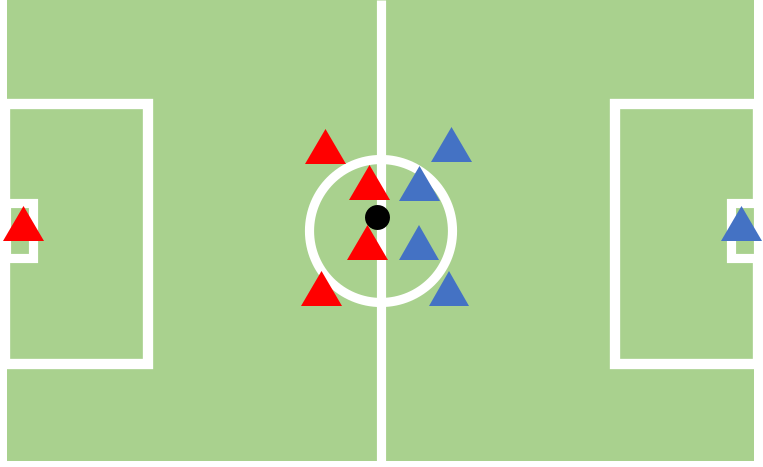}}
	  \end{minipage}          
        \\
      \cline{1-3} \cline{5-5}
        {\textit{3 vs 1 with keeper}}
        &
        {\textit{pass and shoot}}
        &
        {\textit{run pass and shoot}}  
        &
        & 
        {\textit{5 vs 5}} 
        \\
      \cline{1-3} \cline{5-5}
        \begin{minipage}[c][0.15\columnwidth][c]{0.2\columnwidth}
		\centering
		\raisebox{-.5\height}{\includegraphics[width=\linewidth]{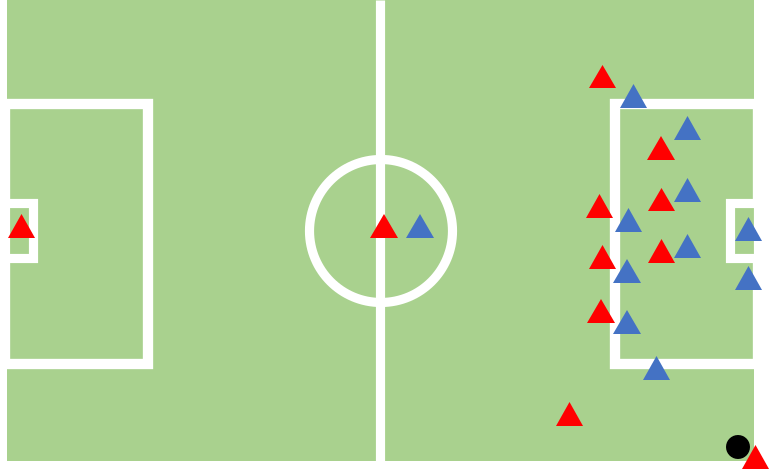}}
	  \end{minipage} 
        &
        \begin{minipage}[c][0.15\columnwidth][c]{0.2\columnwidth}
		\centering
		\raisebox{-.5\height}{\includegraphics[width=\linewidth]{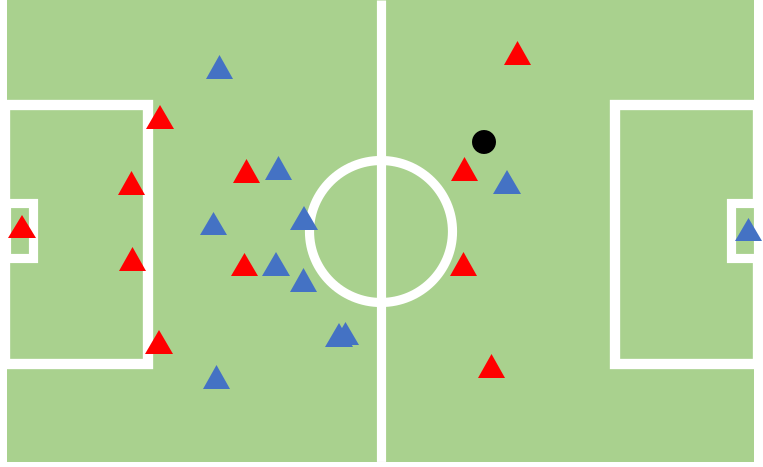}}
        \end{minipage}
        &
        \begin{minipage}[c][0.15\columnwidth][c]{0.2\columnwidth}
    	\centering
    	\raisebox{-.5\height}{\includegraphics[width=\linewidth]{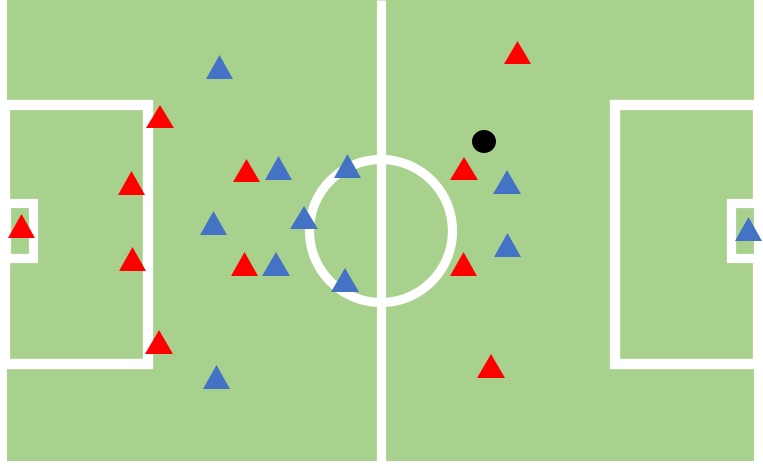}}
    	\end{minipage}
        &
        &
        \begin{minipage}[c][0.15\columnwidth][c]{0.2\columnwidth}
		\centering
		\raisebox{-.5\height}{\includegraphics[width=\linewidth]{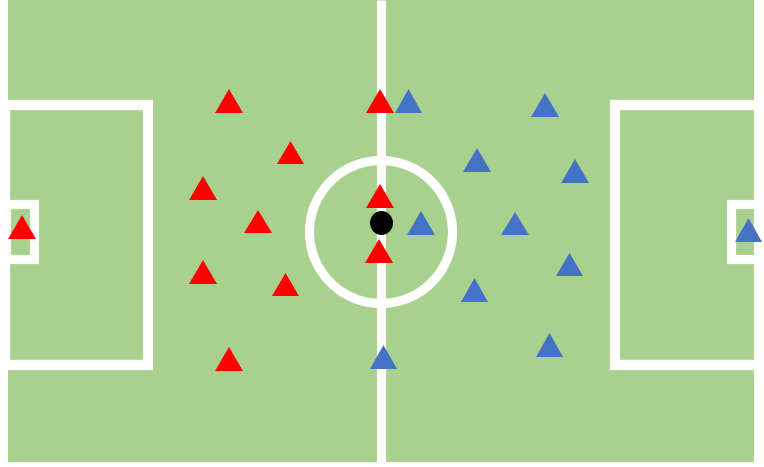}}
	    \end{minipage}
        \\
      \cline{1-3} \cline{5-5}
        {\textit{corner}}
        &
        {\textit{counterattack-easy}}
        &
        {\textit{counterattack-hard}} 
        &
        &
        {\textit{11 vs 11}}
        \\
      \hlineB{2}
    \end{tabularx}
  \end{spacing}
\end{table}


\section{Multi-Agent Cooperative Learning Benchmark}
\label{section:Experiments}

\revision{
\subsection{Problem Formulation}

We formulate any GRF cooperation task as a Multi-Agent Markov Game. If we adopt algorithms with both centralized training and execution, It can also be formulated directly as a Single-Agent Markove Game.  Readers can refer to Appendix \ref{appendix: problem} for the complete formulation.}

\subsection{Algorithms}\label{sec:algos}
\vspace{-2mm}
\revision{
We benchmark some representative MARL algorithms on our scenarios. They are either classic or state-of-art in their own categories. Our categorization and selection follow the general way adopted by many other MARL benchmarks \cite{Papoudakis2020BenchmarkingMD,pan2022mate}, the details of which can be viewed from Table \ref{table:algorithms}. Note that the GRF environment uses discrete actions, so algorithms for continuous-action space, like MADDPG \cite{lowe2017multi} and FACMAC \cite{peng2021facmac}, are not included in our benchmark.
}

\begin{table}[htbp]
    \caption{Overview of algorithms selected for our benchmark. Agent Update Scheme \cite{wang2023order} means whether multiple agents are updated simultaneously or updated with a sequential order in a single optimization step.}
    \label{table:algorithms}
  \centering
  \begin{spacing}{1.5}
    \begin{tabularx}{1.0\textwidth} {
         >{\centering\arraybackslash\hsize=.6\hsize\linewidth=\hsize}X
         >{\centering\arraybackslash\hsize=.5\hsize\linewidth=\hsize}X
         >{\centering\arraybackslash\hsize=.6\hsize\linewidth=\hsize}X
         >{\centering\arraybackslash\hsize=.7\hsize\linewidth=\hsize}X
         >{\centering\arraybackslash\hsize=.5\hsize\linewidth=\hsize}X
         >{\centering\arraybackslash\hsize=.5\hsize\linewidth=\hsize}X
        }
      \hline
      \textbf{Algorithm} & \textbf{Off-Policy / On-Policy}  & \textbf{Value-Based / Policy-Based} & \textbf{Agent Update Scheme} & \textbf{Training Mode} & \textbf{Execution Mode}  \\
      \hline
QMIX  \cite{Rashid2018QMIXMV}     & Off-Policy     & Value-Based         & Simultaneous     & Centralized   & Decentralized  \\
QPLEX \cite{wang2021qplex}     & Off-Policy     & Value-Based         & Simultaneous     & Centralized   & Decentralized  \\
IPPO \cite{de2020independent}       & On-Policy      & Policy-Based        & Simultaneous     & Decentralized & Decentralized  \\
MAPPO \cite{yu2022surprising}      & On-Policy      & Policy-Based        & Simultaneous     & Centralized   & Decentralized  \\
HAPPO \cite{kuba2022trust}     & On-Policy      & Policy-Based        & Sequential       & Centralized   & Decentralized  \\
A2PO \cite{wang2023order}      & On-Policy      & Policy-Based        & Sequential       & Centralized   & Decentralized  \\
MAT  \cite{wen2022multi}      & On-Policy      & Policy-Based        & Simultaneous     & Centralized   & Centralized    \\
      \hline  
    \end{tabularx}
  \end{spacing}
  \vspace{-3pt}
\end{table}

\revision{
\subsection{Experiment Settings}

\subsubsection{Feature Engineering}
We compare two feature encoders in our experiments: the \textit{Simple} and the \textit{Complex}. The \textit{Simple} features refer to the \textit{simple-115} features provided by the original GRF paper \cite{kurach2020google}, which encodes the location and motion information of all players and the ball. To enrich the feature representation, we also design \textit{Complex} features which include additional information such as the relative position, the closest teammate and the closest opponent. The detailed design can be found in Table \ref{tab:feature}.

\subsubsection{Reward Shaping}
We study two reward functions in our experiments: the \textit{Sparse} and the \textit{Dense}, which are based on the official SCORING and CHECKPOINT rewards introduced in the original GRF paper \cite{kurach2020google}. Both of them are simple and have been used in previous works. The SCORING rewards with $+1$ when we score and penalizes with $-1$ when we lose a score, while the CHECKPOINT gives positive feedback whenever our player moves the ball to a checkpoint that is closer to the opponent's goal. SCORING reward can be hard to obtain but CHECKPOINTS reward is easily attainable. Our \textit{Sparse} reward refers to using only the SCORING and \textit{Dense} reward refers to the sum of both SCORING and CHECKPOINT.

\subsubsection{Parameter Sharing} 
Parameter sharing is also studied in our experiments. With parameter-sharing, all agents share a single copy of network parameters. Without parameter sharing, each agent needs to maintain its own network parameters. Parameter sharing has been shown to provide more efficient learning \cite{Papoudakis2020BenchmarkingMD} but may cause high resemblance between behaviors of individual agents \cite{li2021celebrating}. To alleviate such an issue, we also include in the \textit{Complex} features a one-hot vector demonstrating the agent's identity.


For each scenario, we simulate the same number of environment steps and compare the win rates of different algorithms. Additional experiment settings can be found in Appendix \ref{appendix: exp setting}.

}


\subsection{Academy Scenarios}
\label{subsection:academy}


We first benchmark performance on relatively simpler academy scenarios and study different
settings, including reward shaping, feature engineering and parameter-sharing. Then we tackle the
full-game scenarios in section \ref{subsection:full_game} following experience drawn from the study of academy scenarios.

\subsubsection{Policy-Based vs Value-Based Algorithms}
\label{subsubsection: policy-based vs value-based}

\begin{figure}[htbp]
    \centering
    \includegraphics[width=1\columnwidth]{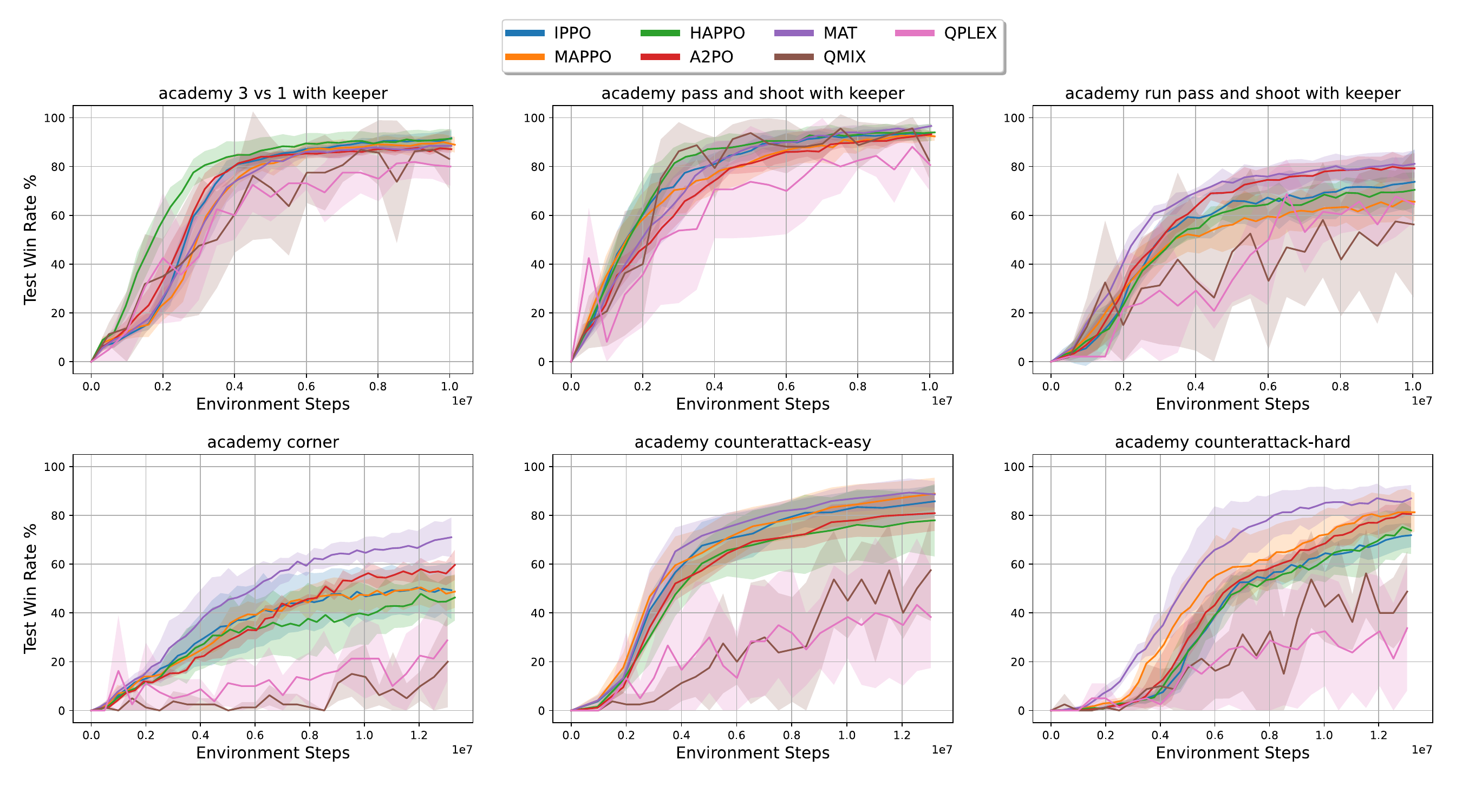}
    \caption{Average win rates on six academy scenarios: \textit{3 vs 1 with keeper}, \textit{pass and shot with keeper}, \textit{run pass and shoot with keeper}, \textit{corner}, \textit{counterattack easy} and \textit{counterattack hard}. Results are averaged over five random seeds and the shaded area represents the standard deviation of the testing win rate. }
    \label{figure:academy}
    \vspace{-3pt}
\end{figure}

\revision{
The final performance of all algorithms on academy scenarios is presented in Table \ref{table:final_performance} and the training curves are illustrated in Figure \ref{figure:academy}. In these scenarios, policy-based methods tend to exhibit overall better performance compared to value-based methods, particularly in \textit{run \& pass}, \textit{corner} and \textit{counterattack}. This discrepancy might be attributed to the curse of dimensionality in state and action spaces, due to the combinatorial nature of multi-agent systems which has been shown to be challenging for value-based algorithms \cite{yang2020overview}. 

}

\revision{
In Figure \ref{figure:academy}, we also find that different policy-based methods could achieve similar performance in small-scale scenarios. For example in \textit{3 vs 1 with keeper} and \textit{pass and shoot with keeper}, all policy-based algorithms manage to attain 85\% testing win rate within 10M environment steps. In \textit{run pass and shoot with keeper} scenario, the performance becomes slightly worse possibly due to higher defensive pressure exerted by the closer opponent (see the line-up in Table \ref{table:scenarios}). In scenarios with a larger number of players (\textit{corner}, \textit{counterattack-easy \& hard}), most policy-based algorithms still manage to achieve over 75\% win rate in counterattack tasks within 13M environment step but struggle to beat the built-in AI in corner tasks. By looking at the line-up in Table \ref{table:scenarios}, we can find that the counterattack tasks allocate only four players to the front-court for participation in offensive maneuvers, with the remaining players situated in the back-court, distanced from the ball.  This configuration simplifies the task complexity, as the MARL learning algorithm can center its attention on the offensive players. Whereas in corner tasks, all players are positioned together and the policy needs to organize all players, which could be relatively harder to learn. Overall, Multi-Agent Transformer (MAT) performs the best in complex scenarios with a larger number of players. This is probably due to the stronger ability of transformers to learn contextual relationships between input data.
}



\begin{table}[h]
    \caption{Final performances of all algorithms on academy scenarios. The final performance is recorded as the maximum mean win rate(standard deviation) scaled by 100.}
    \label{table:final_performance}
  \centering
  \begin{spacing}{1.5}
    \begin{tabularx}{1.0\textwidth} {
         >{\centering\arraybackslash\hsize=1.0\hsize\linewidth=\hsize}X
         >{\centering\arraybackslash\hsize=.5\hsize\linewidth=\hsize}X
         >{\centering\arraybackslash\hsize=.5\hsize\linewidth=\hsize}X
         >{\centering\arraybackslash\hsize=.5\hsize\linewidth=\hsize}X
         >{\centering\arraybackslash\hsize=.5\hsize\linewidth=\hsize}X
         >{\centering\arraybackslash\hsize=.5\hsize\linewidth=\hsize}X
         >{\centering\arraybackslash\hsize=.5\hsize\linewidth=\hsize}X
         >{\centering\arraybackslash\hsize=.5\hsize\linewidth=\hsize}X
        }
      \hline
      \textbf{Task} & \textbf{IPPO}  & \textbf{MAPPO} & \textbf{HAPPO} & \textbf{A2PO} & \textbf{MAT}  & \textbf{QMIX} & \textbf{QPLEX}  \\
      \hline
      \textit{pass \& shoot} &   93.7(0.9) & 92.9(2.6)  & 94.0(3.6)  & 93.2(1.2) & \textbf{96.6}(0.8)   & 95.6(5.8) & \revision{88.1(8.2)} \\
      
      \textit{run pass \& shot} &   73.7(13.4) & 66.0(6.3)  & 70.4(7.2)  & 79.9(6.0) & \textbf{81.1}(5.7)  & 58.1(23.6) & 68.8(14.2)  \\
      \textit{3 vs 1} &  \textbf{91.7}(3.1) & 90.0(3.2) & 91.4(3.9) & 87.6(1.4) & 88.5(2.0) & 86.9(8.2) & 81.9(6.7) \\ 
      \textit{corner} &  50.4(10.3) & 50.5(7.2) & 47.9(9.9) & 59.7(6.2) & \textbf{71.0}(8.1)  & 20.0(18.7) & 28.8(16.7) \\ 
      \textit{ct-easy} &   85.7(6.6) & 88.9(6.6)  & 78.0(14.8)  & 80.9(7.3) & \textbf{89.3}(5.8)  & \revision{57.5(18.9)} & 43.3(27.2)  \\
      \textit{ct-hard} & 71.6(4.9) & 81.3(9.6) & 75.2(10.6) & 80.7(4.2) & \textbf{87.0}(6.0)   & \revision{56.3(18.2)} & 33.8(25.6) \\
      \textit{5 vs 5} &  99.1(0.6) & 96.0(1.8) & 98.0(1.3) & 95.0(2.9) & \textbf{99.3}(1.2) & 0.0(0.0) & 0.0(0.0) \\ 
      \textit{11 vs 11} &  52.7(2.4) & 45.4(2.7) & 52.1(4.8)  & 50.1(3.6)& \textbf{59.7}(3.6)  & 0.0(0.0) &  0.0(0.0)  \\
      \hline  
    \end{tabularx}
  \end{spacing}
  \vspace{-3pt}
\end{table}

\begin{figure}[htbp]
    \centering
    \includegraphics[width=\linewidth]{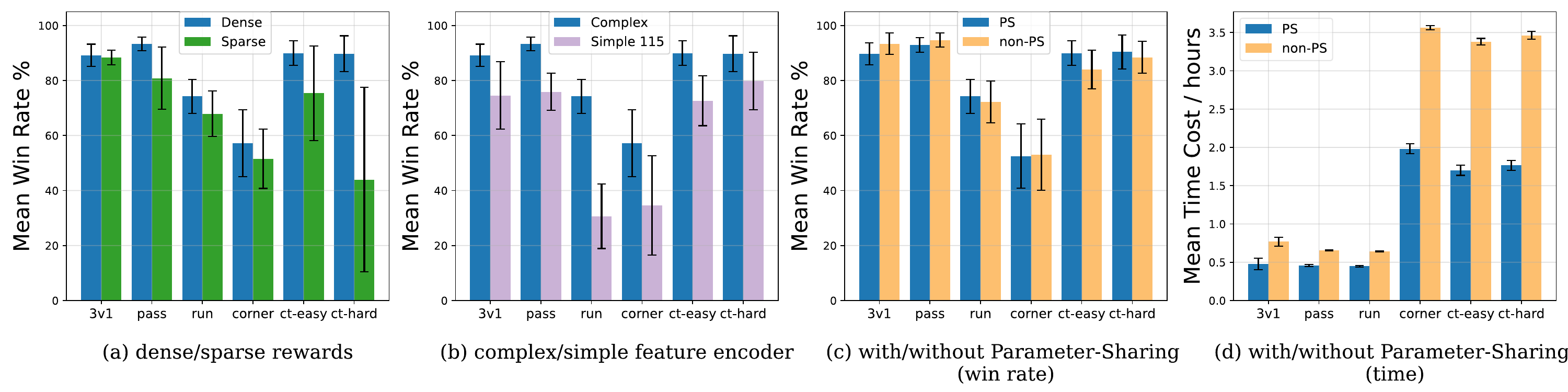}
    \caption{Comparisons under different experiment settings. \textbf{(a), (b)} \& \textbf{(c)}: win rate in each scenario averaged over all \revision{policy-based} algorithms with dense/sparse rewards; with complex/simple feature encoders, and with/without parameter-sharing. \textbf{(d)}: Time cost in each scenario averaged over all \revision{policy-based} algorithms with/without parameter-sharing. Error bars show the standard deviation. 3v1: \textit{3 vs 1 with keeper}; pass: \textit{pass and shoot with keeper}; run: \textit{run pass and shoot with keeper}; ct-easy: \textit{counterattack-easy}; ct-hard: \textit{counterattack-hard}.}
    \label{figure:academy_ablation}
    \vspace{-10pt}
\end{figure}

\subsubsection{An Analysis on Different Algorithm Settings: Rewards, Feature Encoders, and Parameter Sharing}

To validate our choices for addressing more challenging full-game scenarios, we conducted further ablation studies on academy scenarios, and the outcomes are presented in Figure \ref{figure:academy_ablation}. \revision{Note that the results exhibited in the figure are averaged over all policy-based algorithms. The two value-based algorithms are not included because they are currently not good enough for solving difficult scenarios as validated in Section \ref{subsubsection: policy-based vs value-based}. The results of individual algorithms and the analysis of their consistency can be founded in Appendix \ref{appendix: academy exp}. }

\revision{Across all six academy scenarios, dense reward (both SCORING and CHECKPOINTS) shows better and more stable performance than sparse reward (only SCORING). The reason might be that the additional CHECKPOINTS reward, which effectively guides our players toward the opponent's goal, substantially reduces exploratory actions. 
In terms of selecting appropriate feature encoders, complex features offer more efficient learning than simple ones. This demonstrates that integrating domain knowledge into features could be very helpful when solving a complex task. 
We also observe from Figure \ref{figure:academy_ablation} that training without parameter-sharing gives slightly better results in some scenarios but worse performance on counterattack tasks. However, the time consumption of non-parameter-sharing is considerably high, making it less suitable for large-scale training.}


\subsection{Full-Game Scenarios}
\label{subsection:full_game}

\begin{figure}[htbp]
    \vspace{-3mm}
    \centering
    \includegraphics[width=0.85\linewidth]{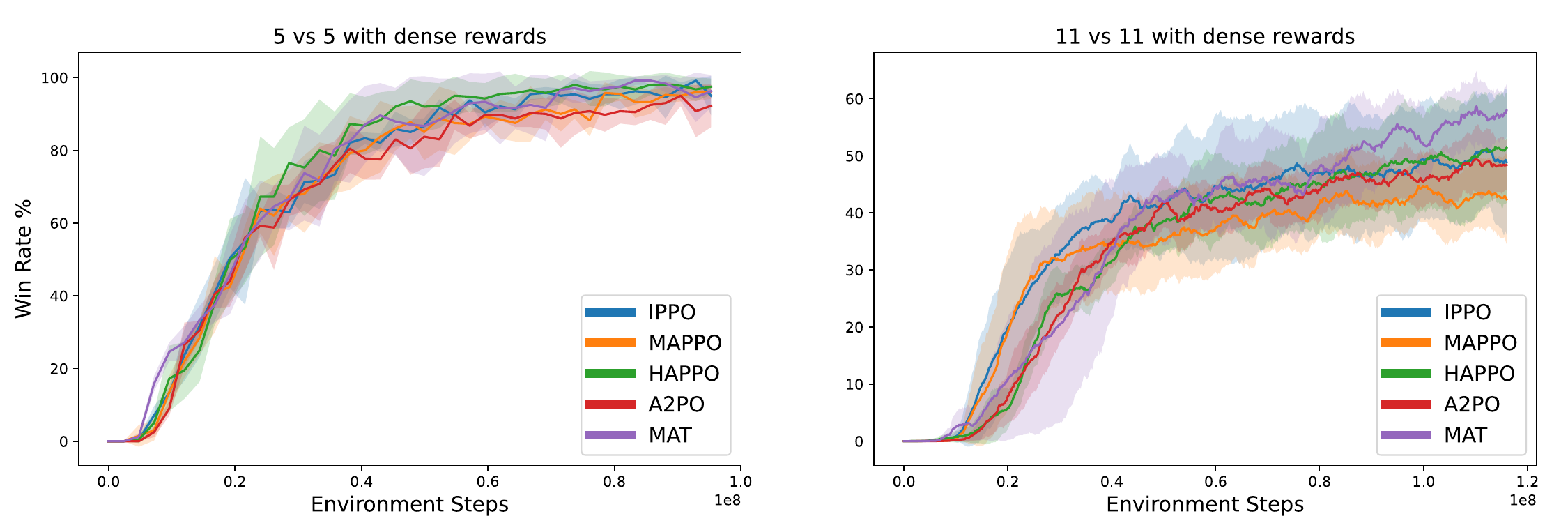}
    \caption{Performance of baseline algorithm in \textit{5 vs 5} (Left) and \textit{11 vs 11} (Right) full-game scenarios. Results are averaged over five and three random seeds. Each \textit{5 vs 5} experiment (\texttt{1e8} steps) takes approximately 8 hours to complete and an experiment of \textit{11 vs 11} (\texttt{1e8} steps) takes around 12 hours on a 200 CPU and $2\times$A100 GPU server).}
    \label{figure:full_game}
    \vspace{-4mm}
\end{figure}

We leverage the insights gained from the academic benchmark to address the challenges posed by full-game scenarios. Our approach involves utilizing dense rewards, the default feature encoder and parameter-sharing. The corresponding performances are documented in Table \ref{table:final_performance}, while the sample-efficiency curves are illustrated in Figure \ref{figure:full_game}. Notably, we observe that value-based methods fail to learn meaningful behaviors within a reasonable number of environmental steps. This observation aligns with our previous findings on academy scenarios, empirically indicating that these value-based methods encounter difficulties in complex multi-agent environments without specialized treatment. Various policy-based algorithms still exhibit similar performance after careful hyper-parameter tuning, with MAT slightly outperforming the others, potentially attributed to its fully-centralized advantage. Although the \textit{5 vs 5} scenario has been largely addressed, the \textit{11 vs 11} scenario remains challenging. Aside from the win rate, given the substantial training time (12 hours for an experiment on the \textit{11 vs 11} scenario with a 128 CPU and 2 A100 GPU server), it becomes crucial to develop algorithms with improved sample efficiency, which is critical because we need to repeatedly compute the best response to the current opponent mixture in self-play. 

\begin{figure}[htbp]
    \centering
    \begin{subfigure}{0.4\linewidth}
        \centering
        \includegraphics[width=0.9\linewidth]{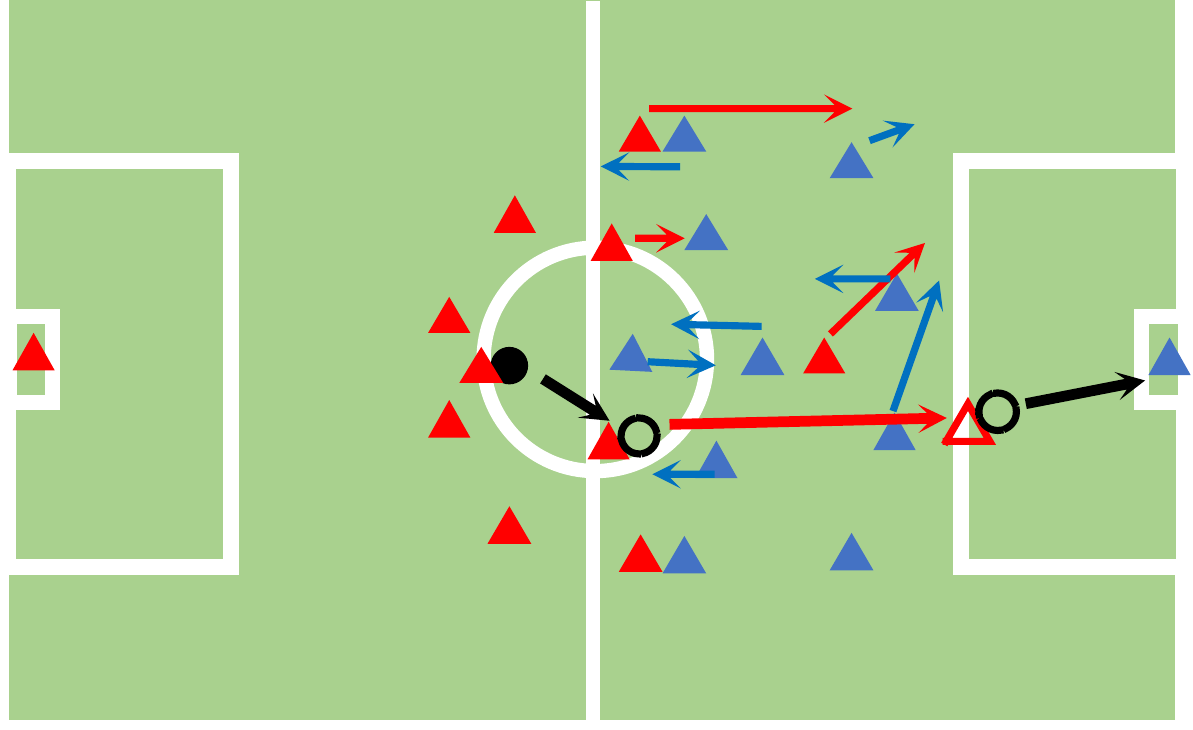}
        \caption{Offensive strategy}
    \end{subfigure}
    \begin{subfigure}{0.4\linewidth}\hspace{1em}
        \centering
	   \includegraphics[width=0.9\linewidth]{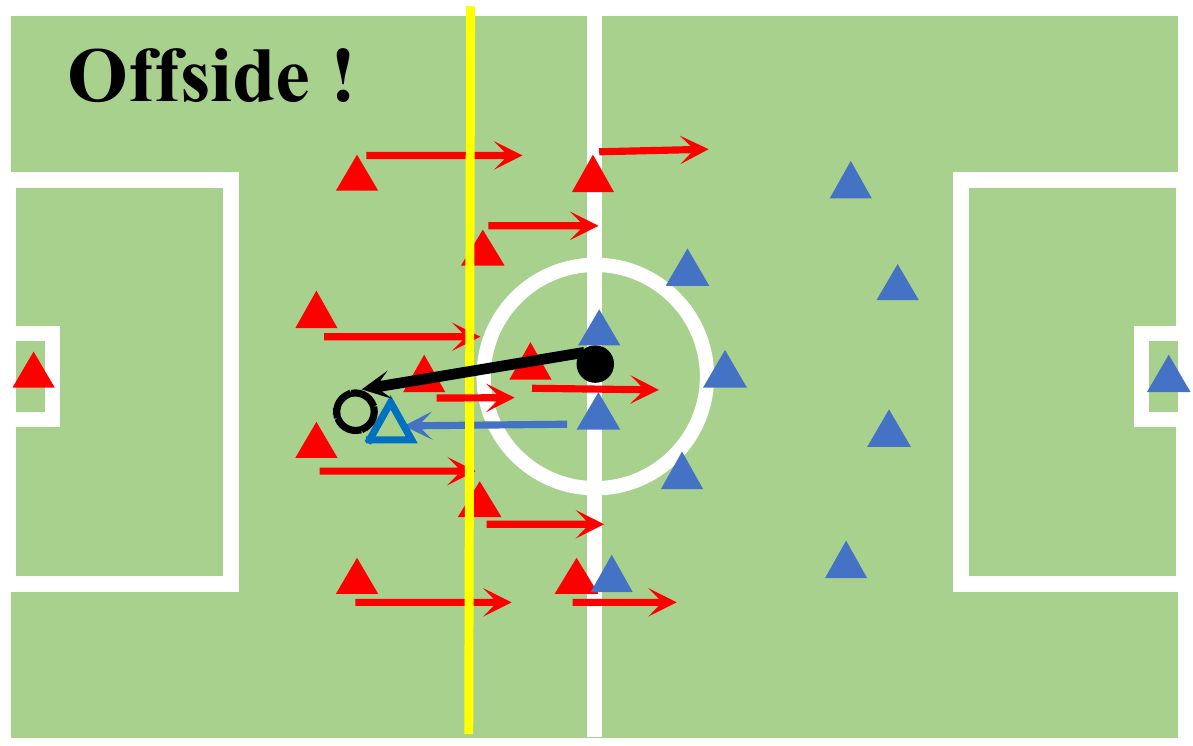}
         \caption{Defensive strategy}
    \end{subfigure}
    \caption{Visualisation for team strategy in 11-vs-11 full-game. (\textcolor{red}{\huge$\blacktriangleup$}: left-team player; \textcolor{blue}{\huge$\blacktriangleup$}: right-team player; {\huge$\bullet$}: ball; \textcolor{red}{\huge$\rightarrow$}: left-team player's movement; \textcolor{blue}{\huge$\rightarrow$}: right-team player's movement; \textcolor{black}{\huge$\rightarrow$}: ball's movement;
    \textcolor{yellow}{\huge$\rVert$}: offside line.)}\label{fig:full-game behavior}
    \vspace{-5pt}
\end{figure}

Additionally, we delve into the behaviors learned by these algorithms in full-game scenarios. Indeed, all algorithms learn similar policies. In both tasks, these policies effectively exploit the weaknesses of the built-in AI and display sophisticated dribbling and shooting skills. During attacks, teammates usually serve as distractions for opponent defenders, while a single key player exploits the weak point of the defense and takes the shot. In defense, all our defenders frequently coordinate their movement forward to force opponent attackers into an offside position. A visualization of these behaviors on the \textit{11 vs 11} scenario is given in Figure \ref{fig:full-game behavior}. However, we aspire to incorporate more diverse playing styles that can adaptively respond to the opponents' styles, as well as foster improved teamwork, such as increased ball passes between teammates. Simply training policies against a fixed opponent often leads to overfitting to a specific playing style, which limits adaptability and versatility. This inspires us to explore methods of building stronger football AI.

\section{Building Stronger Football AI} 
\label{section:Building Stronger Football AI}

In Section \ref{subsection:full_game}, it is evident that the policy trained solely on the benchmark scenarios lacks sufficient intelligence and robustness. 
\revision{In the following, we will focus mainly on our training framework and online ranking system, which are provided as tools for building stronger AI in the future. We also provide a match-decomposition data structure (Appendix \ref{appendix:data structure}) for better analysis and a single-step visual debugger (Appendix \ref{appendix:debugger}) for easier investigation.}

\subsection{Population-based Self-Play Framework and Pre-trained Policies}

\begin{figure}[htbp]
    \centering
    \begin{subfigure}{0.33\linewidth}
    \centering
    \includegraphics[width=\linewidth]{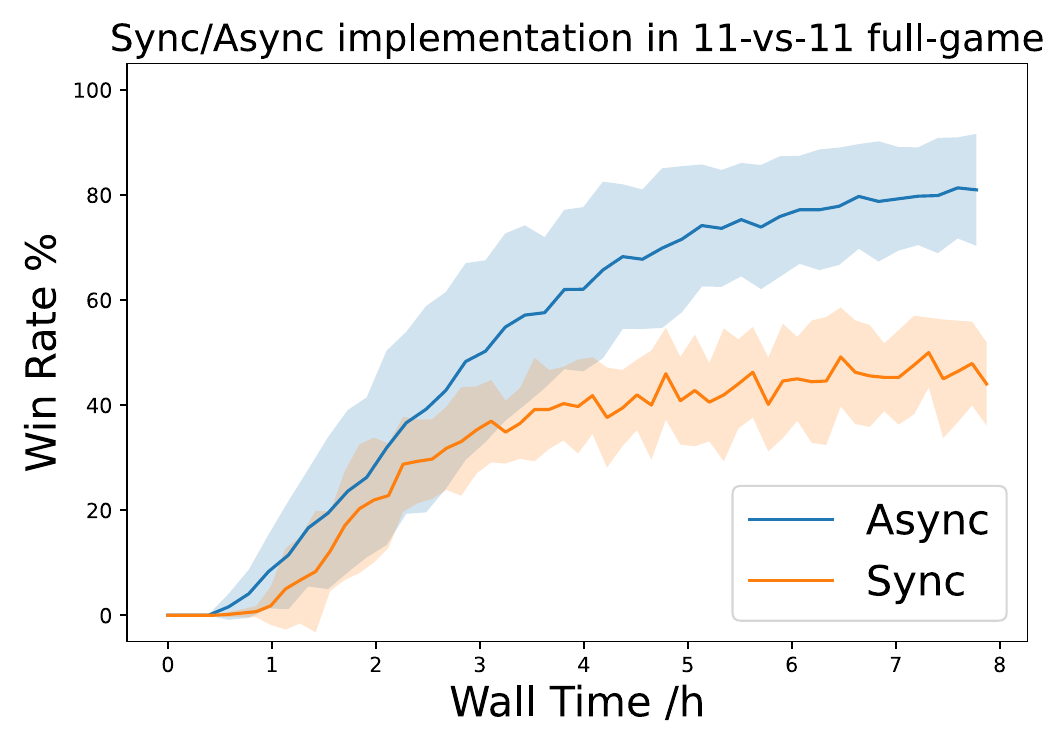}
    \caption{Time analysis of sync/async implementation}
    \label{fig:win-time 11v11 async}
    \end{subfigure}\hspace{3mm}
    \begin{subfigure}{0.25\linewidth}
    \centering
    \includegraphics[width=\linewidth]{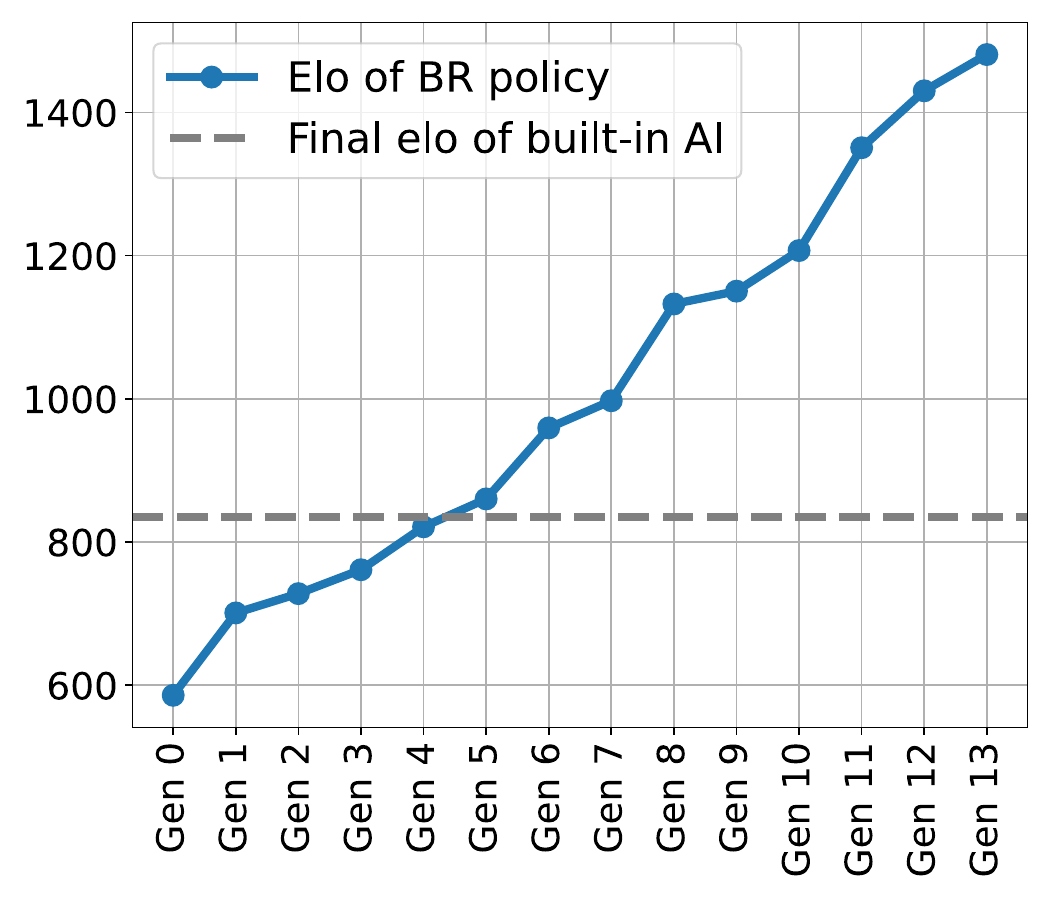}
    \caption{Evolution of Elo rating in a PSRO trial}
    \label{fig:pbt elo}
    \end{subfigure}    \hspace{3mm}
    \begin{subfigure}{0.33\linewidth}
    \centering
    \includegraphics[width=\linewidth]{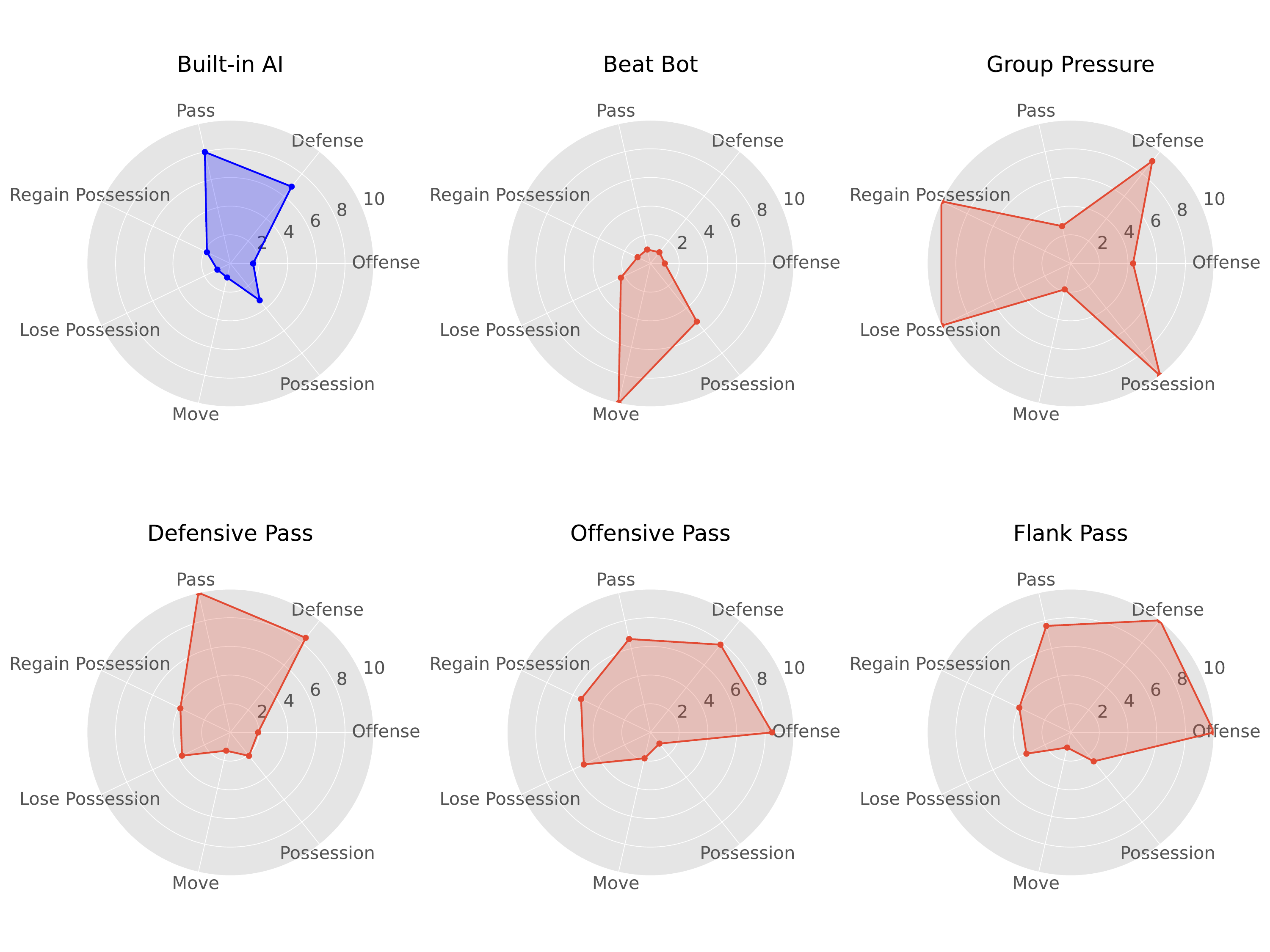}
    \caption{Styles of pre-trained policies in \textit{11 vs 11}}
    \label{figure:radar}
    \end{subfigure}
    \caption{\textbf{(a)} Time comparison of synchronous and asynchronous implementations averaged over all policy-based algorithms discussed in Section \ref{sec:algos} in the \textit{11 vs 11} scenario against the strongest built-in AI. \textbf{(b)} In a PSRO trial of the \textit{5 vs 5} scenario, using our proposed framework, the newest best-response (BR) policy keeps improving as the iterative training proceeds. \textbf{(c)} We use radar plots to depict diverse styles of pre-trained policies in the \textit{11 vs 11} scenario. The performance is evaluated by simulating matches and measured in football-specific metrics with normalization.}
    \vspace{-5mm}
\end{figure}

\revision{One approach to achieving a stronger and less exploitable policy is to utilize more advanced self-play algorithms. To ease the efforts for other researchers, we release our distributed and asynchronous population-based training framework, which implements Policy Space Response Oracle (PSRO) \cite{psro} and League Training \cite{alphastar}. So far as we know, it is the first publicly available training framework that not only beat the hardest built-in AI on full-game scenarios with only one round of best-response computation, but also continue to improve policies with population-based self-play (Figure \ref{fig:pbt elo}). The asynchronous implementation in our framework greatly outperforms the synchronous implementation in terms of both the learning speed and the final performance, as illustrated in Figure \ref{fig:win-time 11v11 async}. Such a speed-up is essential in self-play pipelines, where repeated best-response computations are required.

We also release some of our trained policies with diverse playing styles (Figure \ref{figure:radar}) which can potentially be used for both training and evaluation in the future. These diverse policies are obtained by performing PSRO and League Training with different setups of reward functions and opponent populations.}

For details of the framework, related experiments, training procedure and pre-trained models, please refer to Appendix \ref{appendix:framework} and \ref{appendix:pretrained policies}.


\subsection{Online Ranking}

A publicly accessible online ranking is widely recognized as having significant benefits for related research, such as various competitions held on Kaggle \cite{Taieb2014AGB, Iglovikov2017SatelliteIF}. \revision{In light of this, we introduce the \textit{Google Research Football Online Ranking} targeting specifically at GRF \textit{5 vs 5} and \textit{11 vs 11} multi-agent full-game scenarios \cite{jidi2021grf} on an online evaluation platform called \textit{JIDI} \cite{jidi}. }



\begin{figure}[htbp]
    \centering
    \includegraphics[width=0.85\linewidth]{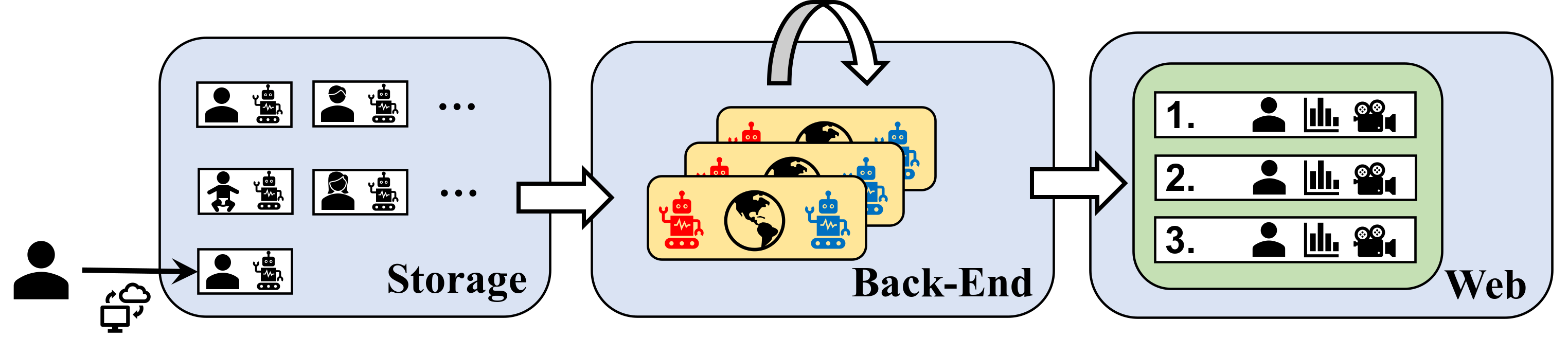}
    \caption{JIDI online Ranking System. When a user submits his customized decision-making agent to the agent storage, the back-end evaluation process executes parallel evaluation tasks distributedly on computing nodes. The evaluation results are updated to the user's score and the online ranking is updated accordingly. Users can view replays of their matches.}
    \label{fig:jidi}
\end{figure}

As illustrated in Figure \ref{fig:jidi}, JIDI automatically simulates football matches between agents submitted by users and continuously updates the live ranking based on match results. This online ranking system allows agents to compete against unseen opponents that are not available during training. This is crucial for studying generalization ability, which is an important aspect of algorithms emphasized by many works \cite{Gorsane2022TowardsAS,leibo2021scalable,ellis2022smacv2,agapiou2022melting}. \revision{Importantly, users are allowed to download the replay of these simulated matches so as to analyze the weakness of their strategies and thus, algorithms.}


Moreover, this ranking system serves as the official platform for the \textit{IEEE Conference on Games Football AI Competition} in both 2022 \cite{cog2022} and 2023 \cite{cog2023}. \revision{In a competition, we often hold multiple competition rounds and apply the \textit{Swiss-system tournament} mechanism \cite{csato2013ranking} to compute Elo-rates of each submission. The scores for each competition round will be accumulated by weights and ranked when the competition ends.}
Table \ref{tab:jidi} presents the current statistics on the number of users, agents, and matches. \textbf{We strongly encourage interested researchers to actively participate in this ranking system and contribute to its growth}.

\begin{table}[htbp]
    \vspace{-2mm}
    \caption{The numbers of users, agents, and matches on JIDI Online Ranking System so far.}
    \label{tab:jidi}
  \centering
  \begin{spacing}{1.2}
    \begin{tabularx}{1.0\textwidth} {
         >{\centering\arraybackslash\hsize=0.7\hsize\linewidth=\hsize}X
         >{\centering\arraybackslash\hsize=0.7\hsize\linewidth=\hsize}X
         >{\centering\arraybackslash\hsize=0.7\hsize\linewidth=\hsize}X
         >{\centering\arraybackslash\hsize=0.7\hsize\linewidth=\hsize}X
        }
      \hline
      \textbf{Scenarios}  & \textbf{Number of Users} & \textbf{Number of Agents} & \textbf{Number of Matches}   \\
      \hline
       5-vs-5 full-game & 109 & 372 & 24520  \\
      \hline  
       11-vs-11 full-game & 119 & 486 & 23562  \\ 
      \hline
    \end{tabularx}
  \end{spacing}
   
\end{table}

\revision{
\section{Limitations and Future Directions}
\label{sec: conclusion}

This paper provides a cooperative MARL benchmark and a set of useful research tools for future studies on the Google Research Football Environment. Nonetheless, several limitations inherent in this work reveal potential future direction:
\begin{enumerate}[topsep=0pt,itemsep=0ex,partopsep=0ex,parsep=0ex]
    \item This work adheres to the foundational settings of the original GRF paper \cite{kurach2020google}, such as the setup of scenarios and reward functions, with few extensions. It would be interesting to study beyond these settings. For example, the existing paradigm allows players to observe nearly all facets of the pitch. We can study the more partially-observable but realistic cases by limiting the vision of players.
    \item  While our work establishes a benchmark for cooperative learning, the absence of an equivalent benchmark for competitive scenarios is conspicuous. In the future, we need to address the reproducibility of complex self-play training pipelines. This often involves the identification of randomness in each phase and the optimization for efficiency as these pipelines require a tremendous amount of computational resources.
    \item Compared to real-life football tactics, our trained policies are still immature and the connections between virtual games and real football matches are still lacking. Bridging the gap is an interesting future direction, including learning from real-world football data to make our agent akin to human\cite{tuyls2021game} and evaluating real players' actions in specific situations \cite{pulis2022reinforcement}.
\end{enumerate}

}

\medskip

\small



\clearpage

\appendix

\section{Benchmark Environments}
\label{appendix: scenario settings}

\subsection{An Introduction to the GRF Environment}
In this section, we provide a brief introduction to the environment. More details can be found in the original paper \cite{kurach2020google}. \\
\\
The \textbf{raw observation} of each player contains information about the game state, the ball and all players on the pitch. The game state includes scores, game modes indicating free kicks, corner kicks or other game stages, and game time represented as steps. The ball information includes its spatial position, direction, and an indicator for ball ownership (identifying the player and team that possess it). The player information comprises position, direction, roles, yellow card records, and more. For players on the opposite team, positions and directions are mirrored.\\
\\
The \textbf{default actions set} comprises 19 actions, including directional movements, 3 various ball passing, ball shooting, sliding, sprinting and others. In the \textit{Action Set V2} version, an additional 20th action is available, which delegates player control to the built-in AI. Throughout our experiments, we utilized the default action set.\\

The \textbf{rewards} provided by GRF include the SCORE and CHECKPOINT rewards. The SCORE reward is +1 when a goal is scored and -1 when a goal is conceded. In our experiment, we referred to the application of the SCORE reward alone as a sparse reward setting. The CHECKPOINT reward is given when the agent moves closer to the opponent's goal, encouraging the progression of the ball toward the goal. When both SCORE and CHECKPOINT rewards are applied, it is considered as the dense reward setting.\\

The \textbf{game dynamics} in GRF closely resemble realistic football games. Players can move or sprint with or without the ball. They can also execute various passes and shots. Besides different actions, complex interactions such as collisions and trips between players are simulated as well. Additionally, the game engine introduces stochasticity in the dynamics, which can influence the passes and shots randomly.

\subsection{Scenario Descriptions}

Table \ref{tab:scenario desc} presents a comprehensive overview of the benchmark scenarios used in our study. These scenarios are characterized by variations in the number of players, initial line-up configurations, and termination conditions. However, it is important to note that our policies do not control the goalkeeper due to limitations with the goalkeeper's action set. By default, we use the winning rate against the opponent as the primary evaluation metric. 

\begin{table}[htbp]
\caption{The detailed description of GRF multi-agent benchmark scenarios. The source code for each scenario can be found in our code-base. (\textcolor{red}{\huge$\blacktriangleup$}: left-team player; \textcolor{blue}{\huge$\blacktriangleup$}: right-team player; {\huge$\bullet$}: ball)}
\label{tab:scenario desc}
\begin{spacing}{1.5}
\begin{tabularx}{\textwidth}{
|>{\raggedright\arraybackslash\hsize=0.2\hsize\linewidth=\hsize}X
|>{\raggedright\arraybackslash\hsize=0.55\hsize\linewidth=\hsize}X
|>{\centering\arraybackslash\hsize=0.25\hsize\linewidth=\hsize}X
|}
	\hline
    \textbf{Scenario Name} & \textbf{Description}  & \textbf{Line-up}\\
 \hline
\textit{pass and shot with keeper} (2v1)    &  A \textcolor{red}{3} vs \textcolor{blue}{2} academy game. \textcolor{red}{Two left-team} players start at the right half, competing against \textcolor{blue}{one right-team} defense player and the goalkeeper.The episode terminates when: \textcircled{1} reaches maximum duration (400 steps) \textcircled{2} ball is out-of-pitch \textcircled{3} one team scores \textcircled{4} ball ownership changes. & \begin{minipage}[b]{0.2\columnwidth}
		\centering
		\includegraphics[width=\linewidth]{scenario_lineup_png/pass_and_shoot.png}
	\end{minipage}\\\hline
\textit{run pass and shot with keeper} (run 2v1)    & Similar to 2v1 but with different position of the \textcolor{blue}{defensive player} & \begin{minipage}[b]{0.2\columnwidth}
		\centering
		\includegraphics[width=\linewidth]{scenario_lineup_png/run_2v1.png}
	\end{minipage}\\\hline
\textit{3 vs 1 with keeper} (3v1)    & A \textcolor{red}{4} vs \textcolor{blue}{2} academy game.  \textcolor{red}{Three left-team} players start at the right half, competing against \textcolor{blue}{one right-team} defense player and the goalkeeper. Same termination condition applies as 2v1. & \begin{minipage}[b]{0.2\columnwidth}
		\centering
		\includegraphics[width=\linewidth]{scenario_lineup/3_vs_1png.png}
	\end{minipage}\\\hline
\textit{corner}    & An \textcolor{red}{11} vs \textcolor{blue}{11} academy game. The \textcolor{red}{left team} starts the ball at the \textcolor{blue}{right team}'s corner. Same termination condition applies as 2v1. & \begin{minipage}[b]{0.2\columnwidth}
		\centering
		\includegraphics[width=\linewidth]{scenario_lineup_png/corner.png}
	\end{minipage}\\\hline
\textit{counterattack-easy} (CT-easy)    & An \textcolor{red}{11} vs \textcolor{blue}{11} academy game. Four \textcolor{red}{left team} players start the ball at the mid-field in the \textcolor{blue}{right team}'s half and only two \textcolor{blue}{right team} players defend in their own half. The rest of the players are at the \textcolor{red}{left team}'s half. Same termination condition applies as 2v1. & \begin{minipage}[b]{0.2\columnwidth}
		\centering
		\includegraphics[width=\linewidth]{scenario_lineup_png/ct_easy.png}
	\end{minipage}\\\hline
\textit{counterattack-hard} (CT-hard)    & Similar to CT-easy but with a slightly difference in \textcolor{blue}{defensive player}'s position. & \begin{minipage}[b]{0.2\columnwidth}
		\centering
		\includegraphics[width=\linewidth]{scenario_lineup_png/counter_attack.png}
	\end{minipage}\\\hline
\textit{5 vs 5 full game} (5v5)   & A \textcolor{red}{5} vs \textcolor{blue}{5} full-game. Four players from each team gather at the center of the field. The \textcolor{red}{left-team} starts the kick-off. The game terminates when the episode reaches the maximum duration (3,000 steps). The second half begins at the 1501st step and two teams will swap sides. & \begin{minipage}[b]{0.2\columnwidth}
		\centering
		\includegraphics[width=\linewidth]{scenario_lineup_png/5v5.png}
	\end{minipage}\\\hline
\textit{11 vs 11 full game} (11v11)   & An \textcolor{red}{11} vs \textcolor{blue}{11} full-game. The \textcolor{red}{left-team} starts the kick-off. The game terminates when the episode reaches the maximum duration (3,000 steps). The second half begins at the 1501st step and two teams will swap sides. & \begin{minipage}[b]{0.2\columnwidth}
		\centering
		\includegraphics[width=\linewidth]{scenario_lineup_png/11v11.png}
	\end{minipage}\\\hline
\end{tabularx}
\end{spacing}
\end{table}

\subsection{Simulation Time Costs}
\begin{figure}[htbp]
    \centering
    \includegraphics[width=0.8\linewidth]{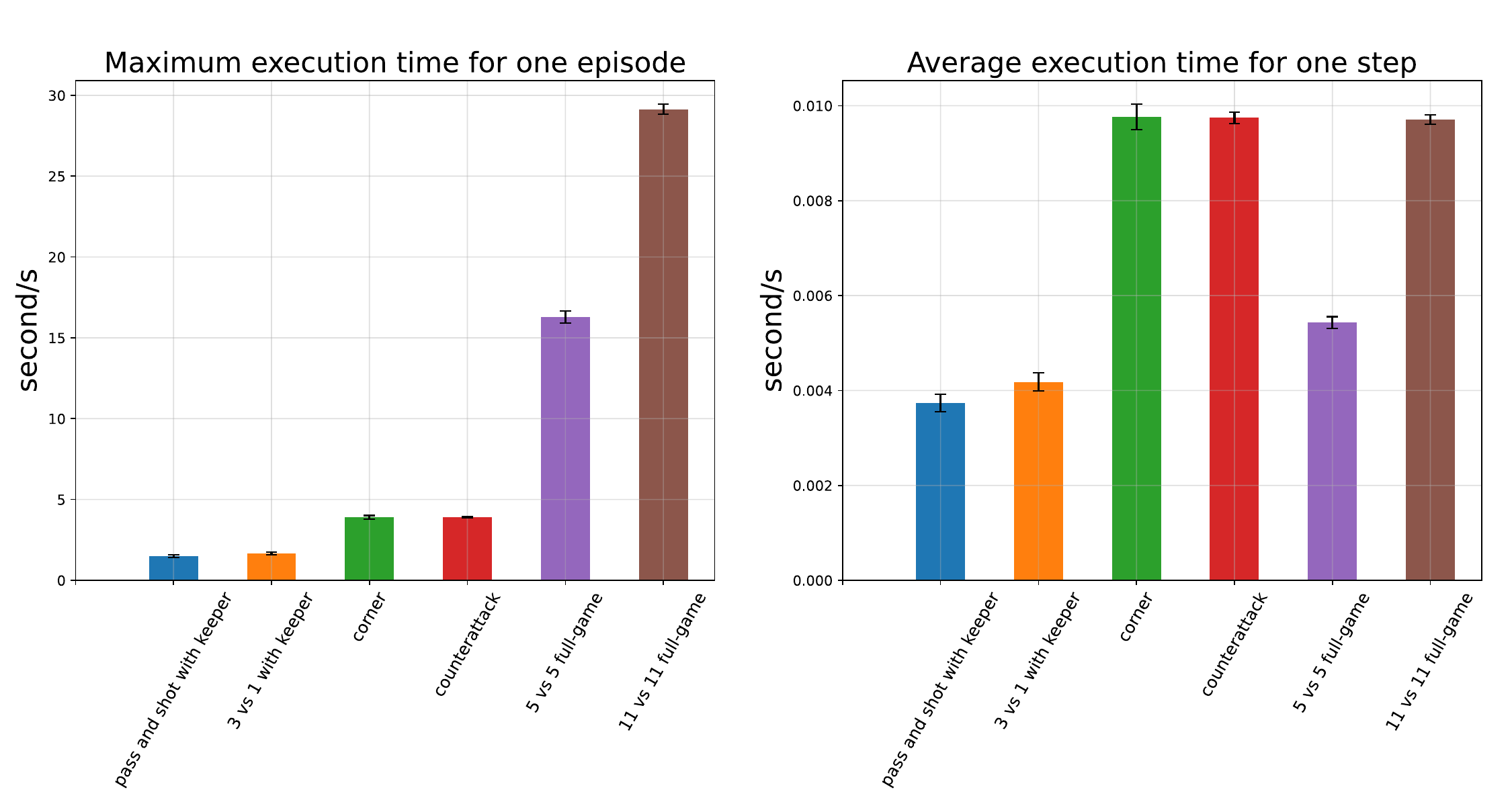}
    \caption{Expected time costs for each GRF scenario (without rendering) measured on a single CPU thread. The error bars show the standard deviation.}
    \label{fig:time}
\end{figure}

Google Research Football is a very time-consuming environment. Figure \ref{fig:time} compares the maximum episode time and the average step time for each scenario. As the episode length and the number of controllable player increase, the step and episode time cost increase greatly, posing a great challenge for training, which motivates us to design more efficient training systems.

\section{Benchmark Experiments}\label{appendix: exp setting}

In this section, we provide a detailed description of our experiments, including problem formulations, feature encoding, action masking, data collection, algorithms setting, hyperparameters selection and computation resources. We also provide supplementary behavior analyses on academy scenarios and ablation studies on full-game scenarios.

\revision{
\subsection{Problem Formulations}\label{appendix: problem}
The multi-agent Google Research Football scenarios can be formulated as a Markov game $G$ with $N$ agents. The game is defined by a set of elements $\langle N, \mathcal{S}, \{\mathcal{A}^{i}\}_{i\in\{1,...,N\}}, P, \{R^i\}_{i\in\{1,...,N\}}, \gamma \rangle$. $\mathcal{S}$ is the set of game states shared by all agents,  $\mathcal{A}^i$ is the set of actions of agent $i$ and we denote $\mathbb{A}:=\mathcal{A}^1\times ... \times \mathcal{A}^N$ as the joint action. $P: \mathcal{S}\times\mathbb{A}\rightarrow \mathcal{S}$ is the transition probability function. $R^i: \mathcal{S}\times \mathbb{A} \times \mathcal{S} \rightarrow \mathbb{R}$ is the reward function for agent $i$. $\gamma\in[0,1]$ is the discount factor that represents the decaying rate. The game also can be viewed under two perspective:
\begin{itemize}
    \item Single-Agent setting: a naive solution to game $G$ is to transform the problem into a Markov Decision Process (MDP) defined by a tuple of elements $\langle \mathbb{S}, \mathbb{A}, P, R, \gamma \rangle$. $\mathbb{S}=\{\mathcal{S}\}, \mathbb{A}=\{\mathcal{A}^1\times ... \times \mathcal{A}^N\}$ are joint state and action of $N$ agents and can be seen as the state and action of a single integrated agent.     The goal of the integrated agent is to solve the MDP, that is to find the optimal joint policy function $\bm{\pi}: \mathbb{S} \rightarrow \mathbb{A}$ such that the discounted cumulative reward is maximised:
    \[
    \mathbb{E}_{s_{t+1}\sim P(\cdot|s_t, \bm{a}_t); \bm{a}_t \sim \bm{\pi}(\cdot|s_t)}\left(\sum_{t\geq 0} \gamma^t R(s_t, \bm{a}_t, s_{t+1}) \bigg| s_0 \right) , s_t, s_{t+1} \in \mathbb{S}, \bm{a}_t\in \mathbb{A}
    \]

    This is also viewed as the paradigm of centralized training with centralized execution (CTCE) ref?. 
    
    \item Fully cooperative setting: this setting can be regarded as a multi-player extension to the MDP where agents are assumed to be homogeneous and interchangeable. Agents also share the same reward function: $R= R^{i}=R^1=...=R^N$. The game proceed as follows: at each time step $t$, the environment has a state $s_t$, each agent executes its action $a_t^i$ simultaneously with all other agents, giving the joint action $A_t = {a_t^1, ..., a_t^N}$. The environment transit to the next state $s_{t+1}\sim P(\cdot|s_t, A_t)$. Then, the environment determines an immediate reward $R^i(s_t, A_t, s_{t+1})$ for each agent. The goal of each agent $i$ is to solve the game by finding an optimal policy $\pi^{i}\in \Pi^i: \mathcal{S}\rightarrow \mathcal{A}^i$ such that the discounted cumulative reward is maximised:
    \[
    \mathbb{E}_{s_{t+1}\sim P(\cdot|s_t, a^{1:N}_t);a_t^i\sim \pi^i(\cdot|s_t)} \left(\sum_{t\geq 0} \gamma^t R_t^{i}(s_t, a^{1:N}_t, s_{t+1}) \bigg| s_0 \right), s_t, s_{t+1} \in \mathbb{S}, a_t^i \in \mathcal{A}^i
    \]
    In this setting, the optimal policy of each agent is influenced by not only its own policy but also the policies of the other agents in the game. This is the fundamental differences between single-agent RL and multi-agent RL.

\end{itemize}}

\subsection{Feature Engineering}
We summarize our encoded features in Table \ref{tab:feature}. Our feature encoder (adapted from \url{https://github.com/seungeunrho/football-paris}) contains information about the current player, the ball, the closest teammate and opponent and the global state of both teams. We have also provided more advanced feature engineering including information such as match details and offside status, which is not included in our experiment but can be used for further studies. All of these features are extracted from the raw observation provided by the game engine.

\begin{table}[htbp]
  \centering
  \caption{Feature encoding used in the experiment. $n$ refers to player number.}\label{tab:feature}
  \begin{spacing}{1.5}
    \begin{tabularx}{\textwidth} {
        >{\centering\arraybackslash\hsize=.5\hsize\linewidth=\hsize}X
        >{\centering\arraybackslash\hsize=.2\hsize\linewidth=\hsize}X
        >{\raggedright\arraybackslash\hsize=1\hsize\linewidth=\hsize}X
        }
      \hline
      \textbf{Feature Name} & \textbf{Dimension}  & \textbf{Description}\\
      \hline
      Player state &  19 & my position, my direction, my speed, my role, etc \\
       \hline
      Ball state &  18 & ball position, ball zone, ball relative position, ball direction, ball speed, ball ownership \\ 
      \hline
    Available actions & 19 & action one-hot vector \\
    \hline
    Closest teammate state &  7 & teammate relative position, teammate direction, teammate speed, distance, if\_tired \\
    \hline
    Closest opponent state &  7 & opponent relative position, opponent direction, opponent speed, distance, if\_tired \\
    \hline
    Teammate state (Global) & 7$n$-1 & teammate relative position, teammate direction, teammate speed, distance, if\_tired \\
    \hline
    Opponent state (Global) &  7$n$-1 & opponent relative position, opponent direction, opponent speed, distance, if\_tired \\
    \hline
    Advanced (optional) & - & match details, offside info \\
    \hline
    \end{tabularx}
  \end{spacing}
\end{table}

\subsection{Action Masking}
Action masking is essential for speeding up training convergence. Our action masking is shown below:
\begin{itemize}
    \item When opponent team owns the ball, \textbf{disable} \texttt{PASS, SHOT} and \texttt{DRIBBLE}.
    \item When no team owns the ball and the ball is far from my team, \textbf{disable} \texttt{PASS, SHOT} and \texttt{DRIBBLE}.
    \item When my team owns the ball, \textbf{disable} \texttt{SLIDE}.
    \item When the ball is too far from the opponent's penalty area, \textbf{disable} \texttt{SHOT}.
    \item When the player is within the opponent's penalty area, \textbf{disable} \texttt{HIGH PASS} and \texttt{LONG PASS}.
    \item When one of our player owns the ball, \textbf{disable} other players' \texttt{PASS}, \texttt{SHOT} and \texttt{DRIBBLE} when the ball is far from them.
    \item Game-mode-related masking in the free kick, corner kick, and penalty kick.
\end{itemize}

\subsection{Data Collection}

The data collection process revolves around the creation, storage, and transmission of data samples for training purposes. In synchronous training, we employ multiple parallel processes to gather trajectory samples, which are subsequently forwarded to the training phase once all processes have completed their respective episodes. For our experiments, the batch size for data collection is uniformly set to 80 across all scenarios except for the \textit{11 vs 11} full-game scenario, the batch size of which is reduced to 60 in order to accommodate the increased GPU memory requirement (caused by larger player numbers and longer horizons) and slower data consumption rate. Each individual process operates on a single CPU thread and establishes its own Google Research Football environment, utilizing a unique random seed. In all academy scenarios, each episode persists for a maximum of 400 steps, while in full-game scenarios, the episode length extends to 3000 steps.

\subsection{Algorithms and Hyperparameters}

\subsubsection{A Brief Overview of Algorithms}
\textbf{QMIX/QPLEX}: QMIX \cite{Rashid2018QMIXMV} is a value-based cooperative algorithm that employs a mixing network to combine individual $Q$ functions into a joint $Q$ function $Q_{joint}$ in a complex non-linear way. The mixing network structurally enforces the Individual-Global-Max (IGM) principle \cite{song2019igm} that the joint-action value should be monotonic in the per-agent values. QPLEX \cite{wang2021qplex} also follows the IGM principle but applies its advantage-based version. It has a $Q$-value mixing network similar to QMIX's but incorporates dueling structures \cite{wang2016dueling} for representing both individual and joint $Q$ functions. In our experiments, we actually adopt their enhanced versions proposed in the CDS paper \cite{li2021celebrating}, where an additional information-theoretical regularization is introduced to promote exploration.



\textbf{Independent PPO (IPPO)}/\textbf{Multi-Agent PPO (MAPPO)}: IPPO \cite{de2020independent} is a fully decentralized variant of the PPO \cite{schulman2017proximal} algorithm, where each agent is trained independently using its local information. General training techniques such as General Advantage Estimate (GAE) \cite{gae} and value target clipping are employed. On the other hand, MAPPO \cite{yu2022surprising} follows the Centralized Training Decentralized Execution (CTDE) paradigm, enabling actors to access local information while critics exploit global state information. Critics with more comprehensive information typically yield better value estimation, thus resulting in improved policies.


\textbf{Heterogeneous-Agent PPO (HAPPO)}/\textbf{Agent-by-agent Policy Optimization (A2PO)}: HAPPO \cite{kuba2022trust} and A2PO \cite{wang2023order} update agents sequentially to prevent the potential conflicting direction updates that may happen in the simultaneous updating. HAPPO leverages the multi-agent advantage decomposition lemma and guarantees a monotonic improvement on the joint policy. A2PO derives an updating formula similar to PPO for each agent and resolves the preceding agent policy shifting issue using a technique called \textit{preceding-agent off-policy correction}, ensuring monotonic improvement for each agent update. A2PO also proposes a novel semi-greedy strategy for sequential agent update ordering.


\textbf{Multi-Agent Transformer (MAT)}: MAT \cite{wen2022multi} treats multi-agent reinforcement learning as a sequence modeling problem, employing Transformer networks \cite{ashish2017transformer} for both actors and critics. Its policy updating algorithm bears a resemblance to IPPO and MAPPO. However, MAT is a fully-centralized algorithm that trains a single joint actor with access to all local observations from each agent, predicting actions in an auto-regressive manner.

\subsubsection{Implementations and Hyperparameters}
For QMix and QPLEX, we use hyperparameters and implementation adopted from the original CDS paper \cite{li2021celebrating}. We implement the policy-based algorithms (IPPO, MAPPO, HAPPO, A2PO, MAT) based on our distributed training framework (subsection \ref{appendix:framework}) following their original papers. For a fair comparison, we implement all policy-based based on the implementation of MAPPO. We keep the same structures for all the algorithms except for MAT. Our default hyperparameter settings for IPPO/MAPPO, HAPPPO/A2PO, and MAT are summarized in Table \ref{table:common hyperparameter}, \ref{table:hyper_ippo_mappo}, \ref{table:hyper_happo_a2po} and \ref{table:hyper_mat} respectively.

\begin{table}[htbp]
    \caption{Common hyperparameters for policy-based algorithms}
    \label{table:common hyperparameter}
  \centering
  \begin{spacing}{1}
    \begin{tabularx}{\textwidth} {
         >{\centering\arraybackslash\hsize=0.3\hsize\linewidth=\hsize}X
         >{\centering\arraybackslash\hsize=0.2\hsize\linewidth=\hsize}X
         |>{\centering\arraybackslash\hsize=0.3\hsize\linewidth=\hsize}X
         >{\centering\arraybackslash\hsize=0.2\hsize\linewidth=\hsize}X
        |>{\centering\arraybackslash\hsize=0.35\hsize\linewidth=\hsize}X
         >{\centering\arraybackslash\hsize=0.2\hsize\linewidth=\hsize}X
        }
      \hline
          Hyperparameters & Values & Hyperparameters & Values & Hyperparameters & Values  \\
         \hline
         orthogonal init & True &  use popart & True & mini batch & 1\\
         gain & 1.0 & popart beta & 0.9999 & feature norm & True \\
         activation & ReLU & optimizer & Adam & entropy & 0.0 \\
         network type & MLP &  actor lr & 5e-4 & clipping & 0.2\\
         use gae & True & critic lr & 5e-4 & other clipping & 0.125  \\
         gae lambda & 0.95 & optim eps & 1e-5 & max grad norm & 10.0 \\
         \hline
    \end{tabularx}
  \end{spacing}
\end{table}

\begin{table}[htbp]
    \caption{Hyperparameters for IPPO/MAPPO}
    \label{table:hyper_ippo_mappo}
  \centering
  \begin{spacing}{1}
    \begin{tabularx}{1.0\textwidth} {
         >{\centering\arraybackslash\hsize=1.5\hsize\linewidth=\hsize}X
         >{\centering\arraybackslash\hsize=.5\hsize\linewidth=\hsize}X
         >{\centering\arraybackslash\hsize=.5\hsize\linewidth=\hsize}X
         >{\centering\arraybackslash\hsize=.5\hsize\linewidth=\hsize}X
         >{\centering\arraybackslash\hsize=.5\hsize\linewidth=\hsize}X
         >{\centering\arraybackslash\hsize=.5\hsize\linewidth=\hsize}X
         >{\centering\arraybackslash\hsize=.5\hsize\linewidth=\hsize}X
        }
      \hline
         &\textbf{2v1}  & \textbf{3v1} & \textbf{Corner} & \textbf{CT} & \textbf{5v5} & \textbf{11v11}\\
         \hline
         Discount Value & 0.99 & 0.99 & 0.995 & 0.995 & 0.995 & 0.995 \\
         Training Epoch & 10 & 15 & 10 & 10 & 20 & 10 \\
         Hidden Dimension &  \multicolumn{6}{c}{MLP[256,128,64]}\\
         \hline
    \end{tabularx}
  \end{spacing}
\end{table}

\begin{table}[htbp]
    \caption{Hyperparameters for HAPPO/A2PO}
    \label{table:hyper_happo_a2po}
  \centering
  \begin{spacing}{1}
    \begin{tabularx}{1.0\textwidth} {
         >{\centering\arraybackslash\hsize=1.5\hsize\linewidth=\hsize}X
         >{\centering\arraybackslash\hsize=.5\hsize\linewidth=\hsize}X
         >{\centering\arraybackslash\hsize=.5\hsize\linewidth=\hsize}X
         >{\centering\arraybackslash\hsize=.5\hsize\linewidth=\hsize}X
         >{\centering\arraybackslash\hsize=.5\hsize\linewidth=\hsize}X
         >{\centering\arraybackslash\hsize=.5\hsize\linewidth=\hsize}X
         >{\centering\arraybackslash\hsize=.5\hsize\linewidth=\hsize}X
        }
      \hline
         &\textbf{2v1}  & \textbf{3v1} & \textbf{Corner} & \textbf{CT} & \textbf{5v5} & \textbf{11v11}\\
         \hline
         Discount Value & 0.99 & 0.99 & 0.995 & 0.995 & 0.995 & 0.995 \\
         Training Epoch & 5 & 5 & 5 & 5 & 5 & 2 \\
         Block Number & 2 & 3 & 2 & 2 & 4 & 5\\
         Hidden Dimension &  \multicolumn{6}{c}{MLP[256,128,64]}\\
         \hline
    \end{tabularx}
  \end{spacing}
\end{table}

\begin{table}[htbp]
    \caption{Hyperparameters for MAT}
    \label{table:hyper_mat}
  \centering
  \begin{spacing}{1}
    \begin{tabularx}{1.0\textwidth} {
         >{\centering\arraybackslash\hsize=1.5\hsize\linewidth=\hsize}X
         >{\centering\arraybackslash\hsize=.5\hsize\linewidth=\hsize}X
         >{\centering\arraybackslash\hsize=.5\hsize\linewidth=\hsize}X
         >{\centering\arraybackslash\hsize=.5\hsize\linewidth=\hsize}X
         >{\centering\arraybackslash\hsize=.5\hsize\linewidth=\hsize}X
         >{\centering\arraybackslash\hsize=.5\hsize\linewidth=\hsize}X
         >{\centering\arraybackslash\hsize=.5\hsize\linewidth=\hsize}X
        }
      \hline
         &\textbf{2v1}  & \textbf{3v1} & \textbf{Corner} & \textbf{CT} & \textbf{5v5} & \textbf{11v11}\\
         \hline
         Discount Value & 0.99 & 0.99 & 0.995 & 0.995 & 0.995 & 0.995 \\
         Training Epoch & 10 & 15 & 10 & 10 & 20 & 10 \\
         Embedding Dimension &  100 & 100 & 128 & 128 & 115 & 128\\
         \hline
    \end{tabularx}
  \end{spacing}
\end{table}

\subsection{Ablation Experiments on Academy Scenarios}\label{appendix: academy exp}
Here we present the details of the ablation studies on feature encoders, reward shaping schemes and whether to use the parameter-sharing technique here.

\begin{figure}[htbp]
    \centering
    \includegraphics[scale=0.48]{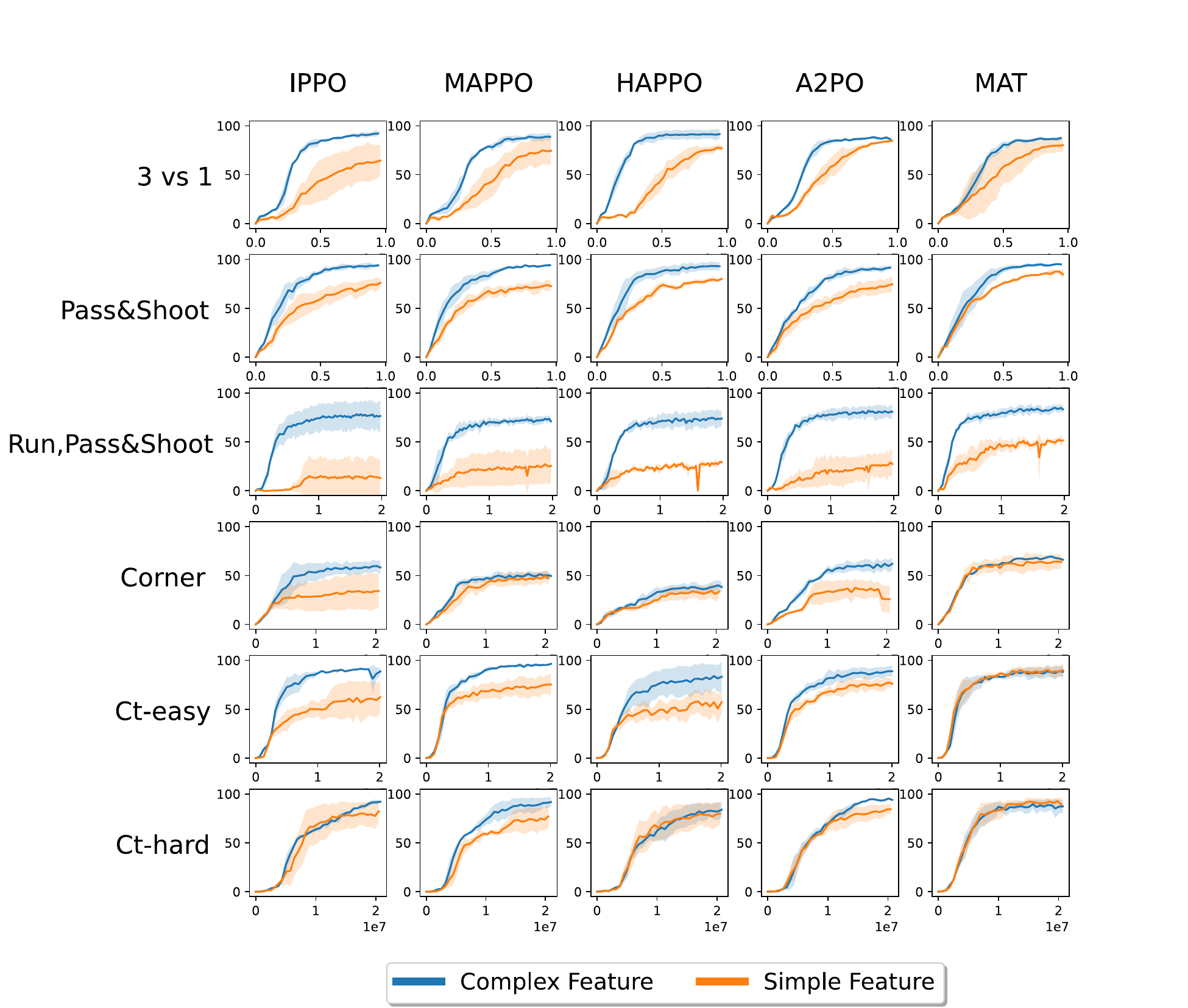}
    \caption{Ablation study on feature encoders. In most cases, the application of Complex feature encoders enables more efficient learning against the built-in AI. It is worth noticing that Multi-agent Transformer with simple features can still achieve similar performance on tasks like corner, counterattack-easy and counterattack-hard.}
    \label{fig:academy_all_FE}
\end{figure}

\begin{figure}[htbp]
    \centering
    \includegraphics[scale=0.48]{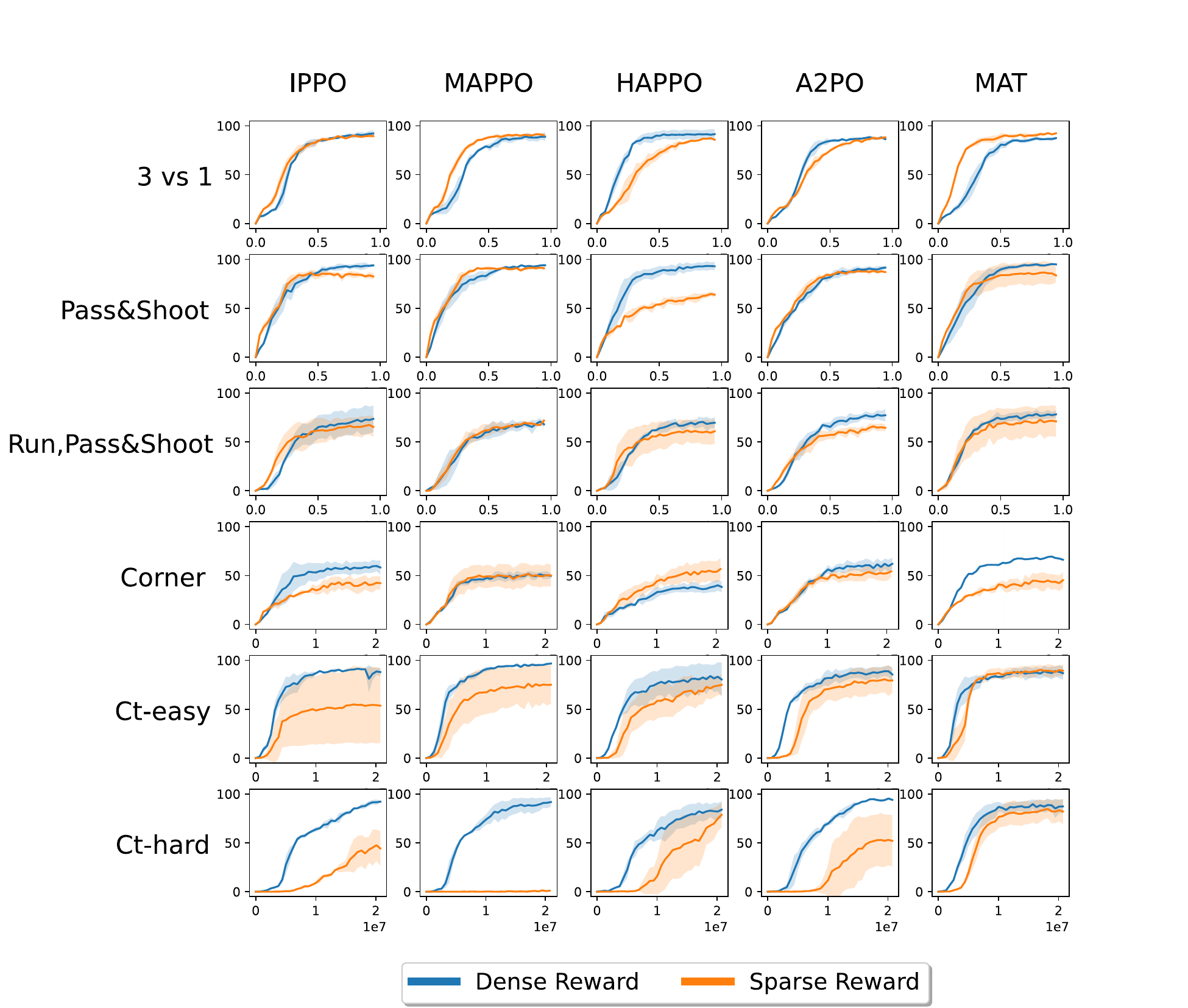}
    \caption{Ablation study on reward shaping. For most cases, applying sparse reward will likely bring more instability and degradation of performance.}
    \label{fig:academy_all_reward}
\end{figure}

\begin{figure}[htbp]
    \centering
    \includegraphics[scale=0.48]{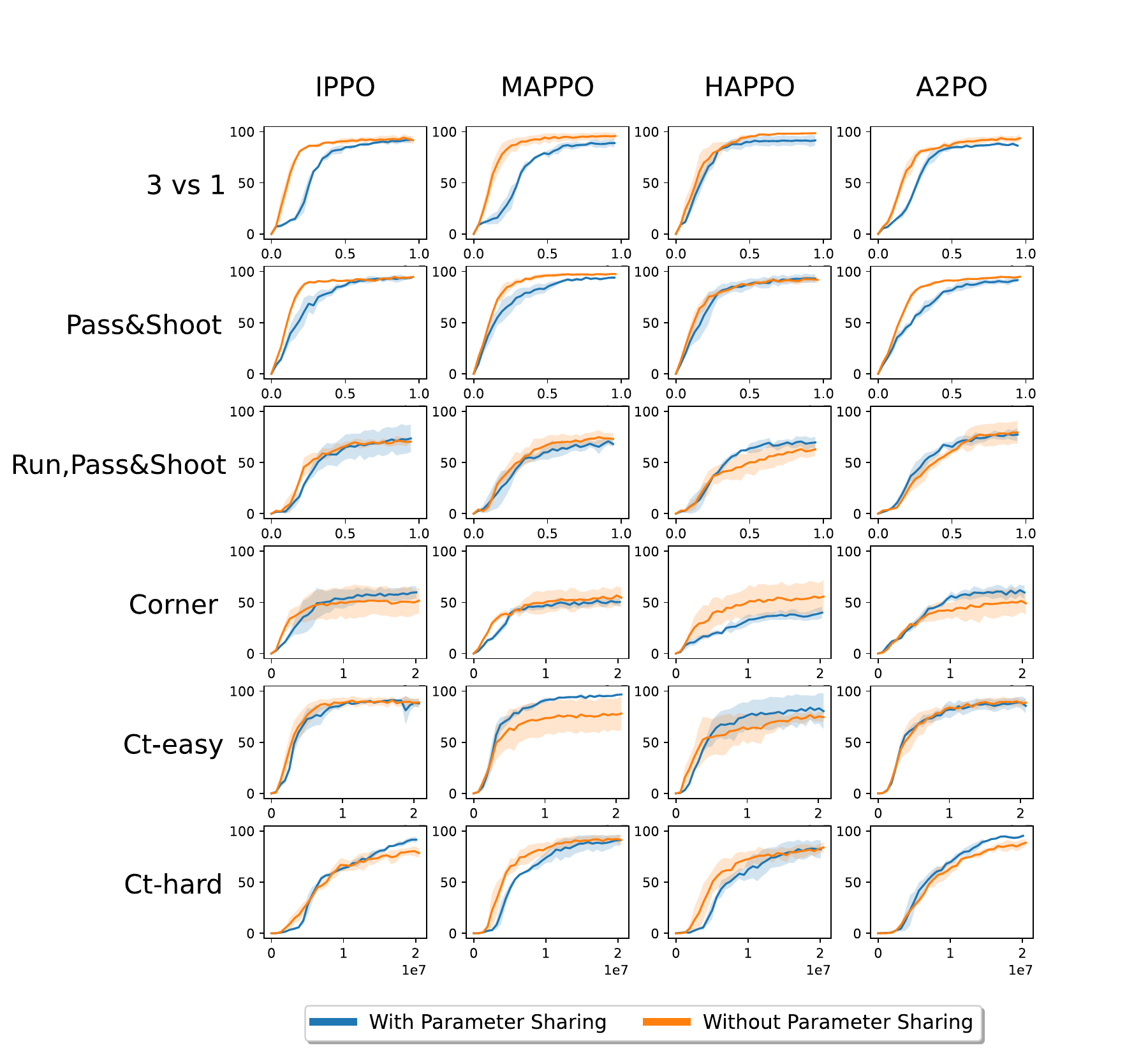}
    \caption{Ablation study on parameter sharing. For small-scale scenarios, not using the parameter-sharing technique will bring slightly better performance. For large-scale scenarios, the gain in performance becomes less obvious. However, according to Figure \ref{figure:academy_ablation} the time cost increase largely.}
    \label{fig:academy_all_PS}
\end{figure}

\revision{
By looking further into ablation studies for each individual algorithm (Figure \ref{fig:academy_all_FE},\ref{fig:academy_all_reward}, \ref{fig:academy_all_PS}), our findings consistently indicate that for most of the cases, the agent with complex encoder and dense reward learns more efficiently. MAT is able to deliver similar performance on the corner and counterattack tasks using both encoders, probably due to its strong ability of feature extraction. With respect to reward shaping, using sparse reward will likely bring instability and sub-optimal performance in most cases. Additionally, training without parameter-sharing is more sample efficient for many policy-based algorithms in small-scaled GRF scenarios but doesn't exhibit distinct advantages in larger tasks.
}


\clearpage

\subsection{Ablation Experiments on Full-Game Scenarios}
\begin{figure}[htbp]
    \centering
    \includegraphics[width=\linewidth]{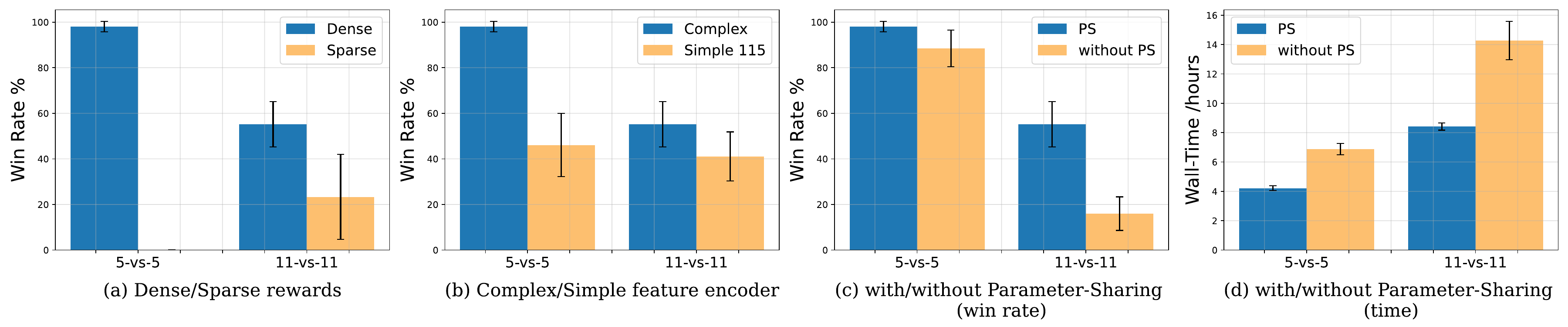}
    \caption{Comparisons under different experiment settings in full-game scenarios. The error bar represents the standard deviation.}
    \label{fig:11v11 additional exp}
\end{figure}

We provide additional experiments on rewards, feature encoders and parameter-sharing in full-game scenarios, as illustrated in Figure \ref{fig:11v11 additional exp}. The results align with our findings in the main paper for academy scenarios. However, there is one notable exception that deserves mention. We cannot easily train a winning policy in the \textit{5 vs 5} scenario with only sparse rewards. The learned policies consistently converge to an overly-conservative one, where players focused solely on retaining possession of the ball without making forward advancements or initiating attacks. This is likely due to the lack of sufficient exploration of goals and early concession of goals during the initial stages of training, which reinforce the conservative behaviors. The problem is alleviated in the \textit{11 vs 11} full-game scenario which has a larger number of controllable players and therefore increases the possibility of exploring a goal. It is also worth noting that similar challenges may arise when training policies against significantly stronger opponents. While complex reward shaping offers a potential solution, we acknowledge that curriculum learning could be another viable approach. However, this falls beyond the scope of this study and is left for future research.




\subsection{Computational Resources}
\revision{
All experiments are run on servers with both CPUs and GPUs. The main types of CPU are Intel(R) Xeon(R) Gold 6330 CPU @ 2.00GHz and AMD EPYC 7773X 64-Core Processors. The main types of GPU are NVIDIA GeForce RTX 3090 and A100. We approximately spend 130,000 CPU hours on all the experiments, including academy and full-game scenarios, reward settings, feature settings, asynchronous implementation, and a PSRO demonstration trial. The time for hyperparameter tuning is not taken into account.

Table \ref{table:computation resources} lists detailed time consumption for each of the experiments on GRF, including synchronous training on academy scenario, full-game scenario and asynchronous training on full-game scenario. Table \ref{table:time prop} lists the time proportion estimate during both rollout and training phase. As for computational cost of QMIX and QPLEX, we use the original implementation and each experiment often requires $2\sim4$ days of training.
}

\begin{table}[htbp]
    \caption{Computational resources consumption for each experiment trial on policy-based algorithm}
    \label{table:computation resources}
  \centering
  \begin{spacing}{1}
    \begin{tabularx}{1.0\textwidth} {
         >{\centering\arraybackslash\hsize=1.5\hsize\linewidth=\hsize}X
         >{\centering\arraybackslash\hsize=.7\hsize\linewidth=\hsize}X
         >{\centering\arraybackslash\hsize=.5\hsize\linewidth=\hsize}X
         >{\centering\arraybackslash\hsize=.5\hsize\linewidth=\hsize}X
         >{\centering\arraybackslash\hsize=1.\hsize\linewidth=\hsize}X
         >{\centering\arraybackslash\hsize=.5\hsize\linewidth=\hsize}X
        }
      \hline
         \textbf{Scenarios}  & \textbf{Env Steps} & \textbf{CPU(s)} & \textbf{GPU(s)} & \textbf{Time(hours)} & \textbf{Sync/Async}\\
         \hline
         \textit{3 vs 1 with keeper}         & $10^7$        & 128 & 1 & $0.43\sim0.87$ & Sync \\
         \textit{pass and shoot}  & $10^7$        & 128 & 1 & $0.43\sim0.77$ & Sync \\
         \textit{run pass and shoot}         & $2\times10^7$ & 128 & 1 & $0.88\sim1.1$ & Sync \\
         \textit{corner}                     & $3\times 10^7$ & 128 & 1 & $3\sim5$ & Sync \\
         \textit{counterattack-easy}         & $3\times 10^7$ & 128 & 1 & $3\sim4.5$ & Sync\\
         \textit{counterattack-hard}         & $3\times 10^7$ & 128 & 1 & $3\sim4.5$ & Sync \\
         \textit{5 vs 5}         & $10^8$        & 128 & 4 & $4\sim6$ & Sync \\
         \textit{11 vs 11}        & $10^8$        & 128 & 4 & $8\sim17$ & Sync\\
         \textit{5 vs 5}          & $10^8$        & 128 & 4 & $2\sim6$ & Async \\
         \textit{11 vs 11}         & $10^8$        & 128 & 4 & $3.3\sim6.2$ & Async\\
         \hline
    \end{tabularx}
  \end{spacing}
\end{table}

\begin{table}[htbp]
    \caption{Time proportion estimate during both Rollout and Training phase using our framework}
    \label{table:time prop}
  \centering
  \begin{spacing}{1}
    \begin{tabularx}{1.0\textwidth} {
         |>{\centering\arraybackslash\hsize=.2\hsize\linewidth=\hsize}X
         |>{\centering\arraybackslash\hsize=.4\hsize\linewidth=\hsize}X
         |>{\centering\arraybackslash\hsize=.3\hsize\linewidth=\hsize}X|
        }
      \hline
          & \textbf{Process} & \textbf{Time proportion} \\
         \hline
         \multirow{4}{*}{Rollout}        & Engine processing       & 73\%  \\
              & Feature encoding       & 10.9\%  \\
        & Policy inference       & 9.5\%  \\
        & Others      & 6.6\%  \\
         \hline
         \multirow{4}{*}{Training}        & Loss computation       & 49.1\%  \\
          & GPU loading      & 44.2\%  \\
           & Data pre-processing      & 5.3\%  \\
            & Others       & 1.3\%  \\
            \hline
    \end{tabularx}
  \end{spacing}
\end{table}


\subsection{Behavior Analysis on Learned Policies}
\begin{figure}[htbp]
    \caption{Behavior illustration in each scenario. \textcolor{red}{\huge$\blacktriangleup$}: left-team player; \textcolor{blue}{\huge$\blacktriangleup$}: right-team player; {\huge$\bullet$}: ball; \textcolor{red}{\huge$\rightarrow$}: left-team player's movement; \textcolor{blue}{\huge$\rightarrow$}: right-team player's movement; \textcolor{black}{\huge$\rightarrow$}: ball's movement
    \textcolor{yellow}{\huge$\rVert$}: offside line. }\label{fig: all behavior}
	\centering
	\begin{subfigure}{0.45\linewidth}
		\centering
		\includegraphics[width=0.85\linewidth]{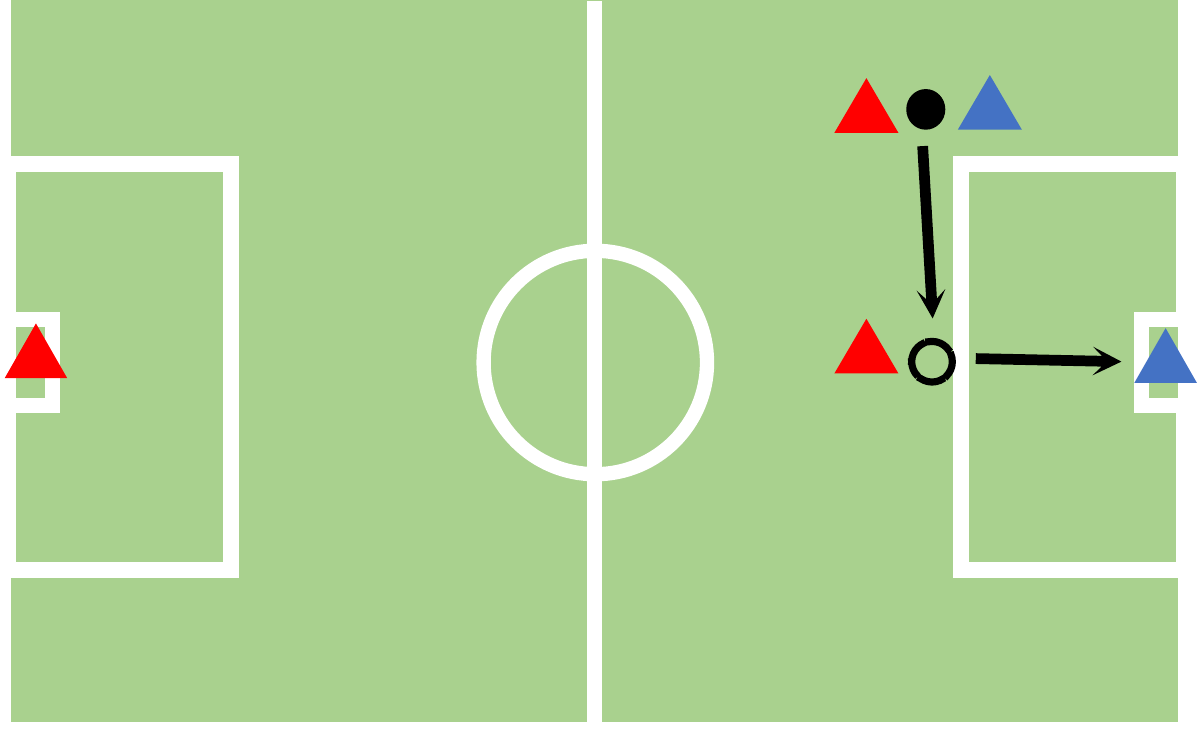}
		\caption{Behavior in \textit{pass and shoot with keeper}}
		\label{behaviour1}
	\end{subfigure}
 	\begin{subfigure}{0.45\linewidth}
		\centering
		\includegraphics[width=0.85\linewidth]{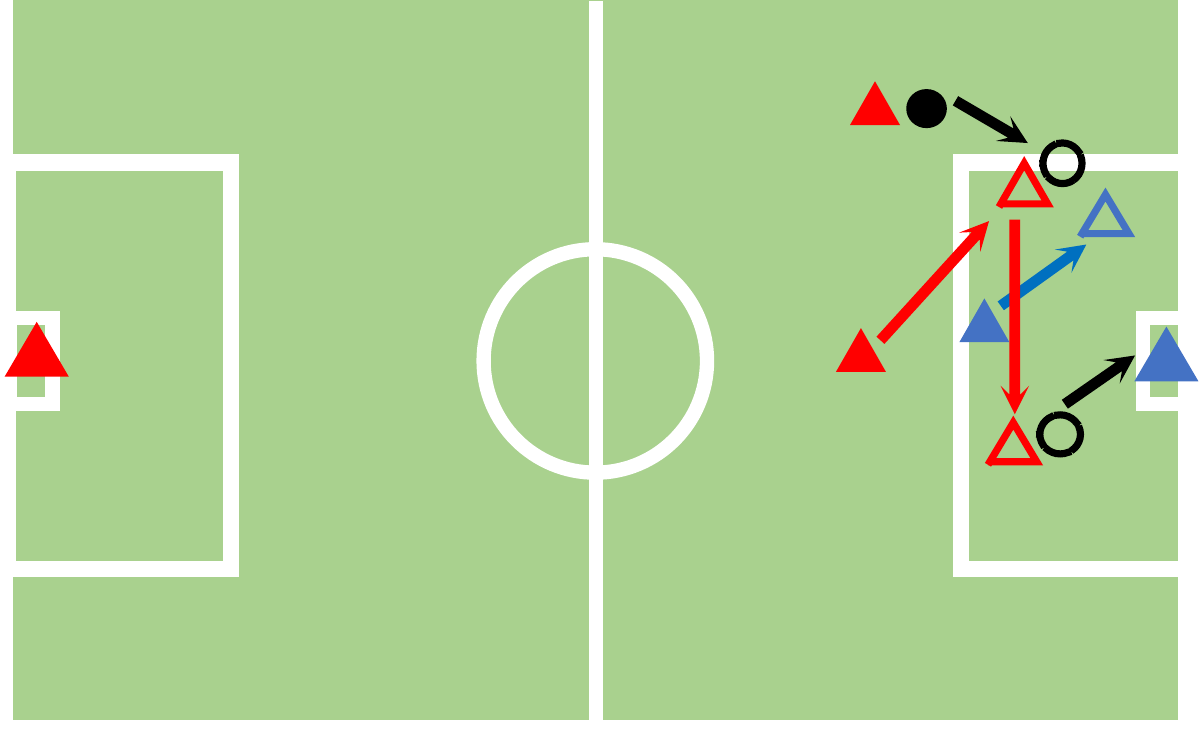}
		\caption{Behavior in \textit{run pass and shoot with keeper}}
		\label{behaviour11}
	\end{subfigure}
	\begin{subfigure}{0.45\linewidth}
		\centering
		\includegraphics[width=0.85\linewidth]{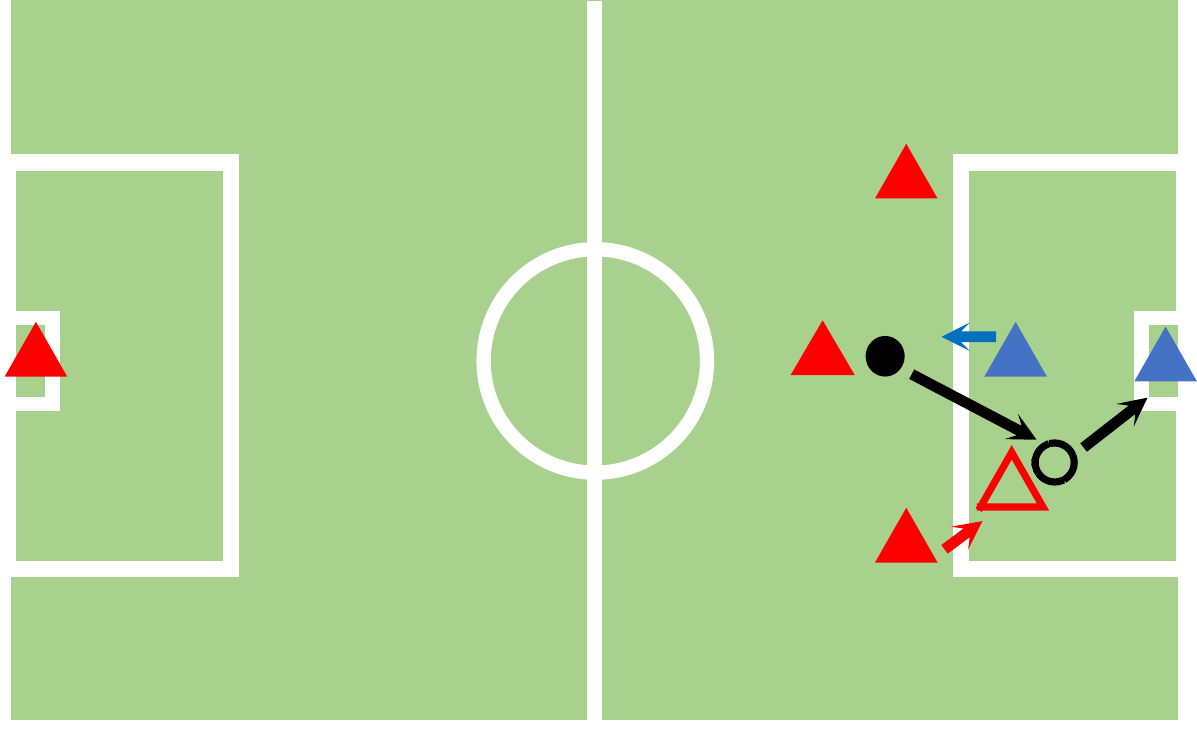}
  		\caption{Behavior in \textit{3 vs 1 with keeper}}
		\label{behaviour2}
	\end{subfigure}
 	\begin{subfigure}{0.45\linewidth}
		\centering
		\includegraphics[width=0.85\linewidth]{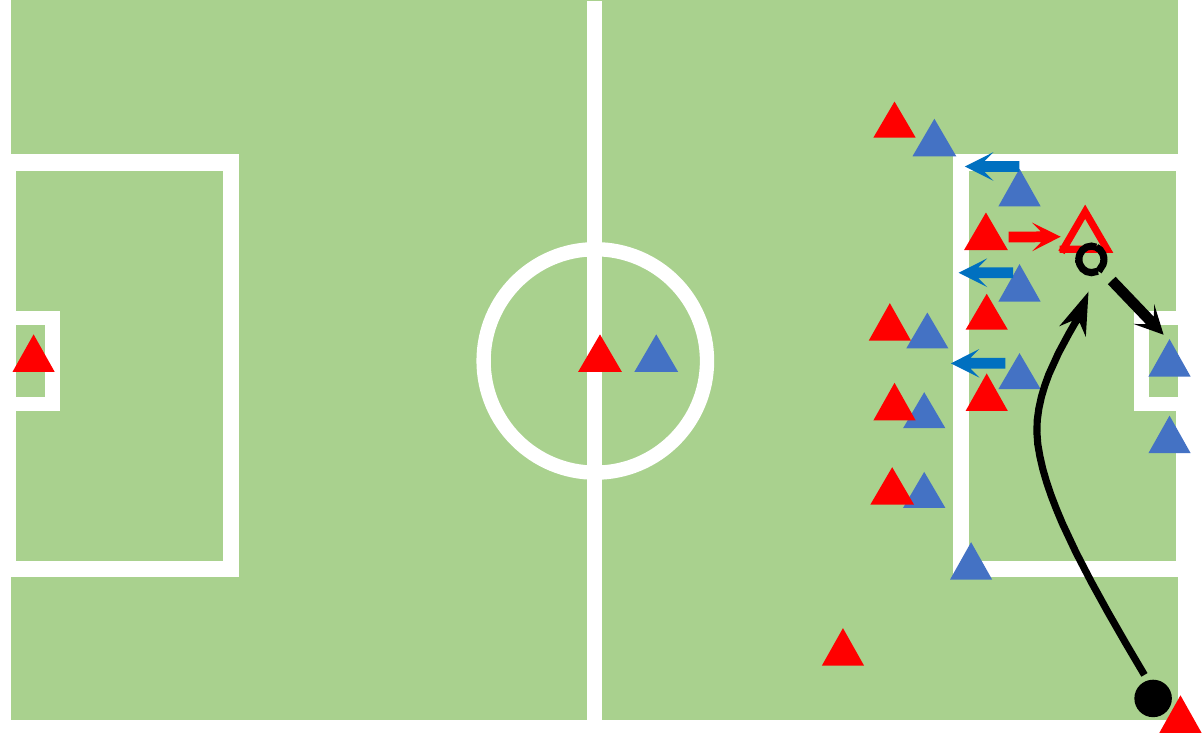}
		\caption{Behavior in \textit{corner}}
		\label{behaviour3}
	\end{subfigure}
 	\begin{subfigure}{0.45\linewidth}
		\centering
		\includegraphics[width=0.85\linewidth]{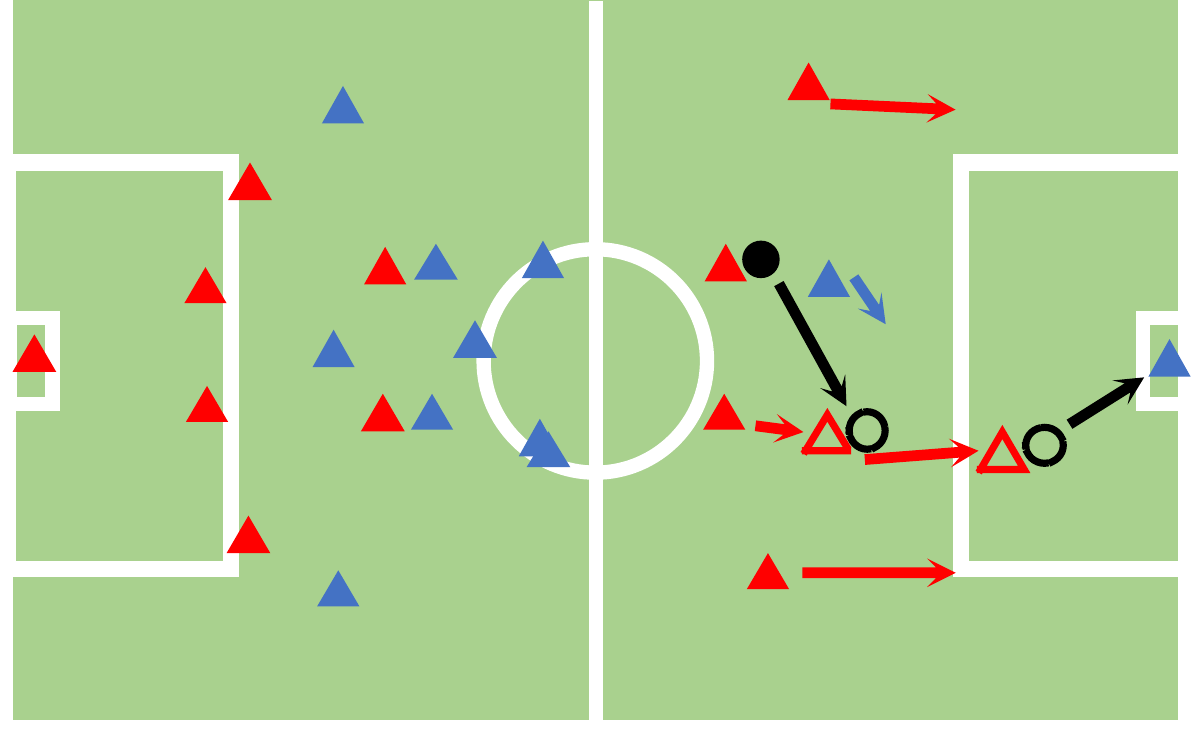}
		\caption{Behavior in \textit{counterattack-easy}}
		\label{behaviour4}
	\end{subfigure}
	\begin{subfigure}{0.45\linewidth}
		\centering
		\includegraphics[width=0.85\linewidth]{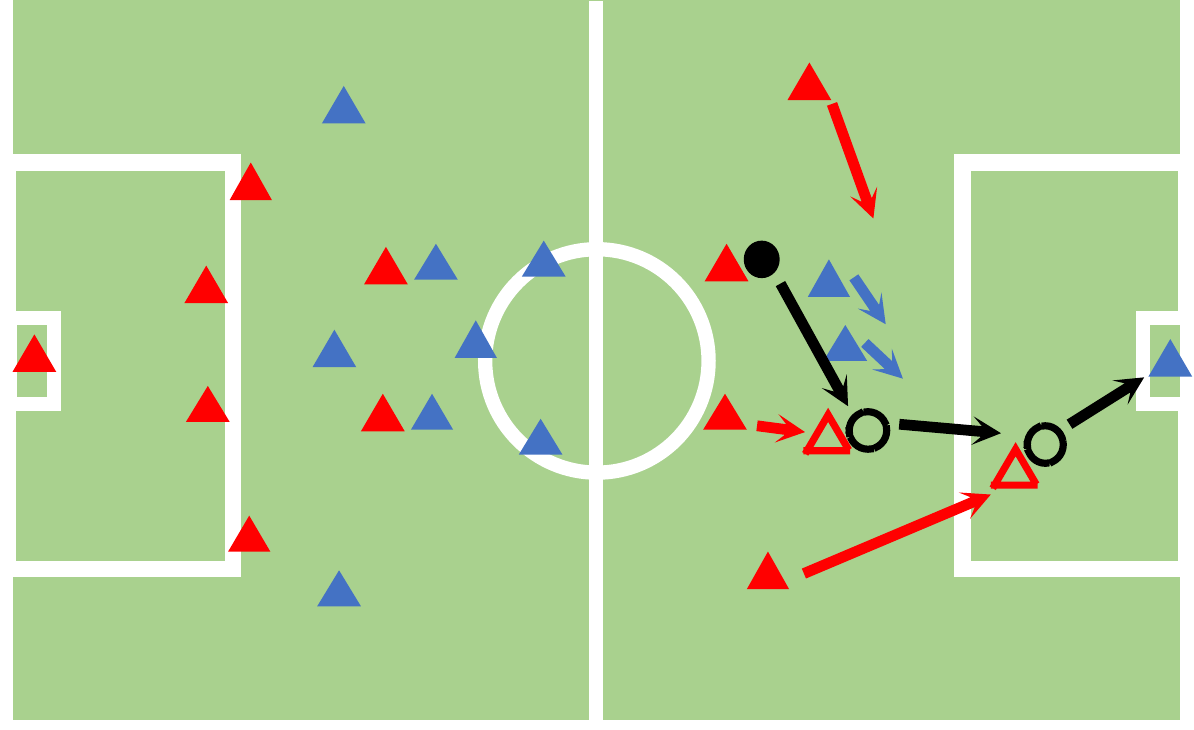}
		\caption{Behavior in \textit{counterattack-hard}}
		\label{behaviour4}
	\end{subfigure}
 	\begin{subfigure}{0.45\linewidth}
		\centering
		\includegraphics[width=0.85\linewidth]{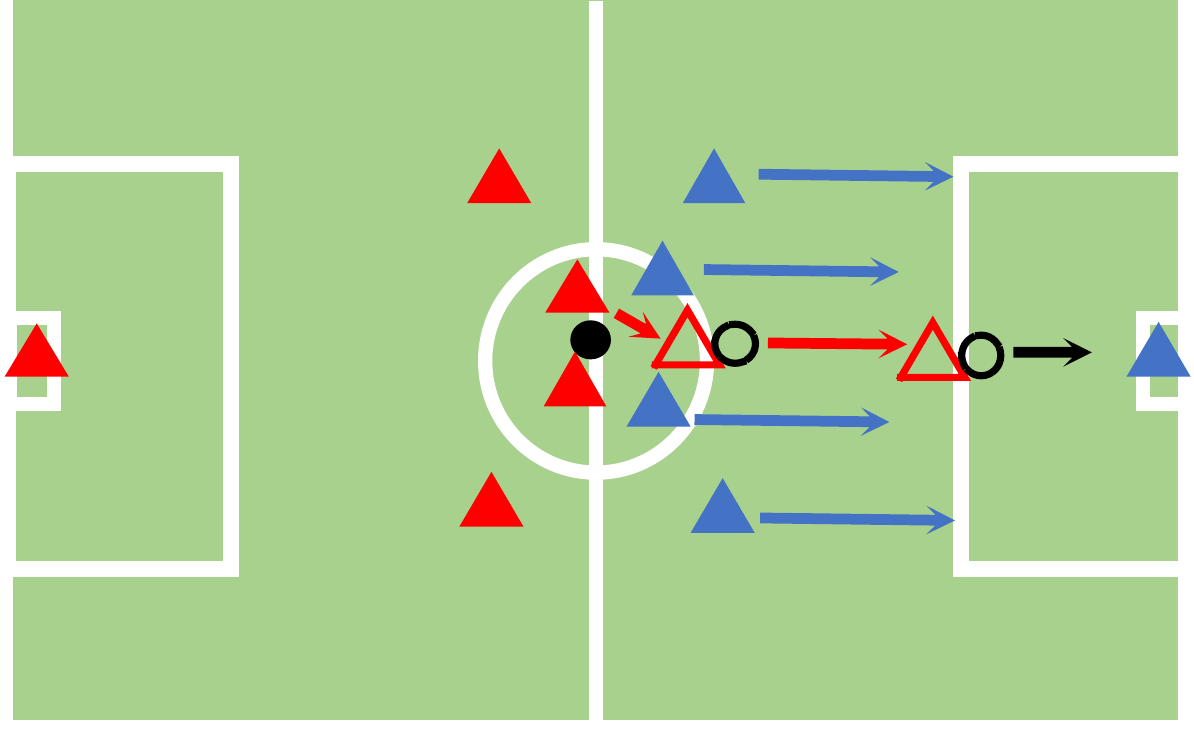}
		\caption{Offensive behavior in \textit{5 vs 5} full game}
		\label{behaviour5}
	\end{subfigure}
	\begin{subfigure}{0.45\linewidth}
		\centering
		\includegraphics[width=0.85\linewidth]{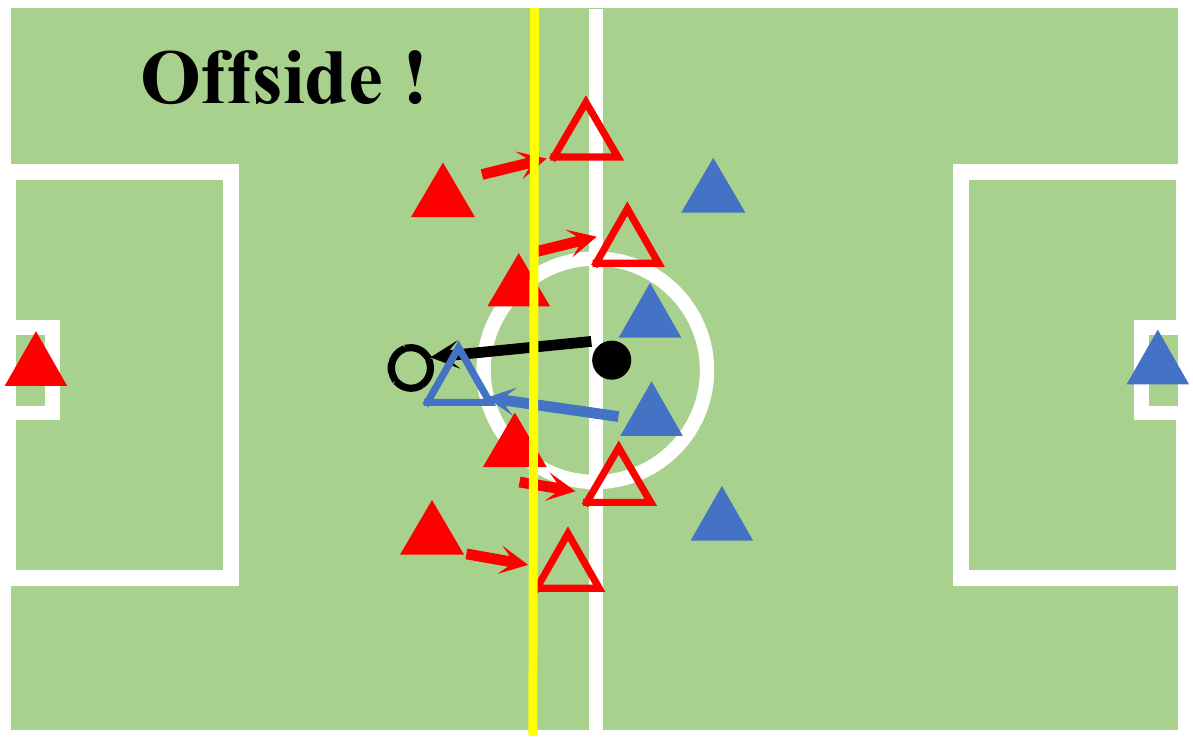}
		\caption{Defensive behavior in \textit{5 vs 5} full game}
		\label{behaviour6}
	\end{subfigure}
  	\begin{subfigure}{0.45\linewidth}
		\centering
		\includegraphics[width=0.85\linewidth]{behaviour_pdf/11v11_offense.pdf}
		\caption{Offensive behavior in \textit{11 vs 11} full game}
		\label{behaviour6}
	\end{subfigure}
	\begin{subfigure}{0.45\linewidth}
		\centering
		\includegraphics[width=0.85\linewidth]{behaviour_pdf/11v11_defense.pdf}
		\caption{Defensive behavior in \textit{11 vs 11} full game}
		\label{behaviour7}
	\end{subfigure}
\end{figure}

The behavior learned by different algorithms converges to a similar strategy in each scenario. Figure \ref{fig: all behavior} summarize and visualize the general strategies in each task. Since we have analyzed behaviors in full-game scenarios in the main paper, we will now focus on the academy scenarios only. In \textit{pass and shoot with keeper}, the player passes the ball to his teammate positioned in the middle, who shoots to the goal. In \textit{3 vs 1 with keeper}, the player passes the ball to either side of the field, allowing the wingers to initiate attacks from the flanks. In \textit{corner}, the player in the penalty area will move forward and seeks space for receiving a high pass from a corner kick and subsequently take a shot at the goal. In \textit{counterattack}, the player engages in consecutive passes until the ball is eventually transferred to the wingers, who then proceed to take shots at the goal.


\section{Research Tools}\label{appendix: research tools}
In this section, we introduce our research tools designed to facilitate large-scale training and in-depth analysis. Our framework supports complex population-based self-play pipelines to explore training beyond overfitting to a fixed opponent. We also provide a match decomposition data structure and a single-step visual debugger, enhancing the analysis of each match at a granular level. Finally, we have fostered a collaborative effort within the community by establishing an online public ranking for football AI. This platform serves as a hub for diverse policies to compete with one another, facilitating a more inclusive and comprehensive evaluation of football AI capabilities.

To the best of our knowledge , our  training framework is the first open-sourced population-based self-play framework tailored to the GRF MARL tasks. Some previous works propose their training framework on GRF but choose not to release it to the public \cite{huang2021tikick,lin2023tizero}. Whereas ours is simple to use and provides detailed tutorials with ready-to-run configurations to build your own football AI.  We also provide novel tools such as the new event data structure and the visual debugger for detailed behaviour analysis on GRF. In addition, the online ranking system is a kaggle-like platform but we are the first to hold GRF MARL competition and provide anytime online evaluation of GRF policies to boost the studies on football AI.



\subsection{The Population-based Self-Play framework}
\label{appendix:framework}

\subsubsection{Motivation}

As discussed in the main paper, training on GRF tasks, particularly in full-game scenarios, is computationally demanding. Additionally, if methods such as self-play are employed, we need to train policies repeatedly. These factors have motivated us to develop a well-designed training framework that prioritizes efficiency and scalability, specifically supporting distributed and asynchronous computing. 

To achieve stronger and less exploitable policies, we have embraced Population-based Self-Play as a general approach. Our framework is designed to seamlessly accommodate various self-play methods, including Naive Self-Play (SP) \cite{alphago}, Fictitious Self-Play (FSP) \cite{fsp}, Policy Space Response Oracle (PSRO) \cite{psro}, and even League Training \cite{alphastar}. By offering such flexibility, we enable researchers to explore and utilize the most suitable self-play method for their specific needs.



\subsubsection{Accessibility and Liscensing}

All code for our distribution population-based self-play framework is available at \url{https://github.com/jidiai/GRF_MARL.git} under the Apache License v2.0.

\subsubsection{Installation and Execution}

The installation procedure of the framework is described as follows:
\begin{enumerate}
    \item Create a Python virtual environment for the experiment;
    \item Clone the repository and installing the required dependencies:\\
    
    \begin{minipage}{5cm}
    \begin{verbatim}
    $ git clone https://github.com/jidiai/GRF_MARL.git
    $ cd GRF_MARL
    $ pip install -r requirements.txt
    \end{verbatim}
    \end{minipage}\\
    
    \item Install the Google Research Football environment following the instruction from the official website: \url{https://github.com/google-research/football};
    \item Install PyTorch dependency for training following instructions from the official website: \url{https://pytorch.org/get-started/locally/}.
\end{enumerate}

After installation, run an example experiment by executing the following command from the home folder of the framework:\\

\begin{minipage}{5cm}\label{code1}
\begin{verbatim}
    $ python3 light_malib/main_pbt.py --config PATH_TO_CONFIG
\end{verbatim}
\end{minipage}\\

Please refer to the repository for more details.

\subsubsection{Workflow}
\begin{figure}[htbp]
    \centering
    \includegraphics[width=0.8\linewidth]{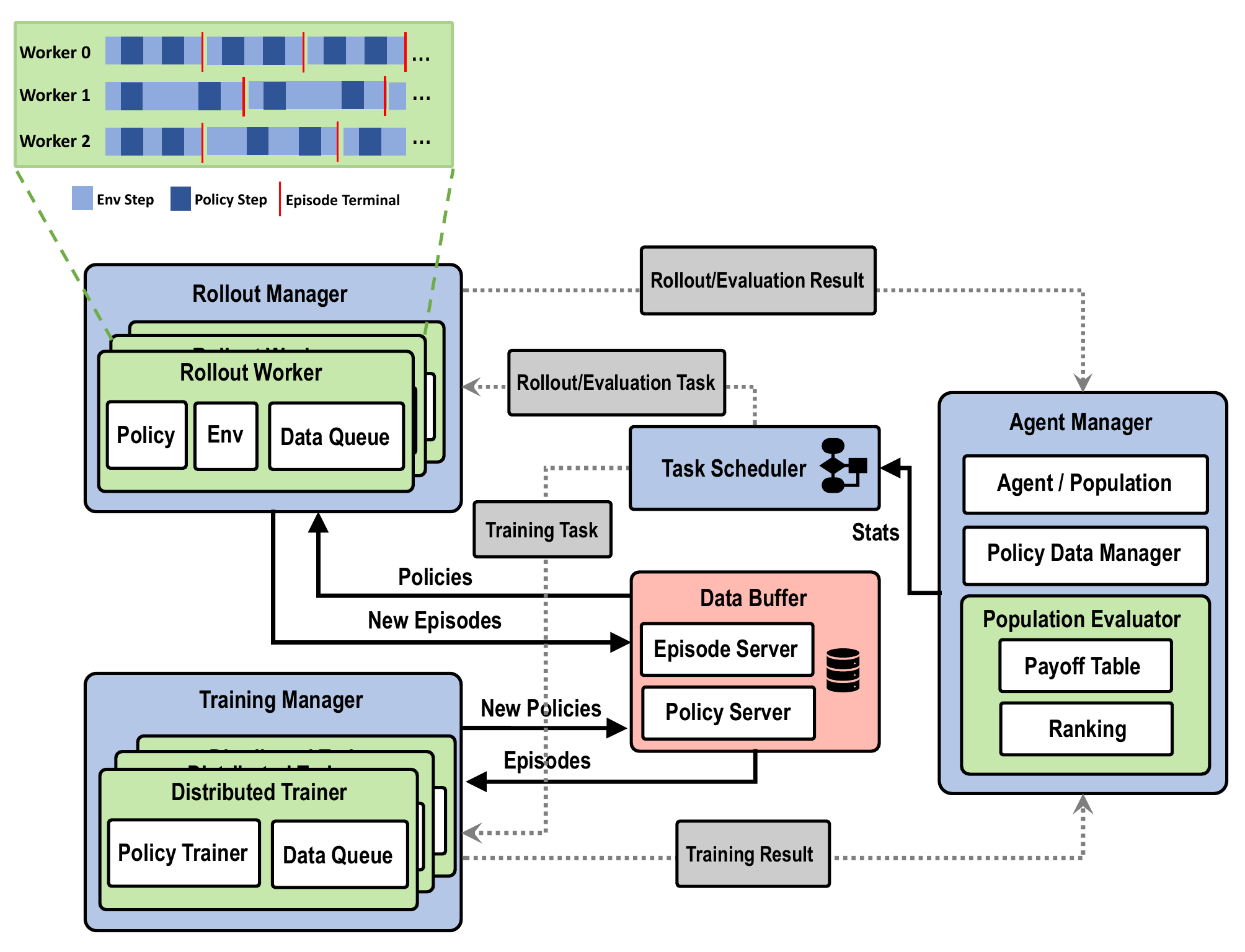}
    \caption{An Illustration of Our Distributed and Asynchronous Population-Based Self-Play Training Framework.}
    \label{figure:framework}
\end{figure}

Our framework design in Figure \ref{figure:framework} draws great inspiration from \textit{MALib} \cite{zhou2023malib} and \textit{RLlib} \cite{liang2018rllib}. It has five major components, each serving a specific role:
\begin{itemize}[topsep=0pt,itemsep=0ex,partopsep=0ex,parsep=0ex]
    \item \textbf{Rollout Manager}: The Rollout Manager establishes multiple parallel rollout workers and delegates rollout tasks to each worker. Each rollout task includes environment settings, policy distributions for simulation, and information pertaining to the Episode Server.
    \item \textbf{Training Manager}: The Training Manager sets up multiple distributed trainers and assigns training tasks to each trainer. Training task descriptions consist of training configurations and details regarding the Policy and Episode buffers.
    \item \textbf{Data Buffer}: The Data Buffer serves as a repository for episodes and policies. The Episode Server saves new episodes submitted by the rollout workers, while trainers retrieve sampled episodes from the Episode Server for training. The Policy Server, on the other hand, stores updated policies submitted by the Training Manager. Rollout workers subsequently fetch these updated policies from the Policy Server for simulation.
    \item \textbf{Agent Manager}: The Agent Manager manages a population of policies and their associated data, which includes pairwise match results and individual rankings.
    \item \textbf{Task Scheduler}: The Task Scheduler is responsible for scheduling and assigning tasks to the Training Manager and Rollout Manager. In each training generation, it selects an opponent distribution based on computed statistics retrieved from the Agent Manager.
\end{itemize}

\subsubsection{Asynchronous Implementation}


Traditionally, most work benchmark algorithms in a synchronous mode to evaluate sample complexity. Namely, the algorithm updates the model after a certain amount of rollout steps. However, in practical large-scale environments such as those in \cite{alphastar, song2023empirical}, rollouts can be very time-consuming. This leads to two bottlenecks: rollouts within a single batch must wait for one another, and trainers must wait for the entire batch of rollouts to complete.

To address these challenges, we have implemented an asynchronous approach that offers improved efficiency during runtime. In the asynchronous mode, rollout and training processes run in parallel, nearly independently, with simple coordination facilitated by a producer-consumer queue. This allows for more efficient utilization of computational resources. Importantly, in the asynchronous mode, samples are allowed to be reused. This means that not only does the waiting time for data decrease, but sample efficiency also increases. By reusing samples, we minimize redundant computations and maximize the utilization of collected data.

\begin{figure}[htbp]
  \centering
  \includegraphics[width=0.8\linewidth]{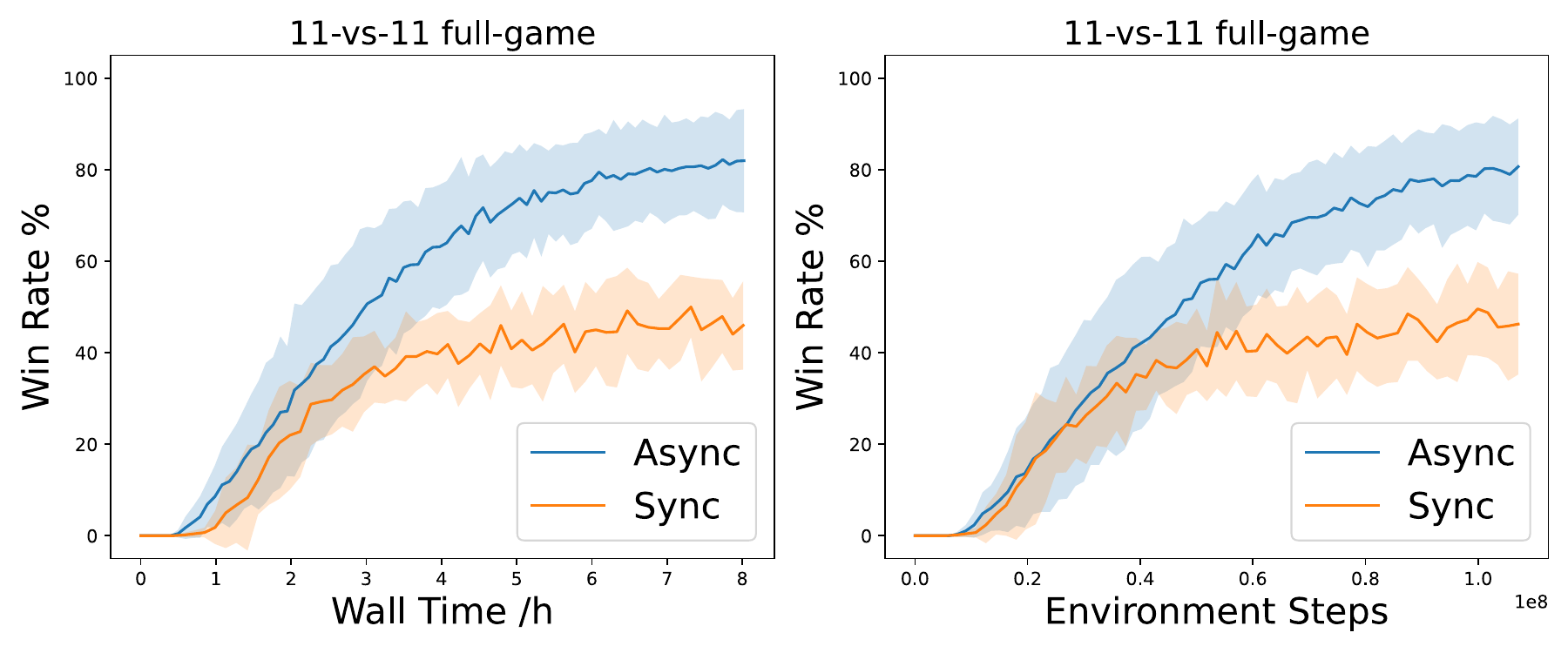}
  \caption{Average win rate with respect to time cost (left) and environment steps (right) under sync/async settings in the \textit{11 vs 11} full-game (with standard deviation). The results are averaged over policy-based algorithms.}
  \label{fig:async_vs_sync}
\end{figure}

To evaluate the effectiveness of the asynchronous implementation, we utilize the \textit{11 vs 11} full-game scenario as our testbed. The results of our experiments in Figure \ref{fig:async_vs_sync} clearly demonstrate the superiority of the asynchronous implementation over the synchronous one in terms of both learning speed and sample efficiency.

\subsubsection{Population-based Self-Play Training}\label{appendix: pbt}

\begin{figure}[htbp]
    \centering
    \begin{subfigure}{0.49\linewidth}
        \raisebox{0.3\height}{\includegraphics[width=\linewidth]{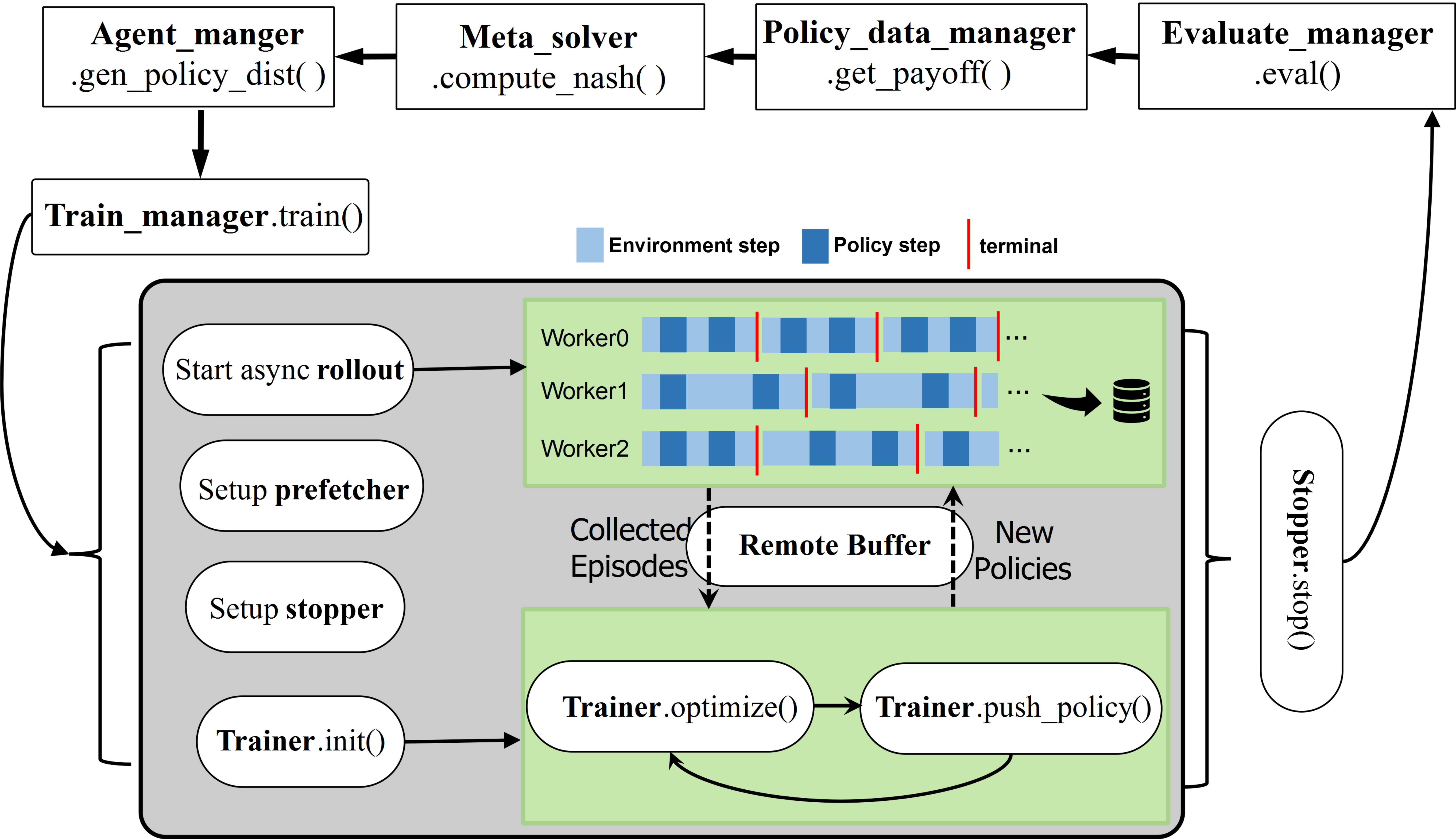}}    
        \caption{An illustration of PSRO training procedure}
        \label{fig:psro depiction}
    \end{subfigure}
    \begin{subfigure}{0.49\linewidth}
            \includegraphics[width=\linewidth]{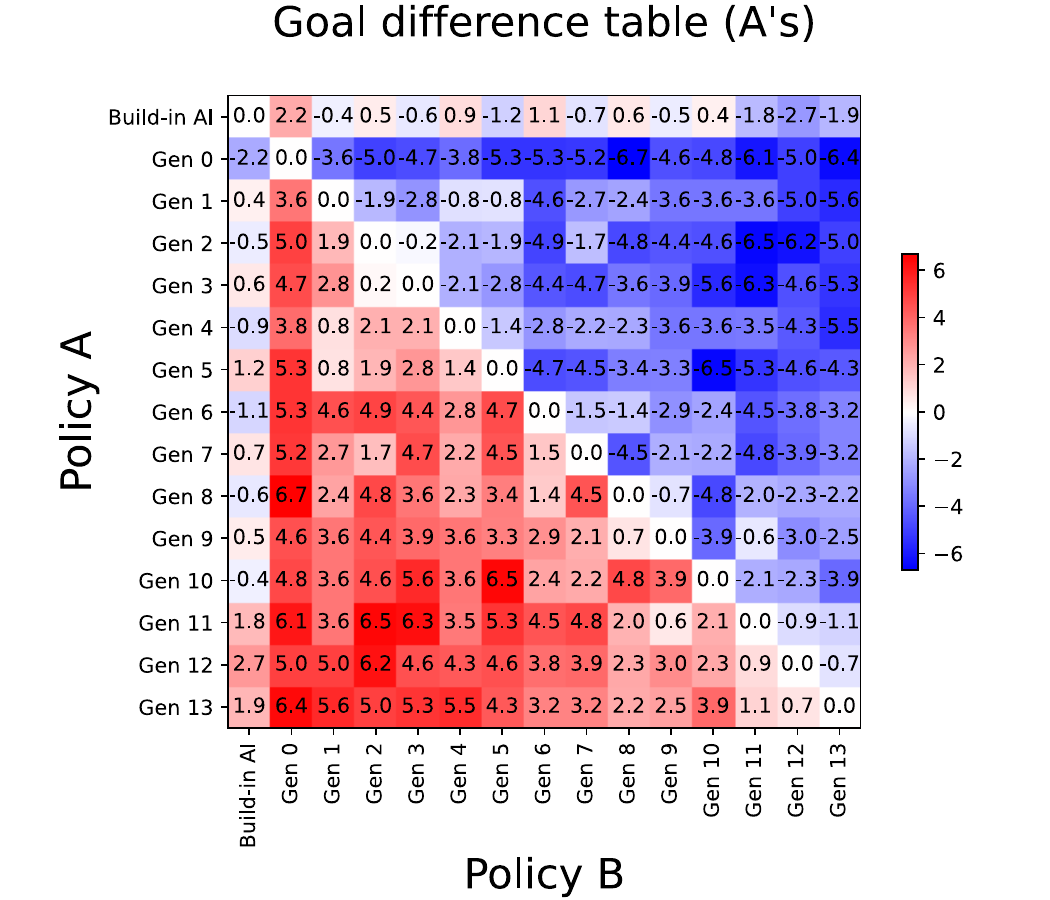}
            \caption{The goal difference table on \textit{5 vs 5} full-game}
            \label{fig:psro payoff}
    \end{subfigure}
    \begin{subfigure}{0.49\linewidth}
        \includegraphics[width=\linewidth]{full_game_pdf/elo.pdf}
        \caption{The Elo scores of all policies in the final population}
        \label{fig:elo}
    \end{subfigure}
    \caption{\textbf{(a)} A depiction of a PSRO example. \textbf{(b)} The goal difference table between built-in AI and best response policies trained at different generations in a PSRO trial. "Gen" is the abbreviation for "generation". The goal difference is calculated by subtracting the score of the column policy from the score of the row policy. A positive goal different often represents higher possibility of winning the game than losing. The training starts with only built-in AI in the population. \textbf{(c)} the Elo scores of all policies in the final population. The policies become stronger as the generation increases in terms of Elo scores.}
    \label{fig:enter-label}
\end{figure}

This section outlines the application of population-based self-play training to enhance the performance of football AI, focusing on the specific example of the Policy-Space Response Oracle (PSRO) algorithm \cite{psro} (Algorithm \ref{algo:psro}). The step-by-step procedure is illustrated in Figure \ref{fig:psro depiction} and can be summarized as follows:

\begin{enumerate}[topsep=0pt,itemsep=0ex,partopsep=0ex,parsep=0ex]
    \item The Evaluation Manager conducts simulations between each pair of policies in the current population. These simulations evaluate the performance of each policy against others, providing valuable information about their relative strengths.
    \item The Policy Data Manager updates the payoff table based on the simulation results. The payoff table captures the performance metrics and outcomes of the policy interactions. Using this information, the manager computes the Nash equilibrium.
    \item The Agent Manager records the simulation results and generates the Nash mixture distribution of opponent policies.
    \item Training and rollout processes are executed according to the framework illustrated in Figure \ref{figure:framework}. The rollout process simulates matches between the policies, while the training process involves updating the policies using the collected data. This process is monitored and terminated by the Stopper component. The Prefetcher component preloads data to expedite the training process.
    \item The trained policy for the current generation is stored in the population. The procedure then returns to step one, initiating the next generation of evaluation and training.
\end{enumerate}



Figure \ref{fig:psro payoff} and Figure \ref{fig:elo} provide visual representations of the policy evolution in a PSRO experiment, showcasing the progression of payoffs and Elo scores. The experiment begins with a simple population consisting solely of the built-in AI. At each generation, a new best response policy is trained and added to the population. We can observe the non-transitivity of the game by analyzing the payoff tables in Figure \ref{fig:psro payoff}. For example, in \texttt{Gen 1}, the policy beats the built-in AI, and in \texttt{Gen 2}, it surpasses \texttt{Gen 1}. However, \texttt{Gen 2} is subsequently defeated by the built-in AI. This non-transitivity implies the complex nature of the game and the difficulty of achieving consistent superiority. However, as the training progresses, the algorithm gradually overcomes the non-transitivity and achieves dominance over every policy in the population by \texttt{Gen 11}. The resulting strategy from this PSRO experiment exhibits similar gameplay characteristics to one of our released pre-trained policies, specifically the \texttt{Group Defense} policy (located at the bottom-left position in Figure \ref{fig:radar_5vs5}). This is because we adopt the reward shaping scheme involving player-ball distance (More details can be found in Appendix \ref{appendix: how to train}).

Furthermore, it's worth noting that our code supports other population-based self-play pipelines, such as the \textit{League Training} algorithm depicted in Algorithm \ref{algo:league}. This flexibility allows for experimentation with different training methodologies and the exploration of diverse AI strategies.

\begin{minipage}{\textwidth}
\begin{algorithm}[H]
	\caption{A Modified PSRO Training Pipeline}     
	 \label{algo:psro}       
    \begin{algorithmic}[1] 
	\Require an initial population $\Phi_{\beta}$ with size $\beta$ (Built-in AI included)   
    \Ensure  a set of best response policies $\pi_{1}, ..., \pi_{G}$   
    \State Initialize $\mathcal{M}_{\beta\times\beta}$ as the payoff matrix recording simulation results of the population   
    \For {$j=1,2,...,G$}
       \State Compute Nash equilibrium policy distribution $p_j$ from payoff $\mathcal{M}_{\beta\times\beta}$
       \State Set the opponent policy $\pi_o$ with a probability $p$ sampled from $p_j$
       \State Train a best response policy $\pi_j=\texttt{BestResponse}(\pi_o)$ initialized from $\pi_{j-1}$ using RL
       \State Save $\pi_j$ into the population and get $\Phi_{\beta+1}$ 
       \State Perform simulation and update payoff $\mathcal{M}_{(\beta+1)\times(\beta+1)}$
   \EndFor
   \State\Return a set of best response policies $\pi_1, ..., \pi_G$ 
   \end{algorithmic} 
\end{algorithm}
\end{minipage}


\begin{minipage}{\textwidth}
\begin{algorithm}[H]
	\caption{A Simplified League Training Pipeline}     
	 \label{algo:league}       
    \begin{algorithmic}[1] 
	\Require an initial population $\Phi_{\beta}$ with size $\beta$ (Built-in AI included), an initial main-agent policy $\pi_{m,0}$   
    \Ensure  a set of exploiter policy $\pi_{e,1},...,\pi_{e,G}$ and a final main-agent policies $\pi_{m,G}$   
    \State Initialize $\mathcal{M}_{\beta\times\beta}$ as the payoff matrix recording simulation results of the population   
    \For {$j=1,2,...,G$}
       \State Compute distribution $p_j$ from  $\mathcal{M}_{\beta\times\beta}$ following Prioritized Fictitious Self-Play (PFSP) \cite{alphastar}. 
       \State Set the opponent policy $\pi_o$ with a probability $p$ sampled from $p_j$
       \State Initialize from $\pi_{m,j-1}$ and train $\pi_{m,j}=\texttt{BestResponse}(\pi_o)$ 
       \State Save $\pi_{m,j}$ into the population $\Phi_{\beta+1}$ 
       \State Perform simulation and update payoff $\mathcal{M}$
       \If {main policy has finished training}
            \State Initialize an exploiter policy $\pi_{e,j}$ 
            \State Set $\pi_{m,j}$ as the opponent
            \State Train the exploiter $\pi_{e,j}=\texttt{BestResponse}(\pi_{m,j})$
            \State Save $\pi_{e,j}$ into $\Phi_{\beta+1}$
            \State Perform simulation and update payoff $\mathcal{M}$
   \EndIf
   \EndFor
   \State\Return a set of best response policies 
   \end{algorithmic} 
\end{algorithm}
\end{minipage}

\subsubsection{How we train stronger policies}\label{appendix: how to train}
\revision{We perform complex training pipeline introduced in Appendix \ref{appendix: pbt} with manual intervention. Our procedure can be summarized as follows:
\begin{enumerate}
    \item Perform Algorithm \ref{algo:psro} to promptly building up sets of policies using different reward shaping schemes;
    \item Perform Algorithm \ref{algo:league} to further improve the generalized abilities of the policies;
\end{enumerate}
In Phase 1, our goal is to discover a group of trained policies that behave differently so that for each style-of-play we have obtained, there is a ready-to-used trained policy that can give a warm-start for Phase 2. The way to do it is by carrying out multiple PSRO trials with different carefully-designed reward shaping schemes (i.e. see Table \ref{table:advanced reward} for details). For example, to learn a high pressuring tactics (i.e. \textit{Group Pressure} in Appendix \ref{appendix:pretrained policies}) we may add the negative distance between players and the ball to the reward function:
\[
R = \texttt{SCORING} + 0.1 \times \texttt{Ball-Player Distance} + \mathbbm{1}_{t=T}\times \texttt{Goal Difference}
\]

or to encourage assistance by backtracking the player who perform a successful assist using data structure introduced in Appendix \ref{appendix:data structure}:
\[
R = \texttt{Role-based SCORING} + 0.3\times \texttt{Assist}
\]

Ideally, the results of Phase 1 are sets of trained policies as representations of each of the reward shaping schemes.

\begin{table}[htbp]
    \caption{More advanced reward shaping for Google Research Football}
    \label{table:advanced reward}
  \centering
  \begin{spacing}{1}
    \renewcommand{\arraystretch}{2} 
    \begin{tabularx}{1.\textwidth} {
         >{\centering\arraybackslash\hsize=.2\hsize\linewidth=\hsize}X
         >{\centering\arraybackslash\hsize=.8\hsize\linewidth=\hsize}X
        }
      \hline
         \textbf{Reward Names}  & \textbf{Description}  \\
         \hline
         \textit{Ball-Player Distance}         & The negative distance between the current controlled player and the ball, optimizing this encourages pressuring.   \\ 
         \textit{Goal Difference}        &  The difference in goals, optimizing this encourages offense. \\
         \textit{Possession}  &  Give penalty $-1$ if the current player loses possession of the ball and give reward $+1$ if the player gains possession, optimizing this encourages ball possession. \\
         \textit{Role-based SCORING}  &  Assigning weighted SCORING reward to each players according to their roles. For example, the forwards receive $50\%$ penalty and the guards receive $50\%$ reward, optimizing this encourage team formation.\\
         \textit{Passing}   &  $+1$ if a pass is successful or $-1$ if a pass fails\\
         \textit{Assist}    & $+1$ if a player pass the ball leading directly to a goal\\
         \hline
    \end{tabularx}
  \end{spacing}
\end{table}

In Phase 2, we use simplified League Training pipeline to continuously improve the \textit{main-agent} policy we wish to develop. The setting up of \textit{exploiter} can potentially discover strategical weakness of the main-agent policy and counter-tactics with new style-of-play. the reward shaping used in Phase 2 is the SCORING reward as we want to avoid any constraint posed on the learned behaviour and only focus on winning the game. Note that the complete iterative training procedure might not be reproducible as it may requires large degree of manual intervention such as tuning the agent to avoid a particularly bad behaviour, or cherry-picking the trained policies for the next trial according to human prior knowledge. All these tricks are to reduce the computational time and guide the large-scale training procedure towards a more human-understandable direction, while complicating the reproducibility. As a result, we choose to release some of the trained policies we have obtained from the pipeline and the codebase as useful tool-kits for potential future attempts. In the future, we will continue working on a more efficient training procedure with better reproducibility.}

\subsection{Pretrained Policies}
\label{appendix:pretrained policies}

\begin{figure}[htbp]
    \centering
    \begin{subfigure}{0.49\linewidth}
        \includegraphics[width=\linewidth]{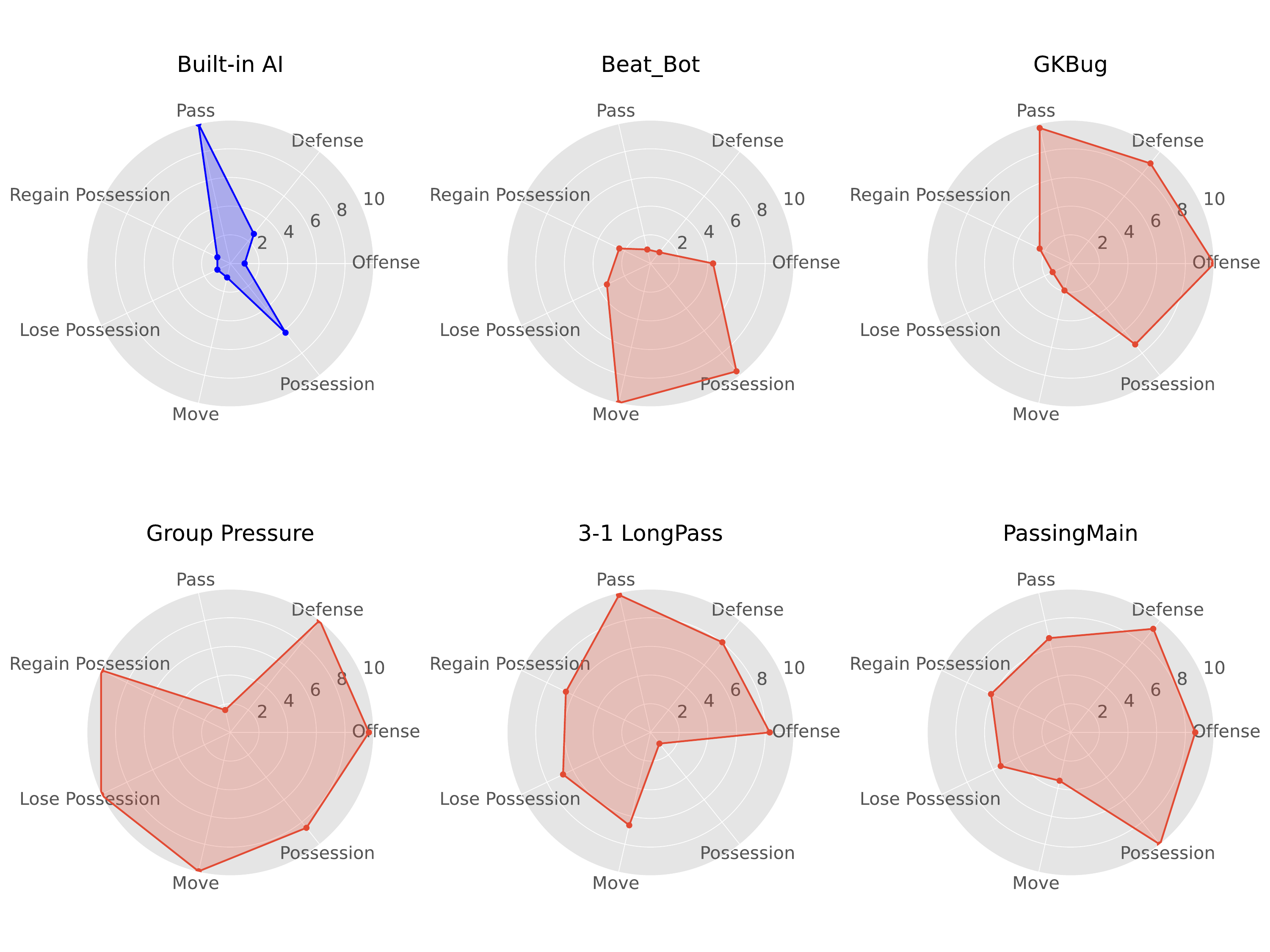}
        \caption{Styles of pre-trained policies in \textit{5 vs 5}}
        \label{fig:radar_5vs5}
    \end{subfigure}
    \begin{subfigure}{0.49\linewidth}
        \includegraphics[width=\linewidth]{full_game_pdf/radar_11v11.pdf}
        \caption{Styles of pre-trained policies in \textit{11 vs 11}}
        \label{fig:radar_11vs11}
    \end{subfigure}
    \begin{subfigure}{0.49\linewidth}
        \includegraphics[width=\linewidth]{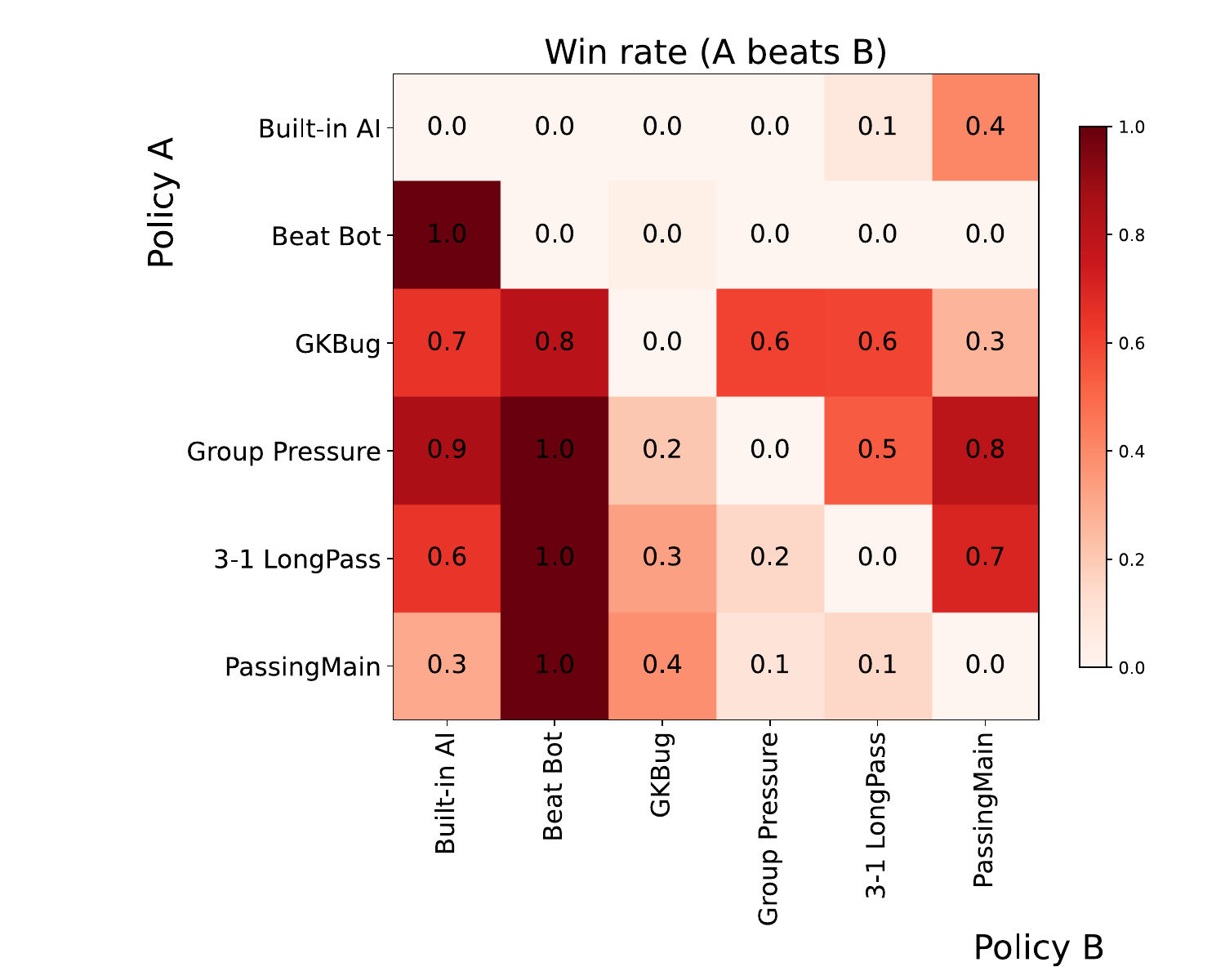}
        \caption{Cross-play win rate of \textit{5 vs 5} pre-trained policies}
        \label{fig:crossplay_5}
    \end{subfigure}
    \begin{subfigure}{0.49\linewidth}
        \includegraphics[width=\linewidth]{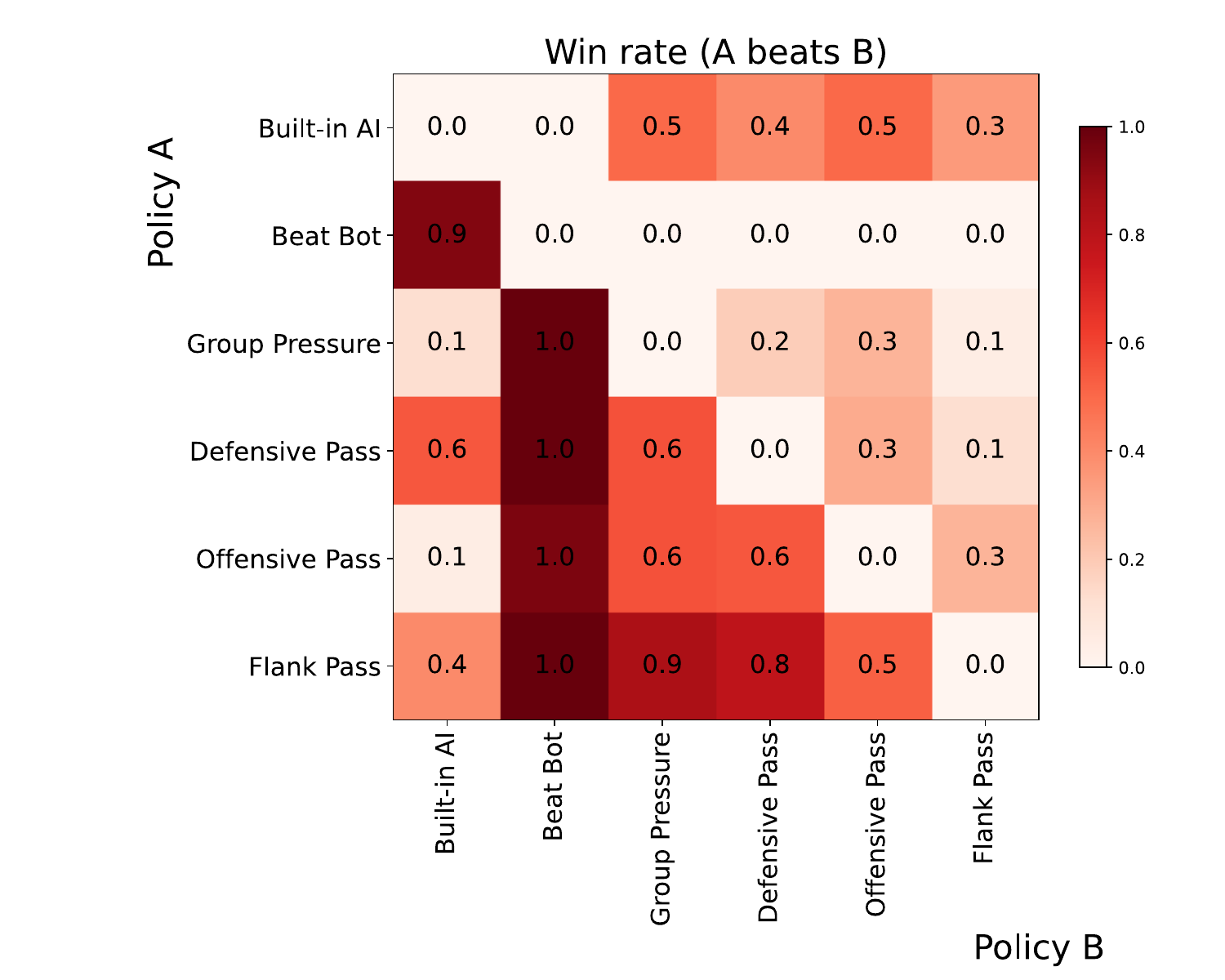}
        \caption{Cross-play win rate of \textit{11 vs 11} pre-trained policies}
        \label{fig:crossplay_11}
    \end{subfigure}
    \caption{Pre-trained policies analysis. \textbf{(a),(b)} The policy covering more dimension in a radar plot is considered generally stronger in the population. Built-in AI is strong in passing and ball possession but weak in other skills. \textbf{(c),(d)} Win rate matrix demonstrating simulation results between pre-trained policies.   }
    \label{fig:radars}
\end{figure}

Figure \ref{fig:radars} provides radar plots and cross-play simulation results depicting diverse policies obtained through population-based self-play training with various configurations. For each full-game scenario, we carefully select representational polices and group them alongside the built-in AI. Subsequently, we conduct massive one-on-one game simulations, collecting football statistics for each policy using a match data structure (discussed in Appendix \ref{appendix:data structure}). Finally, we normalize the football skills exhibited by each policy and visualize them through radar plots. To foster collaboration and further research, we have made these pre-trained policies open-source. They can be used as opponent during training and evaluation, or as valuable resources for model initialization and imitation learning \cite{Ho2016GenerativeAI}. Each policy is also accompanied by a brief description that highlights its key characteristics in Table \ref{table:pretrained}.

\revision{The differences in style-of-play are due to reward shaping and training configurations. In the game of \textit{5 vs 5} full-game, policy \textit{Beat Bot} is trained against the built-in AI directly from scratch. Policy \textit{Group Pressure} is obtained from Algorithm \ref{algo:psro} with a negative distance between the player and the ball added to the reward function. Policies \textit{3-1 LongPass} and \textit{PassingMain} are initially exploiter agents that have learned good ball passing skills and are selected as the main agent to further improve their passing skills. Policy \textit{GKBug} is a main agent that keeps improving as the League Training pipeline proceeds and learns to distract the opponent's goalie to easily score. In the game of \textit{11 vs 11 full-game}, policy \textit{Group Pressure} is obtained in the same way as the one in \textit{5 vs 5 full-game}. Policies \textit{Defensive Pass} and \textit{Offensive Pass} are also exploiter-agent at the beginning and learn to improve their ball-passing skills in a League Training trial. Policy \textit{Flank Pass} is the main agent from the start.}

\begin{table}[htbp]
    \caption{Pretrained policies description}
    \label{table:pretrained}
  \centering
  \begin{spacing}{1.5}
    \begin{tabularx}{1.0\textwidth} {
         >{\centering\arraybackslash\hsize=0.25\hsize\linewidth=\hsize}X
         >{\centering\arraybackslash\hsize=0.45\hsize\linewidth=\hsize}X
         >{\centering\arraybackslash\hsize=1\hsize\linewidth=\hsize}X
        }
      \hline
           & \textbf{Policy Name} & \textbf{Description} \\
         \hline
         \multirow{13}{*}{\textit{5 vs 5}} & \texttt{Group Pressure 5} & All four players group together and chase the ball, pressuring the opponents\\
         \cline{2-3}
         & \texttt{PassingMain} & Players short-pass the ball a lot and prefer a 1-3 formation with three players at the front court to organize offense\\
         \cline{2-3}
         & \texttt{3-1-LongPass} & Players good at long-pass and prefer a 3-1 formation with three players at the back-court to perform good defense and one player at the front to attack when receiving long pass\\
         \cline{2-3}
         & \texttt{GKBug} &  Similar to 3-1-LongPass but the back-court player long-pass to goal while the player at the front-court runs to distract the opponent's goalkeeper so that the ball will not be saved. \\
         \hline
         \multirow{9}{*}{\textit{11 vs 11}} &  \texttt{Group Pressure 11} & All ten players group together and chase the ball, pressuring the opponents.\\
         \cline{2-3}
         & \texttt{Defensive Pass} & The team is good at passing the ball and focuses more on defense. \\
         \cline{2-3}
         & \texttt{Offensive Pass} & The team is good at passing the ball and focuses more on offense.\\
         \cline{2-3}
         & \texttt{Flank Pass} & The team is good at formation and attacks mainly by dribbling the ball to the sides and then high-pass the ball to the center. \\
         \hline
    \end{tabularx}
  \end{spacing}
\end{table}

\subsection{Match Decomposition Data Structure}
\label{appendix:data structure}

\begin{figure}[htbp]
    \centering
    \includegraphics[width=0.9\linewidth]{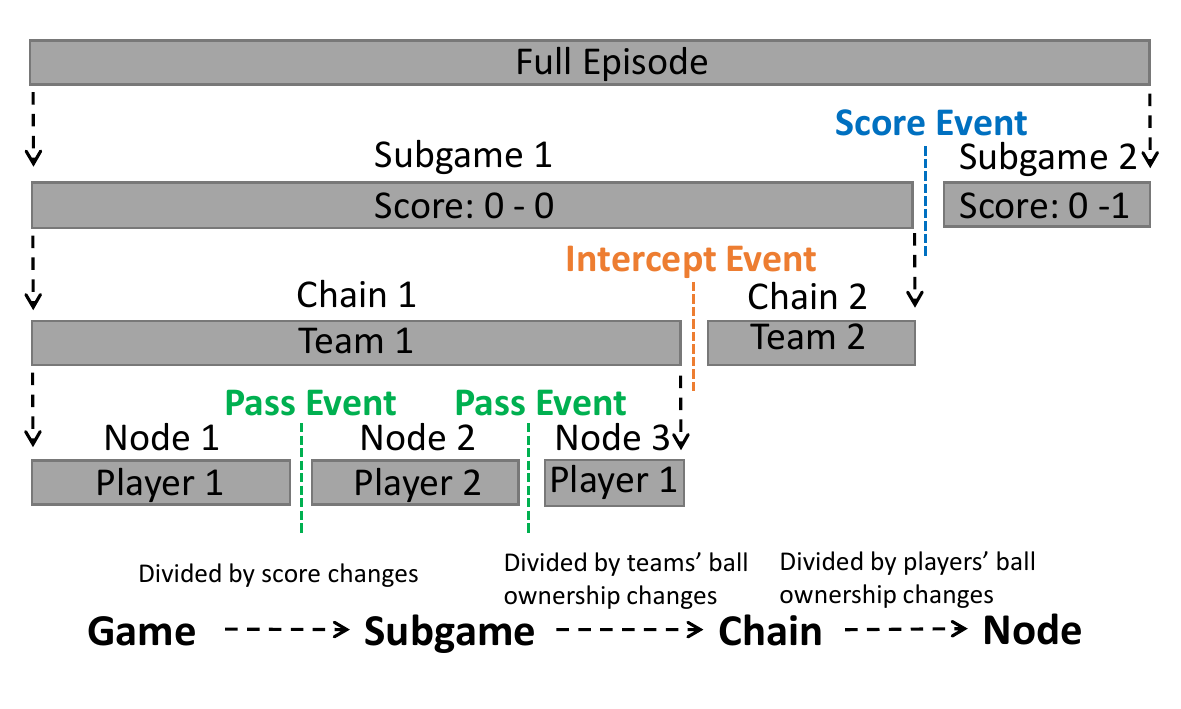}
    \caption{The match decomposition data structure.}
    \label{fig:data_structure}
\end{figure}

In order to enhance event detection and statistical calculations, we introduce a match decomposition data structure (Figure \ref{fig:data_structure}) analogous to complex analysis in real football matches.  \revision{There have been previous attempts on automatic event detection \cite{VidalCodina2022AutomaticED} and behaviour analysis software \cite{McGuckin2020DeterminantsOS} in real football analytics. However, to the best of our knowledge, no such attempts have been made publicly to the game of GRF and the default game replay tools are inconvenient to used for in-depth analysis.} In our new data structure, the complete game episode is initially segmented based on score changes, as when a team scores, both teams restart the game from the center circle and the on-pitch situation is essentially reset. Each segment is referred to as a \textit{subgame} in our framework. Further division of a subgame occurs through ball ownership changes between teams, resulting in \textit{chains}. Consequently, within each chain, the ball is consistently possessed by a single team, passing between teammates. Each chain is constituted by multiple nodes representing individual players in possession of the ball. By default, during the ball's transit to another player, we assume ownership remains with the passer. 


This approach renders certain simple yet crucial events readily discernible. For instance, an intercept event occurs between two chains within a subgame, while a passing event occurs between two nodes within a chain. Consequently, we can effortlessly count these events as indicators of our policies' performances. Moreover, we can compute more intricate statistics, such as the number of assists or the average length of passing sequences. Note that an assist is counted when a pass from the second-to-last node to the final node precedes a goal.


The benefit of such a data structure is that we can more formally and clearly define the computation of these statistics. They can be employed not only for playing style analysis but also for the development of sophisticated reward systems.

\subsection{Single-Step Visual Debugger}\label{appendix:debugger}

To facilitate algorithm debugging and policy analysis, a valuable approach is to review replays of simulated episodes. While the original GRF environment offers this feature by generating a 3D view video, we have encountered certain inconveniences. Firstly, it is challenging to accurately navigate to key frames using conventional video players. Secondly, there is a desire to delve into specific details regarding players and the ball. Lastly, some of the data provided by the GRF environment, such as normalized positions and speeds, are not easily interpretable by human observers. As a result, we have developed a visual debugger to enhance the replay functionality.


\begin{figure}[htbp]
    \centering
    \includegraphics[width=0.9\linewidth]{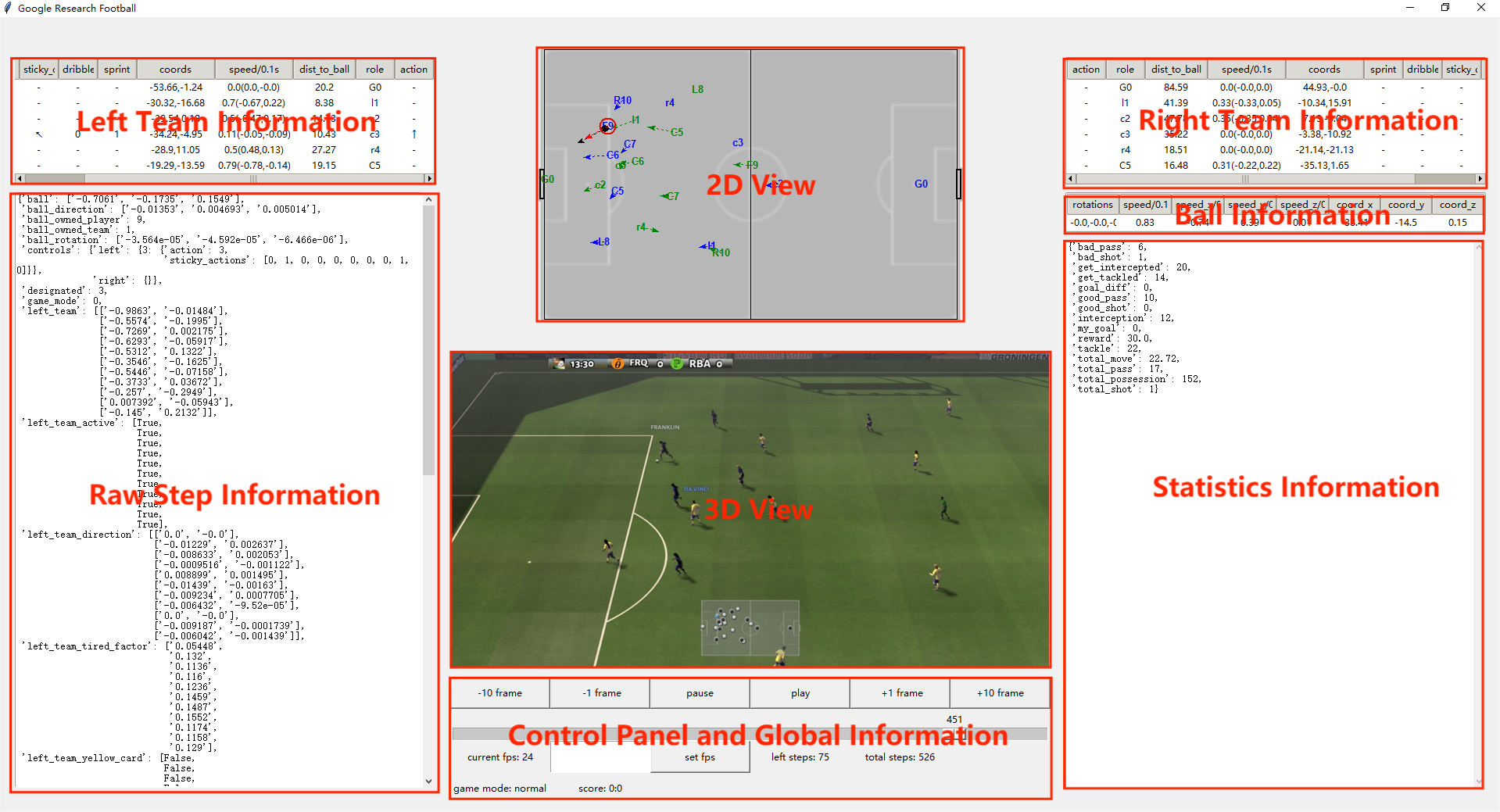}
    \caption{The single-step visual debugger with panel annotations.}
    \label{fig:debugger}
\end{figure}

Our visual debugger, as depicted in Figure \ref{fig:debugger}, comprises a comprehensive set of 8 panels. The top-left and top-right panels present information regarding the players in the left and right teams, respectively. The smaller panel beneath the top-right panel provides details about the ball. To ensure user-friendliness, the information in these panels is presented in an unnormalized format, enabling easy comprehension based on common understanding. The primary left and right panels display the step information provided by the environment and the corresponding cumulative statistics, respectively. Positioned in the top-middle section, we have a bird's-eye view of the pitch. Here, players from the right and left teams are represented by green and blue circles, respectively, while the ball is depicted as a black circle. The size of the ball circle conveys the height of the ball, with larger sizes indicating greater altitude. To identify the player in possession of the ball, a red circle is employed. Additionally, arrows are incorporated into the graph, with their directions and lengths indicating the movement direction and speed of the objects. The central panel shows the original 3D view, while the bottom-center panel primarily serves as a control interface for playing frames. It also provides global information, such as scores and game modes. It remains possible to view the replay like a video. Moreover, users can manually adjust the slider to a precise frame and navigate forward or backward by 1 or 10 steps by clicking the corresponding buttons.


Overall, the debugger significantly aids users in swiftly locating keyframes and scrutinizing the events occurring at each frame.

\subsection{JIDI GRF Online Ranking}

\begin{figure}[htbp]
    \centering
    \begin{subfigure}{0.45\linewidth}
        \includegraphics[width=\linewidth]{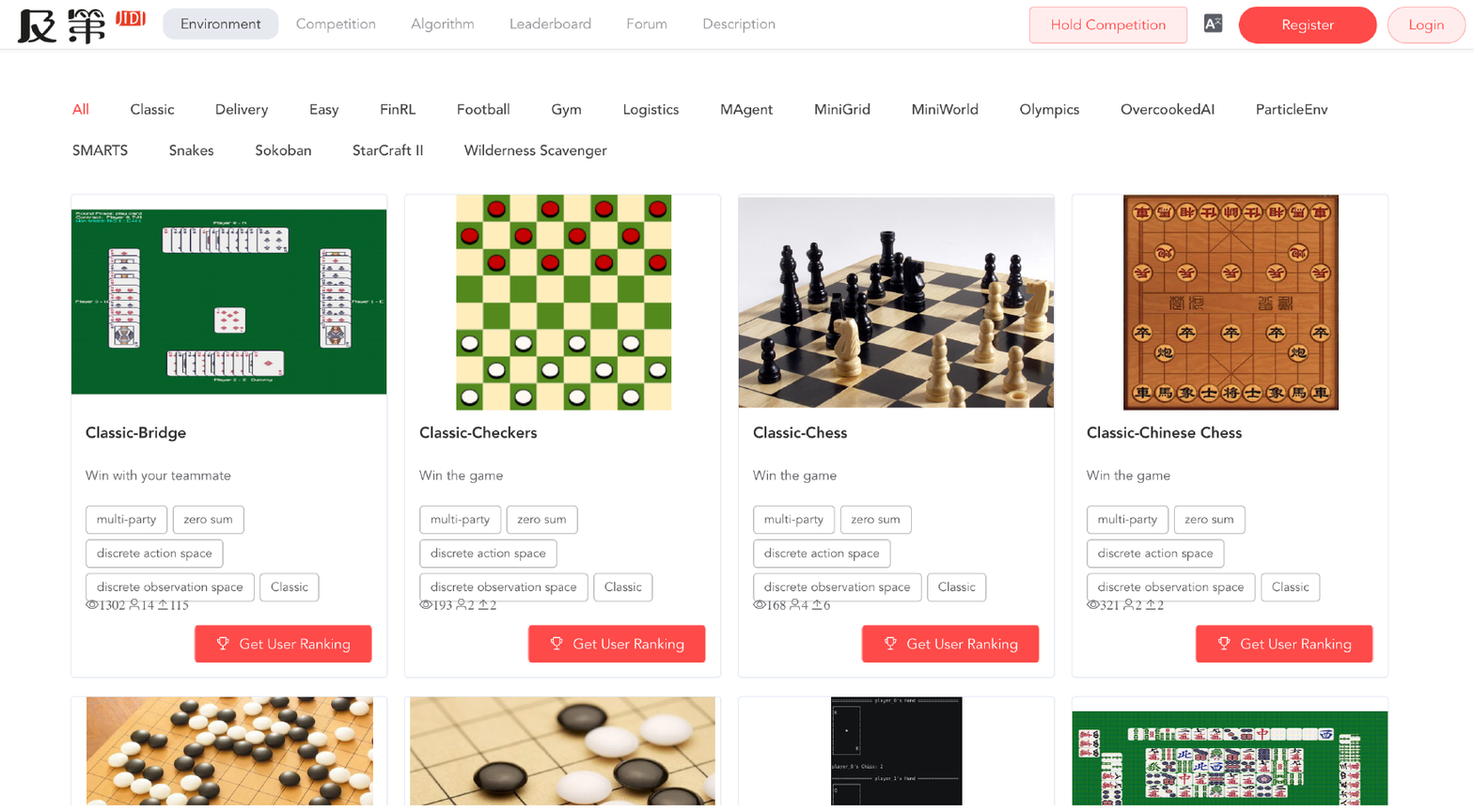}
        \caption{JIDI environment-specific ranking}\label{fig:jidi website 1}
    \end{subfigure}
    \begin{subfigure}{0.45\linewidth}
        \includegraphics[width=\linewidth]{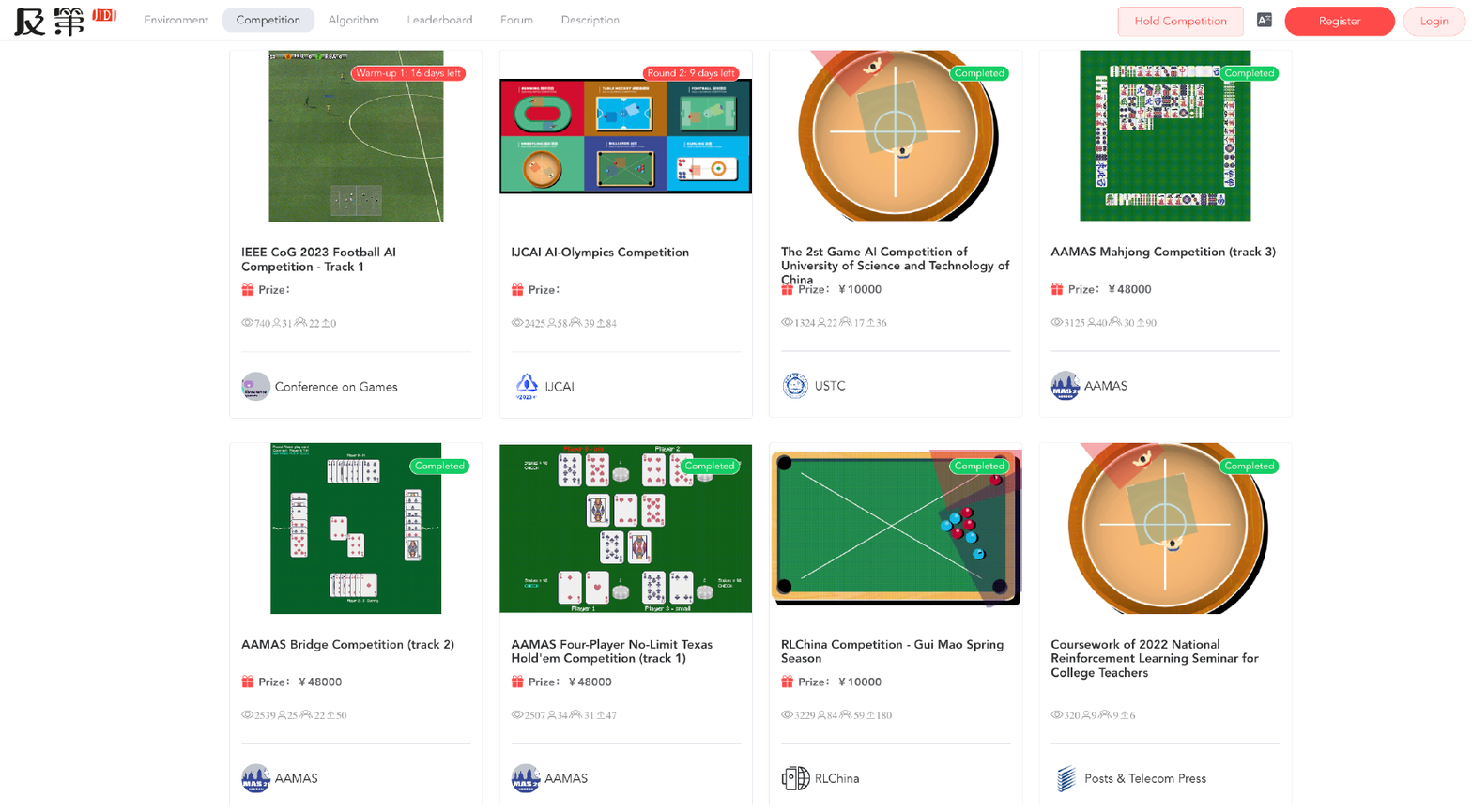}
        \caption{JIDI competition-specific ranking}\label{fig:jidi website 2}
    \end{subfigure}
    \medskip
    \begin{subfigure}{0.45\linewidth}
        \centering
        \includegraphics[width=\linewidth]{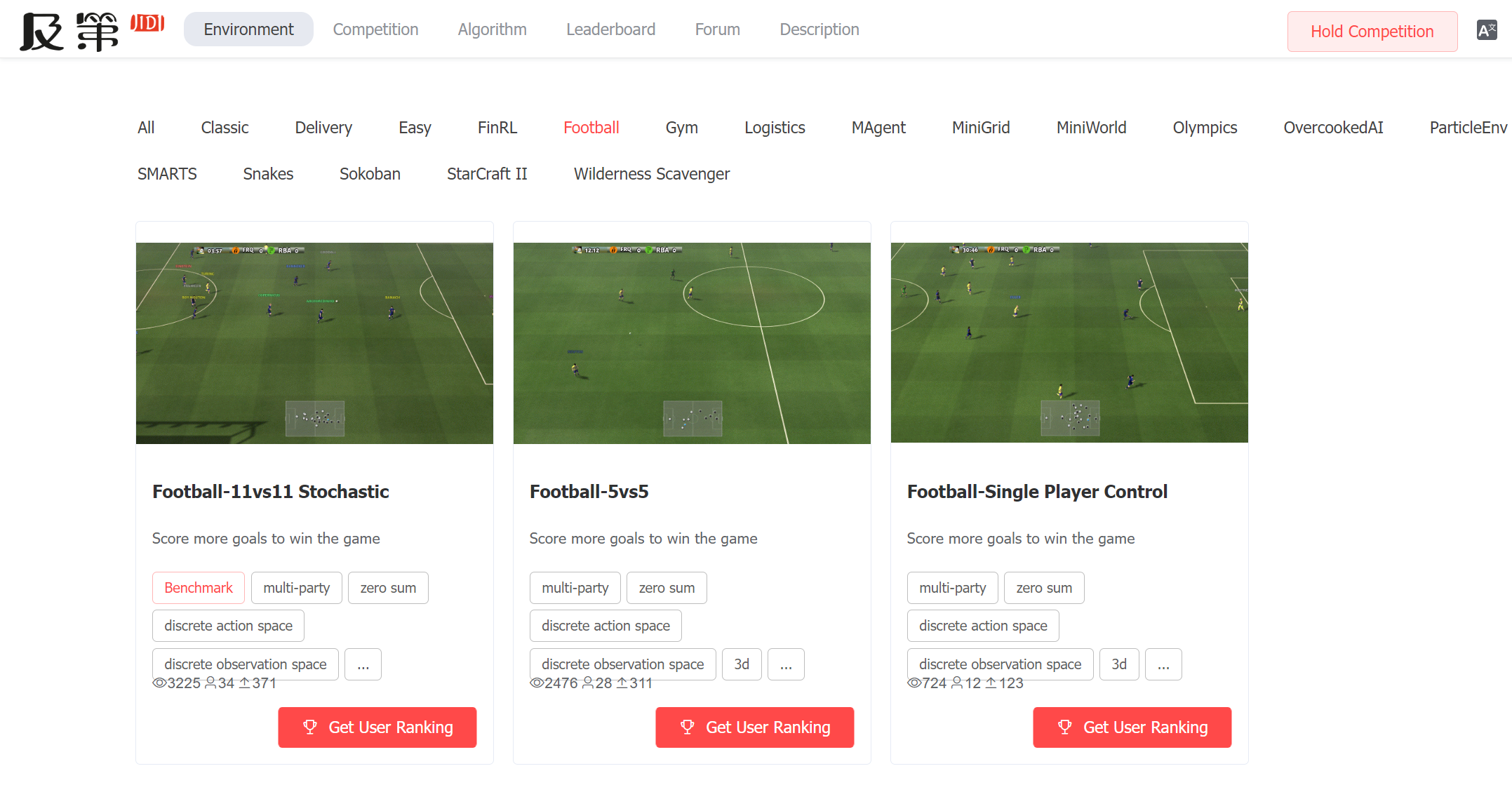}
        \caption{JIDI GRF ranking on \textit{11 vs 11} full game, \textit{5 vs 5} full game and single agent full-game scenarios}\label{fig:jidi website 3}
    \end{subfigure}
    \begin{subfigure}{0.45\linewidth}
        \centering
        \includegraphics[width=\linewidth]{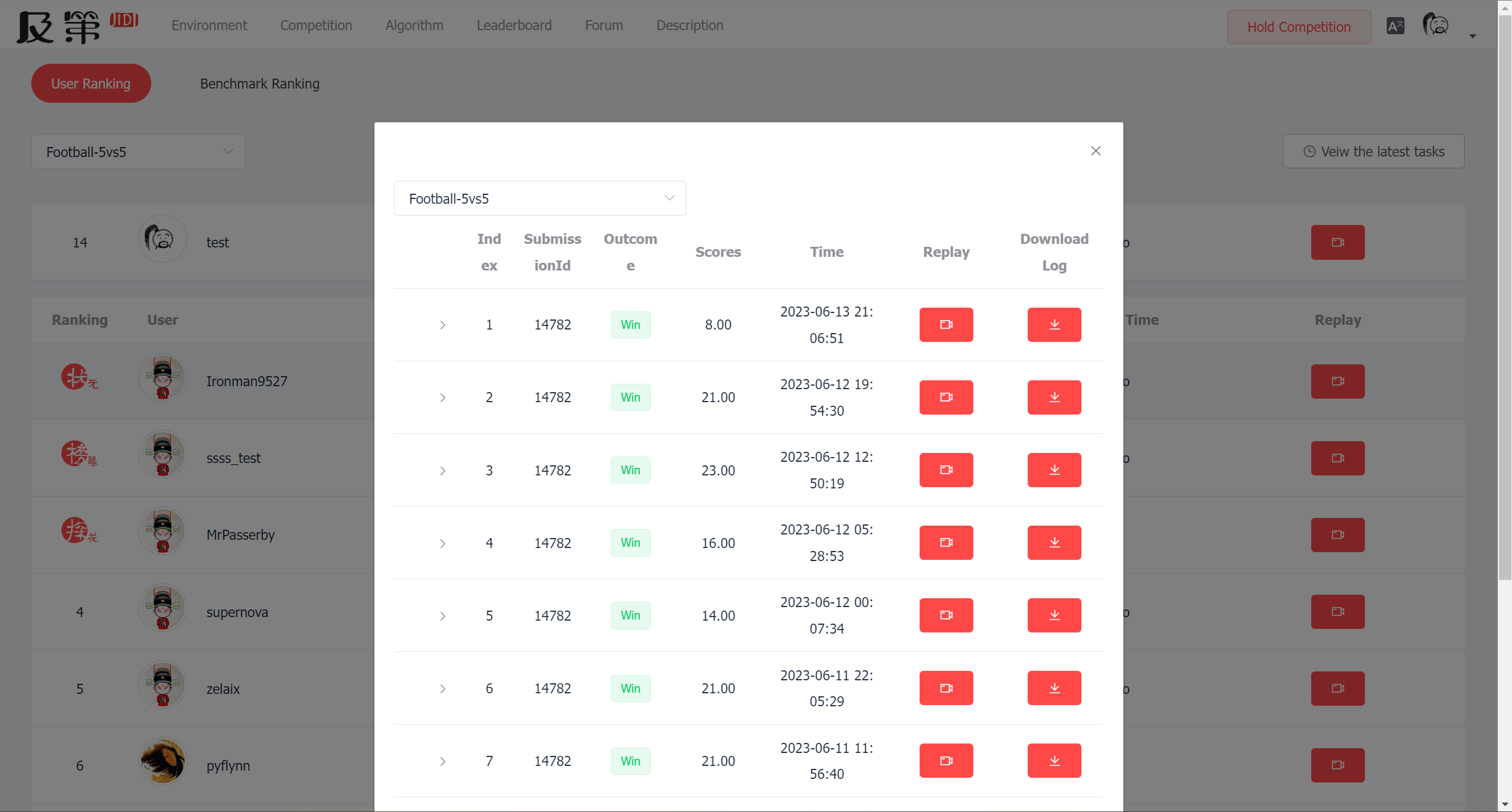}
        \caption{JIDI GRF ranking details page}\label{fig:jidi website 4}
    \end{subfigure}
    \caption{JIDI Online Evaluation Platform. \textbf{(a)} shows the web page of Daily Evaluation where JIDI has a specific ranking for each environment we support, such as the chess and card game set shown in the figure. By clicking the \texttt{Get User Ranking} button you will be directed to the corresponding ranking web page. \textbf{(b)} shows the web page of Competition Evaluation where JIDI holds multiple Kaggle-like AI competitions that are open to the public. JIDI also offer coursework competition for a university program. \textbf{(c)} shows three particular rankings we design for the Google Research Football environment including \textit{11 vs 11}, \textit{5 vs 5} multi-agent full game and \textit{11 vs 11} single-agent full game. \textbf{(d)} shows the detailed page for each ranking. Participants can check their match results and download a replay for offline evaluation. They can also do offline evaluation using the single-step visual debugger we have provided (See Appendix \ref{appendix:debugger}). Website for JIDI: \url{www.jidiai.cn}}
    \label{fig:jidi webpage}
\end{figure}

The JIDI Open-Source Evaluation Platform offers the JIDI GRF online ranking as part of its comprehensive services. Recognizing the lack of public leaderboards in the field of decision-making, JIDI aims to address this issue by providing online evaluation and ranking services for a wide range of reinforcement learning environments. Currently, JIDI supports nearly 100 environments, encompassing both single-agent games like classic-control games in the \textit{gym} framework, as well as multi-agent games such as \textit{Texas Hold'em} and Google Research Football (Figure \ref{fig:jidi website 1} and \ref{fig:jidi website 3}).

The platform offers two types of evaluation: Daily Evaluation and Competition Evaluation. In the Daily Evaluation, users can submit their agents at any time, triggering the background evaluation processes. This allows for continuous assessment and ranking of agents based on their performance. On the other hand, Competition Evaluation usually associates with academic conferences or university coursework. The background evaluation processes are initiated once the deadline is reached. Figure \ref{fig:jidi website 2} provides a glimpse of the JIDI AI competitions, which have attracted a significant number of participants.

JIDI supports evaluation on Google Research Football scenarios for both types of evaluation. In addition to ranking and scores displayed on the leaderboard, users have access to replays and logs of their matches against other submitted agents. This functionality, depicted in Figure \ref{fig:jidi website 4}, enables users to download and analyze their matches, facilitating the identification of weaknesses in their policies. By competing against a diverse range of opponents submitted by different users, participants can gain valuable insights, leading to the improvement of their policies and algorithms.


\end{document}